\documentclass{article}
\usepackage{hyperref}
\pdfoutput=1
\usepackage[a4paper, total={6in, 8in}]{geometry}
\usepackage[utf8]{inputenc}
\usepackage{amsmath}
\usepackage{amssymb}
\usepackage{amsthm}
\usepackage{cite}
\usepackage{xcolor}
\usepackage{graphicx}
\usepackage{subcaption}
\usepackage{enumerate}
\usepackage{float}
\usepackage{comment}
\usepackage{titlesec}
\usepackage[ruled,vlined]{algorithm2e}
\SetKw{KwBy}{by}
\SetAlFnt{\small}
\titlelabel{\thetitle.\quad}
\usepackage{bm}
\newcommand\emilie[1]{\textcolor{black}{#1}}

\theoremstyle{mydef}
\newtheorem{mydef}{Definition}
\theoremstyle{remark}
\newtheorem*{remark}{Remark}

\title{Rank-based Bayesian variable selection for genome-wide transcriptomic analyses}

\author{Emilie Eliseussen$^1$ \and Thomas Fleischer$^2$ \and Valeria Vitelli$^1$}
\date{%
    $^1$Oslo Centre for Biostatistics and Epidemiology, Department of Biostatistics, University of Oslo, Oslo, Norway\\%
    $^2$Department of Cancer Genetics, Institute for Cancer Research, Oslo University Hospital, Oslo, Norway\\[2ex]%
    E-mail: e.e.odegaard@medisin.uio.no\\[2ex]%
    \today 
}

\begin{document}

\maketitle

\begin{abstract}
    Variable selection is crucial in high-dimensional omics-based analyses, since it is biologically reasonable to assume only a subset of non-noisy features contributes to the data structures. However, the task is particularly hard in an unsupervised setting, and a priori ad hoc variable selection is still a very frequent approach, despite the evident drawbacks and lack of reproducibility. We propose a Bayesian variable selection approach for rank-based unsupervised transcriptomic analysis. Making use of data rankings instead of the actual continuous measurements increases the robustness of conclusions when compared to classical statistical methods, and embedding variable selection into the inferential tasks allows complete reproducibility. Specifically, we develop a novel extension of the Bayesian Mallows model for variable selection that allows for a full probabilistic analysis, leading to coherent quantification of uncertainties. Simulation studies demonstrate the versatility and robustness of the proposed method in a variety of scenarios, as well as its superiority with respect to several competitors when varying the data dimension or data generating process. We use the novel approach to analyse genome-wide RNAseq gene expression data from ovarian cancer patients: several genes that affect cancer development are correctly detected in a completely unsupervised fashion, showing the usefulness of the method in the context of signature discovery for cancer genomics. Moreover, the possibility to also perform uncertainty quantification plays a key role in the subsequent biological investigation.
\end{abstract}

\section{Introduction}\label{sec:intro}
Over the last decades, technological advances have generated an explosion of data in a variety of fields. High-throughput technologies have made -omics data more accessible to researchers, allowing for a systematic exploration of the genetic and epigenetic basis of cancer. However, -omics data pose several significant analytic challenges, including high-dimensionality, high complexity and non-normality and thus, standard statistical approaches are not applicable. 
To overcome these complications in high-dimensional omics-based analyses, it is considered biologically reasonable to assume that only a small piece of information is relevant for prediction or subtyping \cite{alexandrov2013, hoadley2014, hoadley2018}, so that a low-dimensional solution can also lead to better results interpretability. However, a frequent approach to variable selection for -omics applications consists in performing a priori ad hoc selection of genes according to in-sample statistics or outer literature information, a strategy that heavily affects the solidity and reproducibility of results \cite{ickstadt2018}. Moreover, it has been established in the last decades, with the explosion of high-dimensional data methods, that variable selection should be performed jointly with the main inferential tasks, to properly estimate the uncertainties and the impact of selection on inference results \cite{buhlmann2011}. Among unsupervised methods for variable selection in the context of -omics data, Bayesian model-based approaches are highly suitable, due to their capability of propagating all uncertainties. While unsupervised methods may lead to inaccurate estimates because of the low signal-to-noise ratio \cite{ideker2011}, especially for high-throughput -omics data, the Bayesian approach allows for incorporating prior biological knowledge when estimating relevant genes from the data. Current methods for Bayesian variable selection for high-dimensional unsupervised problems include probabilistic sparse principal component analysis \cite{bouveyron2018, agrawal2020}, and Bayesian graphical structure learning \cite{li2010, kundu2018}, all with applications to -omics data. In the case of ranking data, there currently exists no Bayesian unsupervised methods capable of tackling the dimensions of -omics data.

Converting continuous data to rankings and then performing rank-based analyses is a practice that has gained much popularity in statistical genomics \cite{deng2014}. Different -omics data layers can have incomparable absolute scales, and the use of ranks leads to robust inference \cite{asfari2014}. Moreover, analysing rankings instead of continuous variables allows for easy integration of multiple heterogeneous data, and enhances the reproducibility and interpretability of results. The analysis of rank and preference data has recently seen a growing interest in the machine learning community \cite{alvo2014}, with the Plackett-Luce \cite{luce1959, plackett1975} and Mallows model \cite{mallows1957} being among the most commonly used models. The Mallows model is an exponential family-type model on the permutation space, where the probability density depends on a distance between rankings. Compared to Plackett-Luce, the Mallows model shows greater flexibility in the choice of the distance between permutations, and it is also more versatile in adapting to different kinds of data. As a probabilistic model for rank data, the Mallows model enjoys great interpretability, model compactness, inference and computational efficiency. Our current work is based on the Bayesian Mallows Model proposed in \cite{vitelli2018}, whose implementing solution, \texttt{BayesMallows}, is described in \cite{sorensen2019}: \texttt{BayesMallows} already provides an inferential computationally feasible approach for the most important choices of distance among permutations, and it shows good accuracy when compared to competitors on datasets of moderate size \cite{liu2019}. However, for this method to be applicable to the common data dimensions in -omics applications, variable selection is crucial. Moreover, from a purely modeling perspective, imposing the Mallows data generating process on the full (genome-wide) ranked list of genes seems highly unrealistic: RNAseq measurements of moderately expressed genes tend to be very noisy, thus making a low-dimensional solution more suited to this kind of data. Imposing a ranking model only on a selection of genes allows focusing on consistently highly expressed genes, since those will be selected as highly ranked items, thus allowing to detect genome-wide high-quality signals in a more robust way. More generally, this leads to an unsupervised variable selection procedure that can learn from the available rankings to distinguish relevant items from background ones in a high-dimensional setting, as well as provide an estimate of the position in the ranking of the selected relevant items. 

The aim of this paper is to develop a novel extension of the Bayesian Mallows model, the lower-dimensional Bayesian Mallows Model (lowBMM), providing a low-dimensional solution for rank-based unsupervised variable selection in high-dimensional settings. To the best of our knowledge, the only proposal similar in scope to ours is \cite{zhu2021}, proposing a partition-Mallows model (PAMA) for rank aggregation that can distinguish relevant items, as well as ranking the relevant ones. This method can handle partial lists, incorporate any available covariate information and can estimate the quality of rankers, as it is based on BARD \cite{deng2014}, which assigns aggregated ranks based on the posterior probability that a certain item is relevant. However, PAMA only considers the Kendall distance for the Mallows model, thus allowing the partition function to have a closed form, while the Bayesian Mallows model can use any distance. Additionally, PAMA is unable to perform individual-level inference, a typical advantage of the Bayesian Mallows model, and it is tested on a relatively complete selection of simulated scenarios, however lacking large-data examples. Concerning rank-based variable selection in -omics analyses, the recent work in \cite{cui2021} converts microbiome data to rankings and develops a Bayesian variable selection model using ranks to identify the best covariates for modeling the outcome. However, this latter method is fully supervised, and moreover, common for both of these methods is the lack of applicability to high-dimensional settings: all data examples are of much smaller dimensions compared to the typical dimension of -omics data ($>10^4$). Other Bayesian implementations to analyse rank data have been proposed to reduce computational burden \cite{koop1994, yao1999, yu2000}, to merge rankings from multiple sources \cite{deng2014, johnson2002}, and to combine rank data with other types of data \cite{barney2015}. \emilie{There exist other methods which are not directly related to lowBMM, but can handle rank data, such as the Markov chain-based methods in \cite{lin2010, linding2009} as well as the Mallows based methods in \cite{irurozki2019, li2020}. Simulation studies show that the proposed approach is superior to existing methods in high dimensions, and has no competitor in \emph{ultra}-high dimensions.}


The remainder of the paper is laid out as follows: in Section \ref{sec:model} we describe the novel lowBMM for variable selection, and its computational implementation. Several simulation studies are described in Section \ref{sec:simulations}, aimed at demonstrating the versatility and robustness of the novel method under different scenarios. \emilie{Included here is also a sensitivity study on the tuning parameters involved in the model, and a comparison with existing methods}. In Section \ref{sec:application}, we present an application of this modeling approach to RNAseq gene expression data from ovarian cancer patients. We conclude with a brief summary and discussion in Section \ref{sec:discussion}.

\section{Lower-dimensional Bayesian Mallows Model}\label{sec:model}
We briefly present the Bayesian Mallows Model (BMM) for complete data as introduced in \cite{vitelli2018} in Section \ref{sec:bmm_mod}, and outline its extension for unsupervised variable selection, lowBMM, in Section \ref{sec:lowbmm_mod}. A description of the inferential procedure can be found in Section \ref{sec:mh}, while Section \ref{sec:off_line_alpha_notation} describes the off-line estimation of the scale parameter $\alpha$.

\subsection{Bayesian Mallows model for complete data}\label{sec:bmm_mod}
Consider a finite set of $n$ items denoted as $\mathcal{A}=\{A_1,A_2,...,A_n\}$. \emilie{A complete ranking is a mapping $\bm{R}: \mathcal{A} \to \mathcal{P}_n$ that assigns a rank $R_i\in\{1,\ldots,n\}$ to each item in $\mathcal{A}$ according to some specified feature\footnote{This feature is unobserved in most recommender system applications of the model, where we assume the users provide their ranking/preference according to the feature levels in their mind. On the other hand, when continuous data are converted to ranks for better modeling (e.g., in transcriptomic data analyses), the feature corresponds to the observed continuous measurement.}, with the item ranked 1 being the most preferred (i.e., showing the largest value of the feature), and the item ranked $n$ being the least preferred (i.e., showing the smallest value of the feature). Thus, a complete ranking $\bm{R}=(R_1,\ldots,R_n)$ lies within the space of $n$-dimensional permutations $\mathcal{P}_n$. Let us assume that $N$ assessors have provided complete rankings of the $n$ items, $\mathbf{R}_j= \{R_{1j},R_{2j},..., R_{nj}\}$, $j=1,...,N$. The Mallows model \cite{mallows1957} is a probabilistic model for complete rankings $\bm{R}\in\mathcal{P}_n$, taking the form
\begin{equation}\label{eq:Mallows}
    P(\bm{R} | \alpha, \bm{\rho}) = \frac{1}{Z_n(\alpha, \bm{\rho})}\exp\left\{-\frac{\alpha}{n} d_n(\bm{R}, \bm{\rho})\right\}1_{\mathcal{P}_n}(\bm{R}),
\end{equation}
where $\alpha>0$ is a positive scale parameter, $\bm{\rho} \in \mathcal{P}_n$ is the latent consensus ranking, $Z_n(\alpha, \bm{\rho})$ is the model partition function, $1_S(\cdot)$ is the indicator function of the set $S$, and finally $d_n(\cdot, \cdot):\mathcal{P}_n\times\mathcal{P}_n\rightarrow[0,+\infty)$ is a right-invariant\footnote{A right-invariant metric $d(\cdot,\cdot)$ is such that, for any $r_1, r_2 \in \mathcal{P}_n$, it holds: $d(r_1,r_2)=d(r_1 r_2^{-1}, \bm{1}_n)$, $\bm{1}_n= \{1,2,...,n\}$ \cite{diaconis1988}} distance function between two rankings. We have here chosen to let the distance function explicitly depend on $n$, so that $d_n(\cdot,\cdot)$ denotes the distance computed on the complete set of items.}

Several possibilities for choosing this distance function exist, such as the footrule distance, the Spearman distance, and the Kendall distance \cite{diaconis1988}. In this paper we choose to use the footrule distance, defined as $d_n(\bm{R}, \bm{\rho})=\sum_{i=1}^n |R_i - \rho_i|$, \emilie{the equivalent of an $\ell^1$ measure between rankings. The choice of this distance is motivated by its greater computational efficiency compared to its competitors, such as the Kendall distance, which is more computationally intensive \cite{sorensen2019}. Moreover, the popularity of the Kendall lies in the fact that a closed form of the model partition function $Z_n(\alpha, \bm{\rho})$ exists in this case; however, we have previously proposed computational solutions to be able to use other distances \cite{sorensen2019}, while also showing that the Kendall is often less accurate \cite{liu2019}. The Spearman distance is also problematic in our case, as it tends to largely penalize extreme data, being the equivalent of an $\ell^2$ measure between rankings. On the other hand, the equivalent of the $\ell^1$ between continuous data seems more suited to -omics applications.}

Since the footrule distance is right-invariant, the partition function $Z_n(\alpha, \bm{\rho})$ is independent of $\bm{\rho}$ and only dependent on $\alpha$. Moreover, we assume in the remainder of the paper that $\alpha$ is fixed and known, and we provide strategies to tune this parameter off--line. \emilie{This simplification is done to avoid having to compute the normalizing constant $Z_n(\alpha)$ at each iteration of the Markov Chain Monte Carlo (MCMC) algorithm (see subsection \ref{sec:mh}), thus greatly lightening the computational burden of the algorithm, at the cost of the tuning of a single scalar parameter, which proves to be straightforwardly carried out in practice (see subsection \ref{sec:off_line_alpha_notation} for further details). Nonetheless, the performance of the approximated approaches to estimate the normalizing constant $Z_n(\alpha)$ introduced in \cite{vitelli2018} greatly depends on the size of $n,$ with the performance degrading for decreasing $n$, and with the approximation becoming increasingly computationally demanding when $n \to \infty $.} Thus, fixing $\alpha$ simplifies the inferential procedure quite substantially. 

Given the model in (\ref{eq:Mallows}) and the choices specified above, the likelihood associated to the observed rankings $\mathbf{R}_1, \ldots, \mathbf{R}_N$ under the Mallows model takes the form:
\begin{align*}
    P(\bm{R}_1, ..., \bm{R}_N | \bm{\rho}) = \frac{1}{Z_n(\alpha)} \exp \left\{ -\frac{\alpha}{n} \sum_{j=1}^N d_n(\bm{R}_j, \bm{\rho})\right\} \prod_{j=1}^N 1_{\mathcal{P}_n}(\bm{R}_j).
\end{align*}
In order to use the Bayesian version of this model as introduced in \cite{vitelli2018}, we need to specify a prior for $\bm{\rho}$. In this paper, we set a uniform prior $\pi(\bm{\rho}) = \frac{1}{n!}1_{\mathcal{P}_n}(\bm{\rho})$ in the space $\mathcal{P}_n$ of $n$-dimensional permutations. The posterior distribution for $\bm{\rho}$ then becomes
\begin{align*}
    P(\bm{\rho} | \bm{R}_1, ..., \bm{R}_N ) \propto \frac{\pi(\bm{\rho})}{Z_n(\alpha)^N} \exp \left\{-\frac{\alpha}{n}\sum_{i=j}^N d_n(\bm{R}_j,\bm{\rho)}\right\}.
\end{align*}
To perform inference, \cite{vitelli2018} proposed a MCMC algorithm based on a Metropolis-Hastings (MH) scheme (see \cite{sorensen2019} for details on the implementation). 

\subsection{Lower-dimensional Bayesian Mallows model (lowBMM) for variable selection}\label{sec:lowbmm_mod}
\emilie{We here introduce a generalization of the BMM presented in Section \ref{sec:bmm_mod} to the scopes of variable selection, which in the context of models for rankings assumes the form of a selection of the \emph{relevant} items, i.e., the items worth being ranked. We are thus thinking of generalizing BMM to make it suitable to situations in which $n$ is large or \emph{ultra-large}, and therefore it becomes unrealistic to assume that all $n$ items can be completely ranked.}
We define $\mathcal{A}^*=\{A_{i_1},..., A_{i_{n^*}}\}$ as an $n^*-$dimensional subset of the original set of items $\mathcal{A}$, with $n^* << n$ and $\mathcal{A}^* \subset \mathcal{A}.$
When defining the lower-dimensional Bayesian Mallows model, the underlying assumption is that only a portion of the data follows the Mallows distribution, while the rest is unranked: this is a very realistic assumption, especially in the large $n$ setting. We then formally assume that items in $\mathcal{A}^*$ are such that $R_j|_{\mathcal{A}^*},$ $j=1,\ldots,N,$ follows a Mallows model over the permutation space of dimension $n^*$, where with the notation $R_j|_{S}$ we indicate the restriction of an $n$-dimensional ranking to only the items belonging to the set $S,$ which is also a ranking in dimension $|S|.$ The remaining items in $\mathcal{A} \setminus \mathcal{A}^*$ are simply irrelevant to the scopes of the analysis, and therefore show no specific pattern in the data. This is equivalent to assuming that $R_j|_{\mathcal{A} \setminus \mathcal{A}^*},$ $j=1,\ldots,N,$ is uniformly distributed over the space of permutations of dimension $n-n^*.$ 

$\mathcal{A}^*$ is then the novel model parameter characterising the lower-dimensional Bayesian Mallows model, the variable selection setting for the Bayesian Mallows model. Not only, lowBMM is assuming the data follow a Mallows model only on the lower-dimensional set of items, so that the consensus ranking parameter is now defined as $\bm{\rho} \in \mathcal{P}_{n^*}$. Therefore, the likelihood of lowBMM is defined on a lower-dimensional space of dimension $n^*$, and it takes the form:
\begin{align*}
    P(\mathbf{R}_1, ..., \mathbf{R}_N | \bm{\rho}, \mathcal{A}^*) = \frac{1}{Z_{n^*}(\alpha)} \exp \left\{ -\frac{\alpha}{n^*} \sum_{j=1}^N d_{\mathcal{A}^*}(\mathbf{R}_j, \bm{\rho})\right\} \prod_{j=1}^N \left\{ U_{\mathcal{P}_{n-n^*}} (\mathbf{R}_j|_{\mathcal{A}\setminus\mathcal{A}^*}) 1_{\mathcal{P}_n}(\mathbf{R}_j) \right\} 
\end{align*}
where $d_{\mathcal{A}^*}(\mathbf{R}_j, \bm{\rho}) := d_{n^*}(\mathbf{R}_j|_{\mathcal{A}^*},\bm{\rho})$. This is the same distance as $d_n(\cdot,\cdot)$ but restricted to the $n^*-$dimensional set of items included in $\mathcal{A}^*$. The vector $\mathbf{R}_j|_{\mathcal{A}^*}$ is the restriction of $\mathbf{R}_j$ to the set $\mathcal{A}^*$: it thus defines a partial ordering from which an $n^*-$dimensional permutation can be derived. Moreover, $U_S$ is the uniform distribution over the domain $S,$ meaning that the term $U_{\mathcal{P}_{n-n^*}} (\mathbf{R}_j|_{\mathcal{A}\setminus\mathcal{A}^*})$ refers to the assumption that $\mathbf{R}_j|_{\mathcal{A}\setminus\mathcal{A}^*}$ includes only noisy unranked data. 

\medskip

\begin{remark}
\emilie{The way the modeling framework is defined does not result in a variable selection procedure that selects the top-ranked items only. Indeed, the items in $\mathcal{A}^*$ do not need to be those that are most often top-$n^*$ ranked in the data, even if this is the most intuitive solution to a lower-dimensional ranking problem. The relevant items could also show a ``consistency'' pattern, i.e., be often ranked in the same respective order across assessors: this latter possibility constitutes a valid data generating process for lowBMM, as proved in the simulation study in Section \ref{sec:sim_data_processes}, as the only requirement for lowBMM to be a suitable model for the data is that the ranks of the items follow a Mallows model in the lower-dimensional space $\mathcal{P}_{n^*}$.} In other words, the model is able to pick out the $n^*$ items on which the assessors agree the most, \emilie{as the model assigns the largest probability to the items whose ranks} show the smallest distance to the consensus, which is constrained to be an $n^*$-dimensional permutation. \emilie{In conclusion, the results from lowBMM will depend on the specific dataset: if there exists a top-rank solution in the data, such a solution will be estimated by lowBMM as the one with the largest posterior. On the other hand, if there exists a pattern in the data showing certain items consistently ranked in a specific order (not necessarily top-ranked), lowBMM will give those items the largest probability in the marginal posterior distribution of $\mathcal{A}^*$, and rank them accordingly in the marginal posterior distribution of $\bm{\rho}$.}
\end{remark}

\medskip

Since the inferential approach is Bayesian, we have to decide priors for all parameters. As we did previously, we set a uniform prior on $\bm{\rho}$, however restricted to the $n^*-$dimensional space of permutations of elements of $\mathcal{A}^*$: $\pi(\bm{\rho}|\mathcal{A}^*)=\frac{1}{n^*!}1_{\mathcal{P}_{n^*}}(\bm{\rho})$. We also set a uniform prior for $\mathcal{A^*}$ over $\mathcal{C}$, $\pi(\mathcal{A}^*)=\frac{1}{|\mathcal{C}|} 1_{\mathcal{C}}(\mathcal{A^*})$, where we define $\mathcal{C}$ as the collection of all $\binom{n}{n^*}$ possible sets of items of dimension $n^*$ chosen from a set of dimension $n$. The posterior distribution of the variable selection for the Bayes Mallows model can then be written as
\begin{align}\label{eq:post}
    P(\bm{\rho}, \mathcal{A}^* | \mathbf{R}_1, ..., \mathbf{R}_N)  \propto & \ \pi(\mathcal{A}^*)\pi(\bm{\rho}|\mathcal{A}^*) \frac{1}{Z_{n^*}(\alpha)} \exp\left\{ -\frac{\alpha}{n^*} \sum_{j=1}^N d_{\mathcal{A}^*}(\mathbf{R}_j, \bm{\rho})\right\} \cdot \nonumber \\
    & \cdot \prod_{j=1}^N \left\{ U_{\mathcal{P}_{n-n^*}}(R_j |_{\mathcal{A}\setminus\mathcal{A}^*}) 1_{\mathcal{P}_n}(\mathbf{R}_j) \right\},
\end{align}
which, by removing the terms not depending on any of the model parameters and by assuming the data are proper rankings, can be simplified to
\begin{equation}\label{eq:postRed}
P(\bm{\rho}, \mathcal{A}^* | \mathbf{R}_1, ..., \mathbf{R}_N)  \propto \exp\left\{ -\frac{\alpha}{n^*} \sum_{j=1}^N d_{\mathcal{A}^*}(\mathbf{R}_j, \bm{\rho})\right\} 1_{\mathcal{P}_{n^*}}(\bm{\rho}) 1_{\mathcal{C}}(\mathcal{A^*}).
\end{equation}

Note that, from the posterior distribution in (\ref{eq:postRed}), marginal posterior distributions of both model parameters, $\bm{\rho}$ and $\mathcal{A}^*,$ can be easily derived. \emilie{Often one is interested in computing posterior summaries of such distributions.} By inspecting the marginal posterior distribution of $\mathcal{A}^*$ estimated by lowBMM, one can for instance a posteriori select the items in $\mathcal{A}$ that maximise this marginal posterior under all possible model reductions of dimension $n^*$, thus practically attaining variable selection via the marginal posterior mode (i.e., the maximum a posteriori -- MAP) of $\mathcal{A}^*$.
\emilie{However, computing the MAP might not be the best way of summarizing the marginal posterior distributions of both $\bm{\rho}$ and $\mathcal{A}^*$, in addition to being a computationally intensive procedure for larger $n$ and $n^*$. Instead, we introduce novel ways for providing posterior summaries of both the consensus ranking of the relevant items, called $\bm{\hat{\rho}}_{\mathcal{A}^*}$, and of the set of relevant items, called $\hat{\mathcal{A}}^*$; these approaches are described in Section \ref{sec:mh_postprocess}.}

Furthermore, the inspection of the marginal posterior distributions of $\bm{\rho}$ and $\mathcal{A}^*$ has much wider implications \emilie{than simply obtaining posterior summaries}: for example, by inspecting the marginal posterior distribution of $\mathcal{A}^*$ one can decide whether variable selection is appropriate to the data at hand (few items are clearly assigned rankings while the rest shows high uncertainty), or rather items are a posteriori ranked in blocks (several items share the same rank with large probability, and only a respective ordering between groups of items can be properly estimated). Scope of the simulation studies in Section \ref{sec:simulations} is to assess the correct behaviour of the model in these different scenarios, together with its gain in efficiency in the large $n$ context. 

\medskip

\begin{remark}
\emilie{Note that we have assumed $n^*$ to be fixed and known. Estimating $n^*$ together with the other model parameters would imply a model selection step, which would make the inferential task substantially more challenging, or require to completely change our inferential approach, for instance towards the use of stick-breaking priors \cite{ishwaran2001gibbs}. Both possibilities are completely out of the scope of the present paper and are left to future speculations. On the other hand, the choice of fixing $n^*$ is typically not problematic in practice, as one can often tune it in connection to the real application (e.g. in genomics, biologists are often interested in a specific ``number'' of relevant genes, even more if we consider that each gene-regulating biological pathway includes on average a hundred genes \cite{ramanan2012pathway}). Another practical possibility is to simply choose $n^*$ as large as allowed by the computing resources at hand. This latter approach is also supported by the simulation studies (see Section \ref{sec:sim_sens_study}), where we realized that choosing $n^*$ larger than necessary would result in a solution including and automatically detecting the lower dimensional relevant subset.}
\end{remark}

\subsection{Metropolis-Hastings algorithm for inference in the lowBMM}\label{sec:mh}
In order to obtain samples from the posterior in equation \eqref{eq:post}, we set up a MH-MCMC scheme. The algorithm iterates between two steps: in one step the consensus ranking $\bm{\rho}$ given the current set $\mathcal{A^*}$ is updated, and in the other the set $\mathcal{A^*}$ is updated given the current consensus ranking. 

In the first step of the algorithm, we propose a new consensus ranking $\bm{\rho}'\in\mathcal{P}_{n^*}$ using the ``leap-and-shift'' proposal distribution described in \cite{vitelli2018}. The acceptance probability for updating $\bm{\rho}$ in the MH algorithm is
\begin{align} \label{eq:accept_prob_rhostar}
    \min\left\{ 1, \frac{P_l(\bm{\rho}|\bm{\rho}')}{P_l(\bm{\rho}'|\bm{\rho})}\exp \left[ -\frac{\alpha}{n^*} \left( \sum_{j=1}^N d_{\mathcal{A}^*}(\mathbf{R}_j,\bm{\rho}') - \sum_{j=1}^N d_{\mathcal{A}^*}(\mathbf{R}_j,\bm{\rho})\right)\right] \right\}.
\end{align}
In equation (\ref{eq:accept_prob_rhostar}) above, $P_l$ denotes the probability mass function associated to moving from one $n^*$-dimensional ranking vector to another according to the leap-and-shift proposal. The parameter $l$ denotes the number of items perturbed in the consensus ranking $\bm{\rho}$ to get a new proposal $\bm{\rho}'$, and is used to tune the acceptance probability. To compute the distances restricted to $\mathcal{A}^*$, the $\bm{R}_j$'s needs to be updated so that they correspond to the items in the current set $\mathcal{A}^*$.  

The second step of the algorithm updates the set $\mathcal{A}^*$. We propose a new set $\mathcal{A}^*_\textnormal{prop}$ by perturbing $L \in \{1,...,n^*\}$ elements in the current $\mathcal{A}^*$, \emilie{selected with uniform probability.} The $L$ items are swapped with $L$ items from the set $\mathcal{A} \setminus \mathcal{A^*}$\emilie{, again uniformly.} Therefore, we can formally write the proposal distribution as $q(\mathcal{A}_\textnormal{prop}^*| \mathcal{A}^*) = |\mathcal{C}_L^{\textnormal{in}} |^{-1} |\mathcal{C}_L^{\textnormal{out}}|^{-1} 1_{\mathcal{C}_L^{\textnormal{in}}}(\mathcal{A}_\textnormal{prop}^* \setminus \mathcal{A}^*) 1_{\mathcal{C}_L^{\textnormal{out}}}(\mathcal{A}^* \setminus \mathcal{A}_\textnormal{prop}^*)$, where we define $\mathcal{C}_L^{\textnormal{in}}$ as the collection of all $\binom{n-n^*}{L}$ possible sets of items of dimension $L$ chosen from a set of dimension $n-n^*$, to be brought ``in'' to the reduced set $\mathcal{A}^*$ in order to obtain $\mathcal{A}^*_{\textnormal{prop}}$. Likewise, we define $\mathcal{C}_L^{\textnormal{out}}$ as the collection of all $\binom{n^*}{L}$ possible sets of items of dimension $L$ chosen from a set of dimension $n^*$, to be brought ``out'' of the reduced set $\mathcal{A}^*$ in order to obtain $\mathcal{A}^*_{\textnormal{prop}}$. Note that such proposal distribution is by definition symmetrical. The move from $\mathcal{A}^*$ to $\mathcal{A}^*_\textnormal{prop}$ is then accepted with probability:
\begin{align*}
    \min\left\{ 1, \frac{\pi(\mathcal{A}^*_\textnormal{prop}) P(\mathbf{R}_1,..., \mathbf{R}_N | \mathcal{A}^*_\textnormal{prop}, \bm{\rho}) q(\mathcal{A}^*| \mathcal{A}^*_\textnormal{prop})}{\pi(\mathcal{A}^*) P(\mathbf{R}_1,..., \mathbf{R}_N | \mathcal{A}^*, \bm{\rho}) q(\mathcal{A}^*_\textnormal{prop}| \mathcal{A}^*)} \right\}.
\end{align*}
Since the prior for $\mathcal{A}^*$ is uniform, the prior terms in the ratio simplify, as do the symmetrical proposal distributions. We are then left with a ratio of likelihoods, that can be written as
\begin{align} \label{eq:accept_prob_Astar}
    \min\left\{ 1, \exp \left[ -\frac{\alpha}{n^*} \left( \sum_{j=1}^N d_{\mathcal{A}^*_\textnormal{prop}}(\mathbf{R}_j,\bm{\rho}) - \sum_{j=1}^N d_{\mathcal{A}^*}(\mathbf{R}_j,\bm{\rho})\right)\right] \right\}.
\end{align}
Note that $d_{\mathcal{A}^*_\textnormal{prop}}(\mathbf{R}_j,\bm{\rho}) = d_{n^*}(\mathbf{R}_j|_{\mathcal{A}^*_\textnormal{prop}},\bm{\rho})$, i.e. the distance is computed on the restricted set of the current $\mathcal{A}^*_{\textnormal{prop}}$, and $\bm{R}_j$ is also restricted to the same set. Therefore, one needs to decide how to evaluate $\bm{\rho}$ on $\mathcal{A}^*_{\textnormal{prop}}$. There are multiple ways of doing this, the most simple being to randomly perturb the current $\bm{\rho}$ from the first step of the algorithm to accommodate the new $L$ items proposed for $\mathcal{A}^*_\textnormal{prop}$. That is, given the current $\bm{\rho}$, we discard the ranks corresponding to the $L$ items that were left out of $\mathcal{A}^*_\textnormal{prop}$, and we randomly re-assign those same ranks to the new items. 

The MH-MCMC algorithm described above is summarized in Algorithm \ref{alg:mcmc_item}. 

\medskip

\begin{algorithm}[H]
\SetAlgoLined 
\scriptsize
\textbf{input:} $\mathbf{R}_1,..,\mathbf{R}_N$, $\alpha$, $d(\cdot, \cdot)$, $l$, $L$, $M$ \\
\textbf{output:} posterior distributions of $\mathbf{\rho}$ and $\mathcal{A}^*$ \\
 \textbf{Initialization:} randomly generate $\mathbf{\rho}_0$ and $\mathcal{A}^*_0$ \\
  \For{$m \gets 1$ \KwTo $M$}{
    \textbf{M-H step: update $\mathbf{\rho}$} \\
        sample: $\rho' \sim \textnormal{LS}(\rho_{m-1}, l)$ restricted on $\mathcal{A}^*_{m}$ and $u \sim \mathcal{U}(0,1)$ \\
        compute: \textit{ratio} $\gets $ equation \eqref{eq:accept_prob_rhostar} with $\rho \gets \rho_{m-1}$, $\mathcal{A}^* \gets \mathcal{A}^*_{m-1}$ \\
        \eIf{$u <$ ratio}{$\rho_m \gets \rho'$}{$\rho_m \gets \rho_{m-1}$}
    
    \textbf{M-H step: update $\mathcal{A}^*$} \\
    sample: $L$ elements in $\mathcal{A}^*_{m-1}$ to get $\mathcal{A}^*_{\textnormal{prop}}$, and $u \sim \mathcal{U}(0,1)$ \\
        compute: \textit{ratio} $\gets $ equation \eqref{eq:accept_prob_Astar} with $\rho \gets \rho_{m}$, $\mathcal{A}^* \gets \mathcal{A}^*_{m-1}$ \\
        \eIf{$u <$ ratio}{$\mathcal{A}^*_m \gets \mathcal{A}^*_{\textnormal{prop}}$}{$\mathcal{A}^*_m \gets \mathcal{A}^*_{m-1}$}
    
    }
 \caption{MH-MCMC scheme for inference in lowBMM} \label{alg:mcmc_item}
\end{algorithm}

\subsubsection{Postprocessing of the MCMC results}\label{sec:mh_postprocess}

\emilie{In order to derive posterior summaries of the consensus ranking of the relevant items and of the set of relevant items, named $\bm{\hat{\rho}}_{\mathcal{A}^*}$ and $\hat{\mathcal{A}}^*$ respectively, we inspect the marginal posterior distribution of $\bm{\rho}$ and $\mathcal{A}^*$ (after burn-in). Suppose $M$ posterior samples are obtained: $\{\bm{\rho}_m, \mathcal{A}_m^*\}_{m=1}^M$ with $\bm{\rho}_m = \{\rho_{mi_1^m}, ..., \rho_{mi_{n^*}^m}\}$ and $\mathcal{A}_m^* = \{A_{mi_1^m}, ..., A_{mi_{n^*}^m}\}$. Given the samples $\{ \mathcal{A}^*_1,..., \mathcal{A}^*_M\}$, let $W \in \mathbb{R}^{M \times n}$ be such that $W_{mi}= 1_{\mathcal{A}_m^*}(A_i)$ for each item $A_i$, $i=1, ...,n$. For the computation of the posterior summaries, we first need the following definitions. }
\begin{mydef}
	Given a vector of real numbers $(x_1, ..., x_n)\in \mathbb{R}^n,$ its corresponding rank vector is obtained as follows:\\
	
	$rank(x_1, ..., x_n) = (r_1,..., r_n), \text{ such that}\quad r_i = \sum\limits_{j=1}^{n}\delta (x_i - x_j)$ for $i = 1, ..., n$, \\
	where $\delta(x) = \left \{
	\begin{aligned}
	&1, && \text{if } x\geq 0 \\
	&0, && \text{if } x < 0
	\end{aligned} \right.$
\end{mydef}  

\begin{mydef}\label{def:hps}
For $n^* \leq k \leq n$, the Highest Probability Set (HPS) of $\mathcal{A}^*$ is defined as follows: \\

 $\mathcal{A}' = \{A_i, i=1,...,n \, | \, rank(\bm{\bar{w}})_i \leq k\}$, \\

    where $\bm{\bar{w}}=(\bar{w}_1, ...,\bar{w}_n)$ and $\bar{w}_i = \frac{1}{M}\sum_{m=1}^{M} W_{mi}$.
\end{mydef}
\emilie{Let $\mathcal{A}'$ be the the HPS of $\mathcal{A}^*$. Based on $\mathcal{A}'$ we compute $\Bar{\mathbf{x}} \in \mathbb{R}^{|\mathcal{A}'|}$, $\Bar{x}_i = \frac{\sum_{m=1}^M \bm{\rho}_{mi} 1_{\mathcal{A}_m^*}(A_i)}{\sum_{m=1}^M 1_{\mathcal{A}_m^*}(A_i)}$ for all $A_i \in \mathcal{A}'$. We now quantify the two posterior summaries of $\bm{\rho}$ and $\mathcal{A}^*$ as follows: 
\begin{align}
 \hat{\mathcal{A}^*} = \left\{A_i \in \mathcal{A}' \, | \, rank(\Bar{\mathbf{x}}) \leq n^*\right\}, \hspace{1.5cm} \bm{\hat{\rho}}_{\mathcal{A}^*} = rank(\Bar{\mathbf{x}})|_{\mathcal{\hat{A}^*}}.
\end{align}
}
\emilie{Note that in all the figures in the paper reporting heatplots of the marginal posterior distribution of $\bm{\rho}$, the items on the x-axis will be ordered according to their respective ranking in $\bm{\hat{\rho}}_{\mathcal{A}^*}$.}

\emilie{In some cases, it might be of interest to obtain a ``top probability selection within the selection'' $\mathcal{\hat{A}}^*_{\textnormal{top}} \in \mathcal{\hat{A}}^*$ for further downstream analysis (for instance in cases when not only $n$ but also $n^*$ is quite large). The strategy for computing $\mathcal{\hat{A}}^*_{\textnormal{top}}$ consists in inspecting the probability distributions $P(A_i \in \textnormal{top-}K, i=1,...,n \, | \, A_i \in \mathcal{\hat{A}^*})$ for varying $K \leq n^*$. From these probability distributions, we would like to choose a $K$ that discriminates the items that have a high probability of being selected compared to the rest of the items. Once the $K$ is specified, the probability distribution will by construction be bimodal, and it will be possible to find a cut-off point $c$ that discriminates its two peaks well (this has explicitly been explained in the context of the case study presented in Section \ref{sec:application}, see Figure \ref{fig:ovarian_violin_hist}). The cut-off $c$ is then used as a lower probability bound for being included in $\mathcal{\hat{A}}^*_{\textnormal{top}}$. The result is a ``top probability selection'' of $|\mathcal{\hat{A}}^*_{\textnormal{top}}|$ items where $\mathcal{\hat{A}}^*_{\textnormal{top}} = \{ A_i \in \mathcal{\hat{A}^*} \, \textnormal{s.t.} \, P(A_i  \in \textnormal{top-}K, i=1,...,n \, | \, A_i  \in \mathcal{\hat{A}^*}) > c \}$.}

\normalsize
\subsection{Off-line estimation of $\alpha$} \label{sec:off_line_alpha_notation}
\emilie{To simplify inference by avoiding the computation of the normalizing constant, we decided to fix $\alpha$. Nonetheless,} we would still need an estimation of its most reasonable value when we are given a new ranking dataset. We propose one possible method for estimating such a value, inspired by \cite{liu2021}. 

As $\alpha$ describes how closely the rankings are distributed around the common consensus, or in other words, how similar the individual rankings are to each other, we can assume that a higher $\alpha$ indicates a higher ``similarity'' between assessors. Therefore, when assuming that the individual rankings of the assessors are drawn from a Mallows with parameter $\alpha$, a higher $\alpha$ results in a lower mean distance between the assessors. To estimate $\alpha$, we assume that the mean distance among $N$ assessors for a real ranking dataset should resemble the mean distance of a simulated ranking dataset of the same dimension, generated by drawing independent samples from the Mallows distribution. Precisely, suppose we want to estimate $\alpha$ for a real ranking dataset with $n$ items ranked by $N$ assessors: we can then generate several ranking datasets of dimension $(N,n)$ over a grid of $\alpha_0$ values, $\bm{R}^{\alpha_0}_1, \ldots, \bm{R}^{\alpha_0}_{N}$ (the chosen consensus ranking $\bm{\rho}_0$ does not matter). Then, we calculate the mean pairwise distance between assessors for each simulated dataset (i.e. for every $\alpha_0$), defined as:
\begin{align}\label{eq:meandistalpha}
    \Bar{d}_{\alpha_0} = \frac{1}{N(N-1)}\sum_{j=1}^{N} \sum_{k \neq j} d(\bm{R}^{\alpha_0}_j, \bm{R}^{\alpha_0}_k).
\end{align}
Finally, we compute the same mean distance between assessors as in (\ref{eq:meandistalpha}) for the dataset we are interested in, and choose the $\alpha_0$ value where the two distance measures intersect. 

The approach to estimate $\alpha$ described so far works for a complete ranking dataset where we assume that all $n$ items are ranked according to independent draws from the Mallows distribution. However, when using the lowBMM, we will be assuming that only $n^*$ items are ranked according to a Mallows model. Hence, when using the method described above we will need to re-scale the estimated $\alpha$ from dimension $(N,n)$ to dimension $(N,n^*)$. Let $\hat{\alpha}_n$ be the optimal $\alpha_0$ value estimated from the datasets of dimension $(N,n),$ as described above. We can then re-scale $\hat{\alpha}_n$ to obtain the optimal $\hat{\alpha}_{n^*}$ value in dimension $n^*$ by matching the terms in the exponent in the Mallows model:
\begin{align*}
    \hat{\alpha}_{n^*} = \hat{\alpha}_n \frac{n}{n^*}\frac{\max_{d_{n^*}}}{\max_{d_{n}}}
\end{align*}
where $\max_{d_n}$ (respectively $\max_{d_{n^*}}$) is the maximum attainable footrule distance between two rankings in dimension $n$ (respectively $n^*$). The method is assessed in a simulation study described in Section \ref{sec:off_line_alpha_simulation}. 

\section{Simulation experiments}\label{sec:simulations}
This section describes how lowBMM performs under different data generating processes, \emilie{how its performance is affected by the various parameters involved, and how it compares to other methods}. The variable selection procedure provides a posterior distribution for the consensus ranking in a reduced dimension, and the type of distribution obtained in turn depends on the data generating process. Therefore, inspecting the solution provided by lowBMM under different scenarios for generating the data allows us to understand the data structure and patterns in real-life situations.

First, we consider a top-rank data generating process, where only a subset of the items is ranked, and assigned the top ranks in reduced dimension, while the rest is noisy: this scenario results in a top-rank solution, i.e., lowBMM correctly selects the top-items following a rank model. Second, we generate data where items are consistently awarded ranks in a specific order: this results in a rank-consistency solution, i.e., items that \emph{can} be ordered are selected and top-ranked. \emilie{A study on how lowBMM works under different data generating processes is presented in Section \ref{sec:sim_data_processes}, and a sensitivity analysis on the tuning parameters involved in lowBMM can be found in Section \ref{sec:sim_sens_study}. We also test the method's robustness to the noise level in the data in Section \ref{sec:sim_noise}, and we provide an extensive study on how lowBMM compares to existing methods in Section \ref{sec:sim_comparison_all_methods}. Finally, we present a performance assessment of the off-line estimation of $\alpha$ in Section \ref{sec:off_line_alpha_simulation}.}

\subsection{Varying the data generating process: top-rank and rank consistency experiments}\label{sec:sim_data_processes}

\emilie{To test the method's applicability to different data structures, we consider two data generating processes: top-rank and rank-consistency.} In the case of top-rank we generate data in the following way: $n^*$ items were given ranks sampled from the Mallows model, $R_j |_{ \mathcal{A}^*} \sim \textnormal{Mallows}(\bm{\rho}_{\mathcal{A}^*}, \alpha)$, with $\bm{\rho}_{\mathcal{A}^*} = (1,...,n^*)$ while the rest of the items were randomly assigned ranks, $R_j |_ {\mathcal{A} \setminus \mathcal{A}^*} \sim \mathcal{U}(\mathcal{P}_{n^*+1,...,n})$. Here the top items are ranked according to the Mallows model, and thus result in a top-rank solution. In the rank consistency experiment on the other hand, the aim is to see whether lowBMM can select consistently ranked items even if not necessarily top-ranked. The setting was the following: items in $\mathcal{A}^*$ were given a respective ranking sampled from a Mallows model in dimension $n^*$; then, for obtaining each assessor-specific ranking, the same items were assigned a random ranking in dimension $n$ in their respective positions; finally, the rest of the items had randomly assigned ranks. Therefore, items in the true $\mathcal{A}^*$ were always ranked in a certain order for each assessor, however their specific ranks were not consistent.

\emilie{For the top-rank data generating process, we performed two data simulation experiments: first a toy example with $n=20$ items, and then a larger example with $n=1000$ items.}

\begin{figure}[!htb]
\minipage{0.49\textwidth}
  \includegraphics[width=\linewidth]{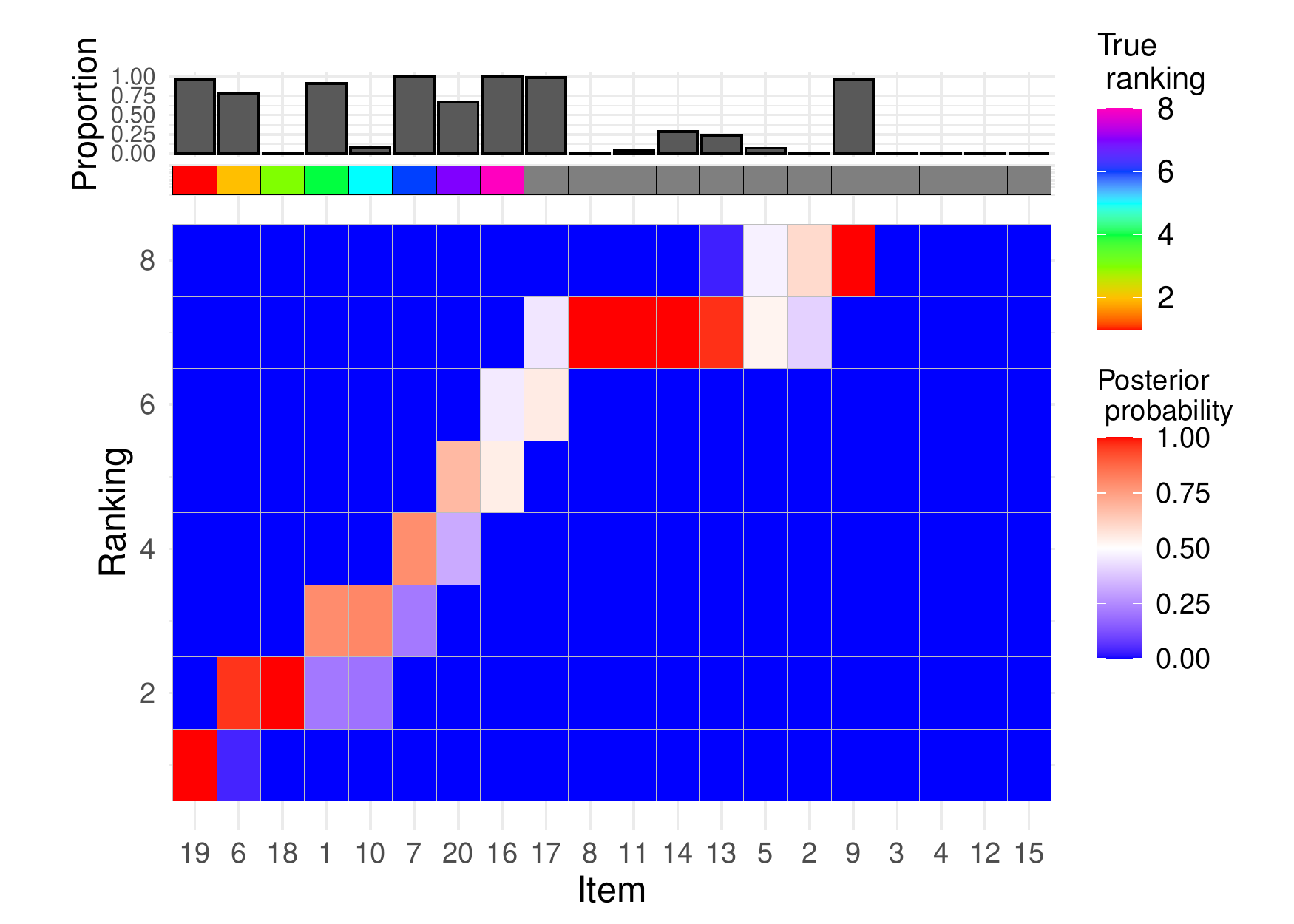}
\endminipage\hfill
\minipage{0.49\textwidth}
  \includegraphics[width=\linewidth]{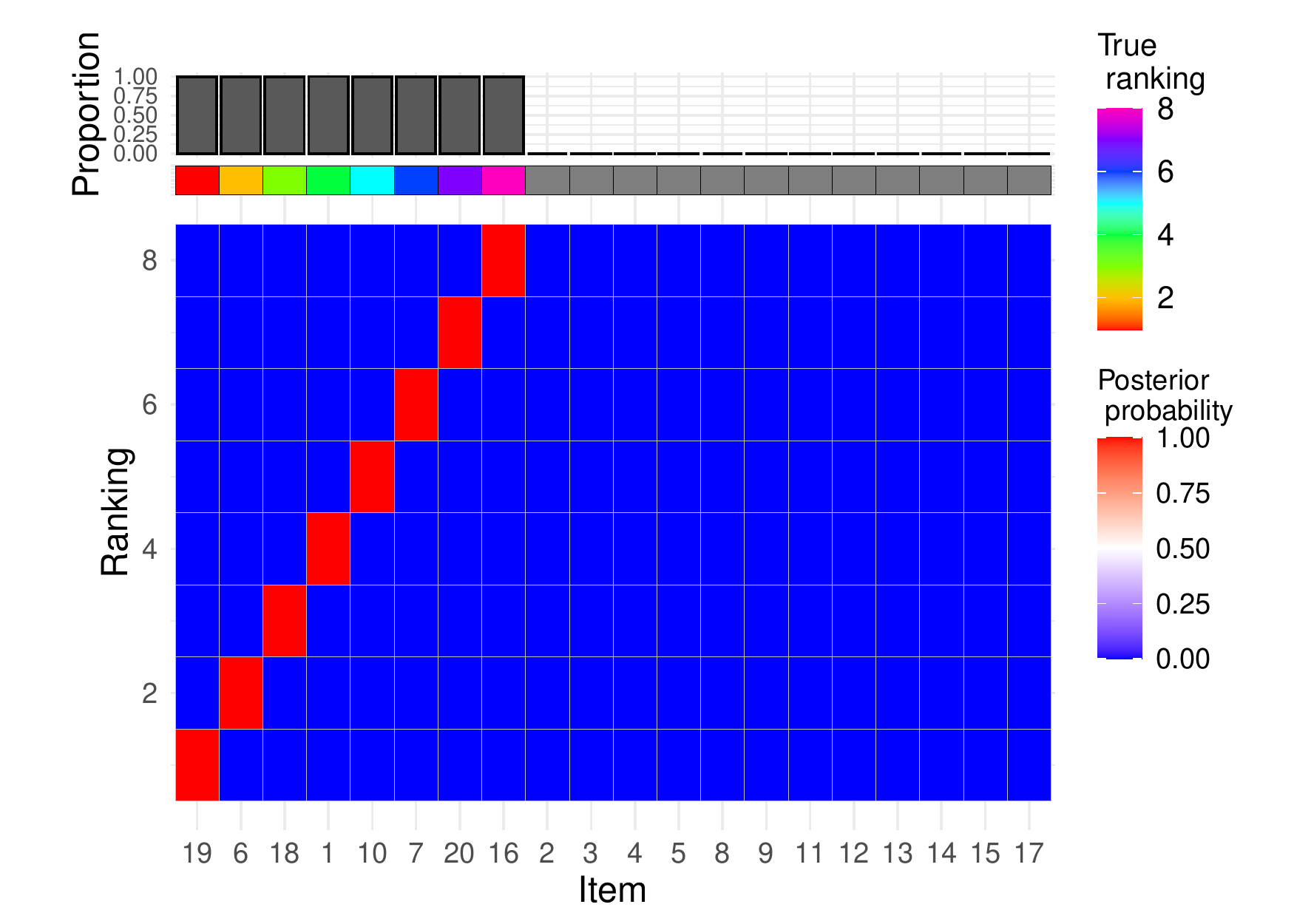}
\endminipage

\caption{Results from the small simulation experiment described in Section \ref{sec:sim_data_processes} with $\alpha=10$, $n=20$, $N=50$, $n^*=8$, $L=1$ and $l=\textnormal{round}(n^*/5)$: both panels display heatplots of the marginal posterior distribution of $\bm{\rho}$ with the items ordered according to $\bm{\Hat{\rho}}_{\mathcal{A}^*}$ on the x-axis. Left: top-rank simulation example, right: rank consistency simulation example. The rainbow grid indicates the true ordering $\bm{\rho}_{\mathcal{A}^*}$, and the bar plot indicates the proportion of times the items were selected in $\mathcal{A}^*$ over all MCMC iterations.}
\label{fig:toy_topk_vs_rankcons_alpha10}
\end{figure}

\emilie{In the small example, we considered a group of $N=50$ assessors, and $n^*=8$ relevant items. For what concerns the tuning parameters, we set $L=1$ and $l=\textnormal{round}(n^*/5)$ as suggested by the tuning parameter study in Section \ref{sec:sim_sens_study}. We used $M=5\cdot 10^3$ MCMC iterations, and the algorithm took 10 seconds to run. The left panel in Figure \ref{fig:toy_topk_vs_rankcons_alpha10} displays the marginal posterior distribution of $\bm{\rho}$ where the items on the x-axis have been ordered according to $\bm{\Hat{\rho}}_{\mathcal{A}^*}$ (see Section \ref{sec:mh_postprocess} for the computation of posterior summaries). The results show that the method is able to correctly select the relevant items to be included in the selection set, as the items whose ranking was simulated from the Mallows model are all top-ranked. However, there is uncertainty associated to their specific rankings: some items seem to ``compete'' for ranks, e.g. item 6 (true rank 2) and item 18 (true rank 3) both show a large posterior marginal probability of being ranked 2; at the same time, their final ordering is correct (the ordering on the x-axis is consistent with the rainbow-grid representing the true ordering $\bm{\rho}_{\mathcal{A}^*}$). This is not an unexpected behaviour, as the variability in the items' true ranks is dependent on the scale parameter, $\alpha$.} As we do not estimate $\alpha$ in the current version of lowBMM, we inspected how the solution changed when varying this parameter. The $\alpha$ parameter regulates how much the assessors agree with each other on the ranking of the items: a large $\alpha$ means a better agreement between the assessors, and a lower $\alpha$ means less agreement. So, when $\alpha$ is large, the items distributed according to the Mallows are easier to pick out, making it less likely that other items are accepted in the proposal set. This is evident in the distribution of the bars on the top of the plots in Figure S1 in the supplementary material, where the larger $\alpha$ value shows less variability in the selected items. Moreover, when the data is ``easier'', the algorithm converges much more quickly to the correct set. The bottom items (whose ranks are essentially noise) are clearly distributed on the bottom for $\alpha=10$, while there is a bit more uncertainty for $\alpha=3$, as expected.

\begin{figure}[!htb]
\minipage{0.49\textwidth}
  \includegraphics[width=\linewidth]{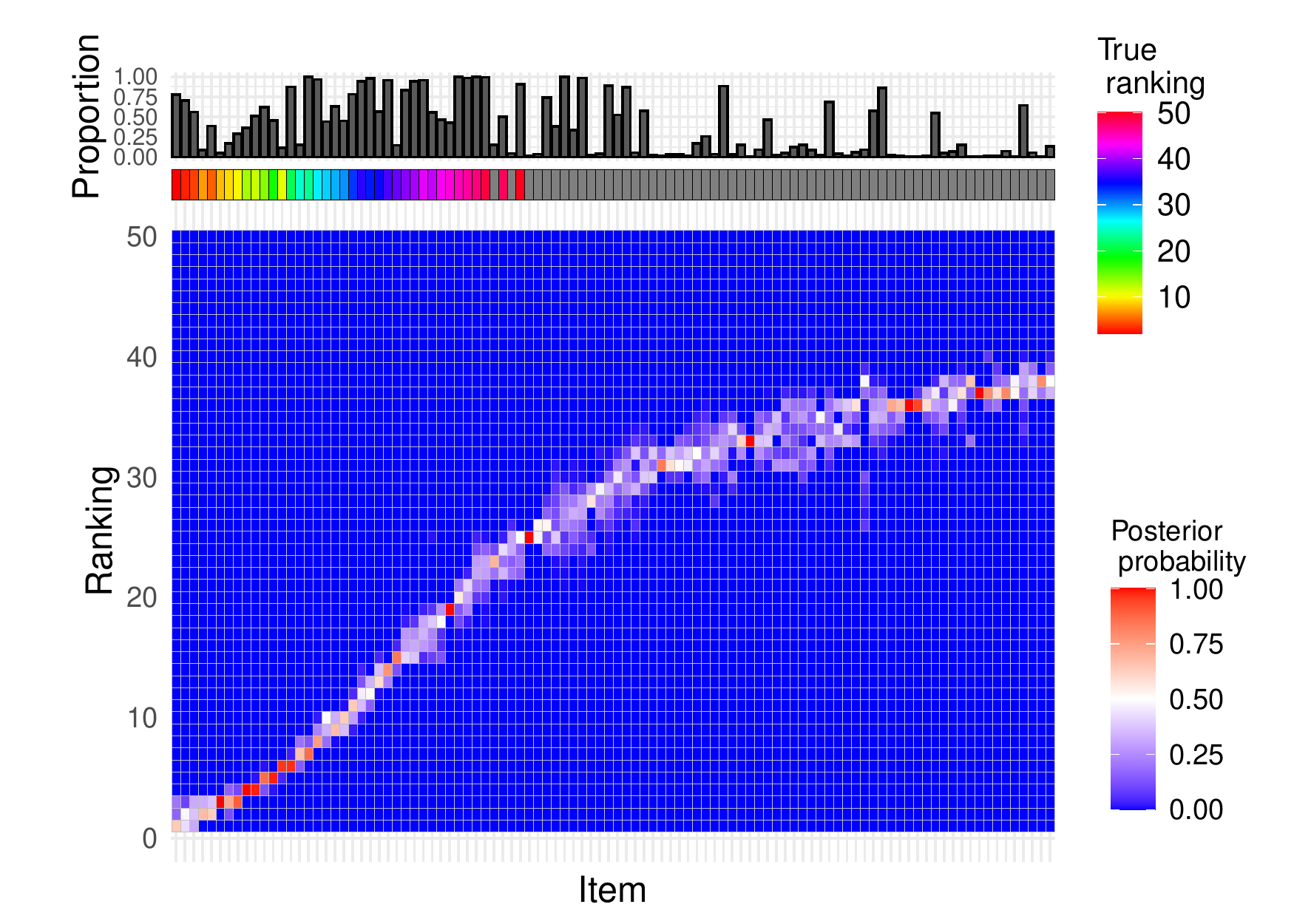}
\endminipage\hfill
\minipage{0.49\textwidth}
  \includegraphics[width=\linewidth]{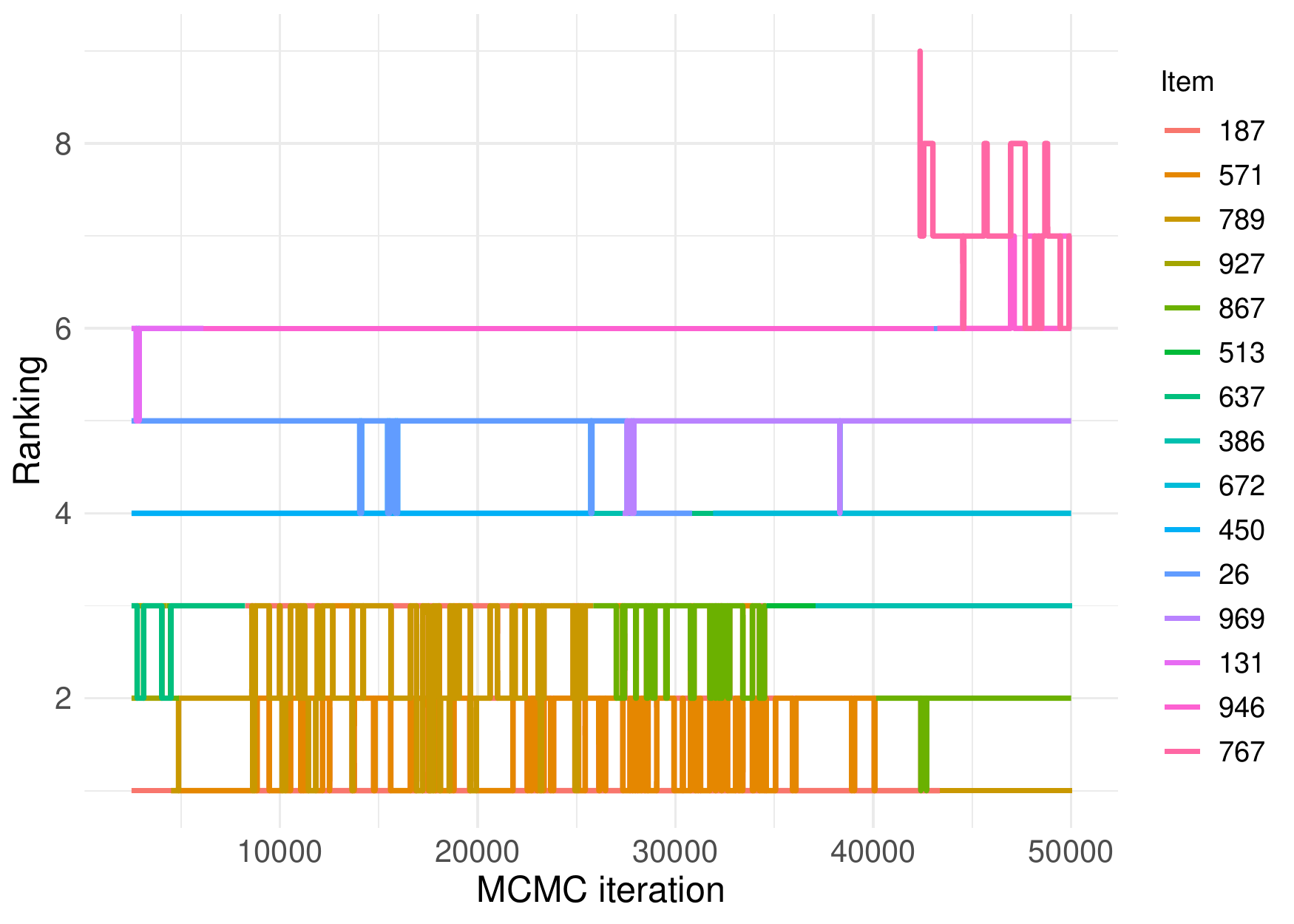}
\endminipage
\caption{Results from the large top-rank simulation experiment described in Section \ref{sec:sim_data_processes} with  $\alpha=10$, $n=1000$, $N=50$, $n^*=50$, $L=1$, $l=\textnormal{round}(n^*/5)$. Left: heatplot of the marginal posterior distribution of $\bm{\rho}$ for the top-100 items ordered according to $\bm{\Hat{\rho}}_{\mathcal{A}^*}$ on the x-axis, where the rainbow grid on top indicates the true ranking of the relevant items $\bm{\rho}_{\mathcal{A}^*}$, and the bar plot indicates the proportion of times the items were selected in $\mathcal{A}^*$ over all MCMC iterations. Right: trace plot of the top-15 items in $\bm{\Hat{\rho}}_{\mathcal{A}^*}$ along MCMC iterations.}
\label{fig:alpha10_sim_big}
\end{figure}

In order to test the method in a more realistic scenario, we simulated the data in a dimension closer to that often encountered in -omics applications. This larger experiment aims to investigate the method performance and its computing capabilities on a more demanding dataset when using the same data generating process as before (top-rank). We set the total number of items to $n=1000$, the number of relevant items to $n^*=50$ and the number of assessors to $N=50$. We also set the tuning parameters to $L = 1$ and $l = \textnormal{round}(n^*/5)$, as previously done. We ran the MCMC for $5 \cdot 10^4$ iterations in this larger experiment, and the algorithm used 155 seconds to run. Note that we used a larger $M$ in the larger examples to ensure good space exploration and convergence of the chains, and the computing times were all scaling well with the increasing dimension. The posterior of $\bm{\rho}$ can be seen in the left plot of Figure \ref{fig:alpha10_sim_big}: more than $75\%$ of the items in the true selection are assigned top ranks with large marginal posterior probability, while the uncertainty associated to the items' rankings increases moving towards the bottom, for the items not in the top-$n^*$. The trace plot on the right in Figure \ref{fig:alpha10_sim_big} demonstrates convergence, with most top-ranked items clearly converging to one final rank, and only a few items competing for the same rank throughout the chain. 
This ``competing'' behaviour should to some degree be expected, however we noticed it to be enhanced by increasing the tuning parameter $L$ (see Figure S9 in the supplementary material), suggesting to keep $L$ small (this is confirmed in the sensitivity study on the tuning parameters in Section \ref{sec:sim_sens_study}). 

\begin{figure}[!htb]
\minipage{0.49\textwidth}
  \includegraphics[width=\linewidth]{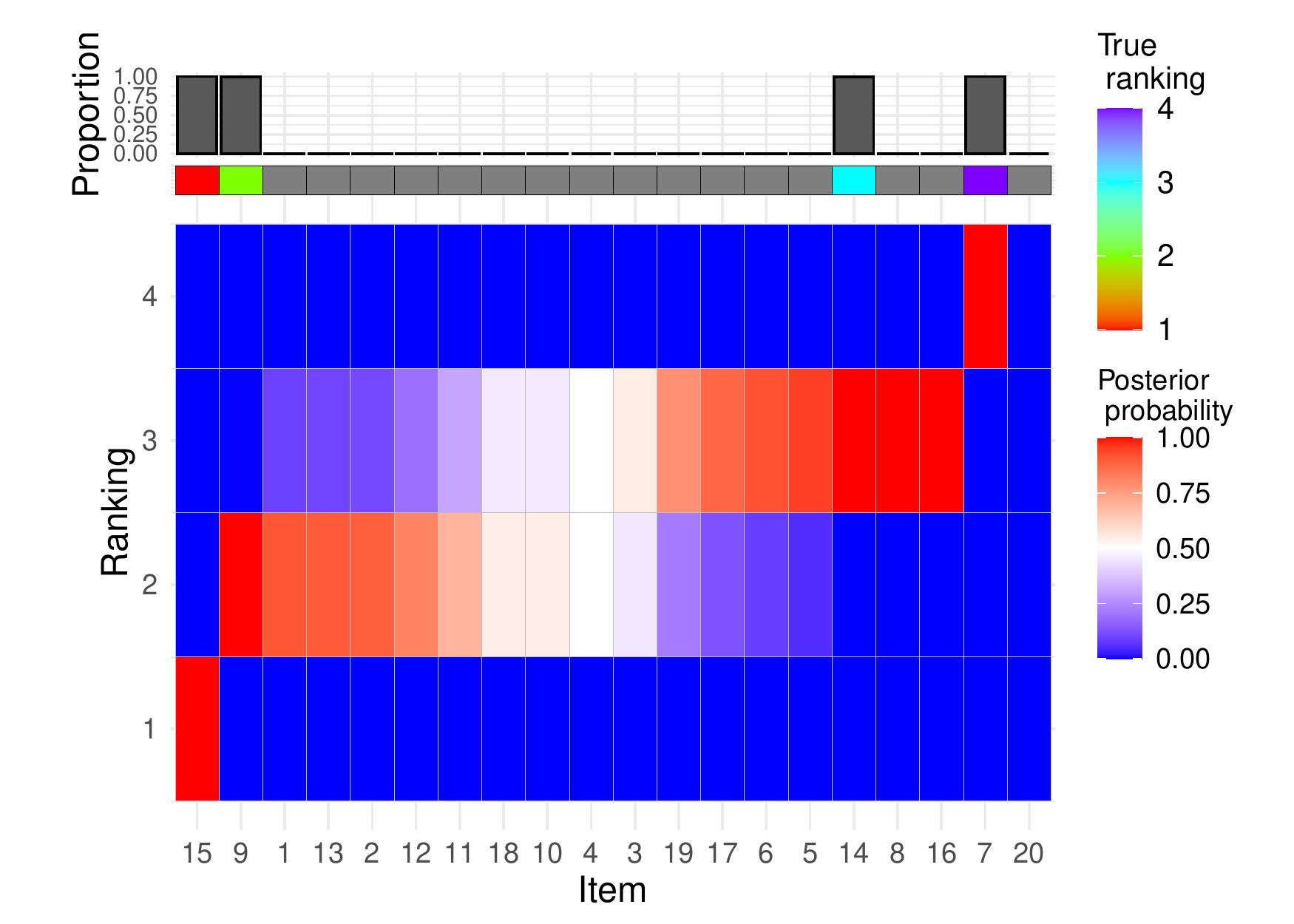}
\endminipage\hfill
\minipage{0.49\textwidth}
  \includegraphics[width=\linewidth]{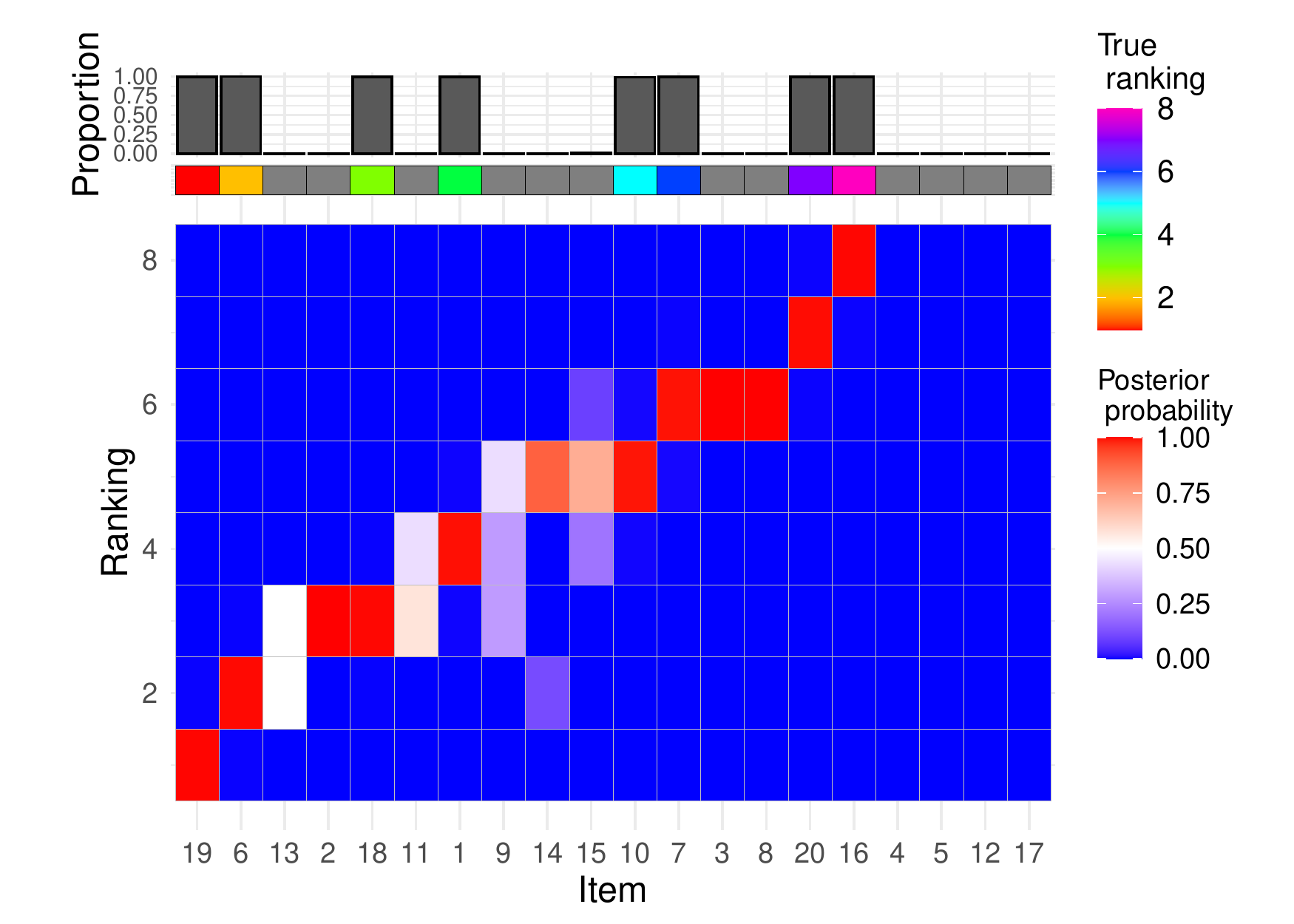}
\endminipage\hfill
\caption{Results from the rank consistency simulation experiment described in Section \ref{sec:sim_data_processes}: heatplot of the marginal posterior distribution of $\bm{\rho}$ with the items ordered according to $\bm{\Hat{\rho}}_{\mathcal{A}^*}$ on the x-axis with $\alpha=0.5$, $n=20$, $N=50$, $L=1$ and $l=\textnormal{round}(n^*/5)$. Left: $n^*=4$; right: $n^*=8$. The rainbow grid indicates the true ranking of the relevant items $\bm{\rho}_{\mathcal{A}^*}$, and the bar plot indicates the proportion of times the items were selected in $\mathcal{A}^*$ over all MCMC iterations.}
\label{fig:sim_rank_consistency}
\end{figure}

\emilie{For the rank consistency data generating process, we carried out a simulation study similar to the small top-rank example, also here focusing on a small experiment with $n=20$, $n^*=8$ and $N=50$.} The right panel in Figure \ref{fig:toy_topk_vs_rankcons_alpha10} presents the marginal posterior distribution of $\bm{\rho}$ where the items have been ordered according to $\Hat{\bm{\rho}}_\mathcal{A^*}$ on the x-axis. Here the performance of lowBMM is evident: the correct items (indicated by the rainbow grid on the top) are given the highest marginal posterior probability in $\mathcal{A}^*$ (as shown by the bar plot on top), and the ordering of the items is also consistent with the true $\bm{\rho}_{\mathcal{A}^*}$. \emilie{This shows that lowBMM is able to handle varying data structures, and quite interestingly, is performing slightly better on a dataset with consistently ranked items, compared to the top-rank data example. The method also seems to be somewhat more robust to the choice of the tuning parameters for this type of data generating process (see Section \ref{sec:sim_sens_study}, Figure \ref{fig:sim_rank_consistency} for details). The rank consistency simulation experiment can be considered a more realistic example of how ranking data could be structured in real-life situations, particularly in genomics: a subset of genes might be consistently ranked in a certain pattern across samples (e.g. consistently more or less expressed), while their individual rankings do not really matter as much. At the same time, such a pattern would be much more interesting to detect than the actual true ranking estimation. Furthermore, experiments making use of this data generating process were overall more challenging to handle for all the other competing methods considered for comparison (see Section \ref{sec:sim_comparison_all_methods}). }

In the rank consistency simulation experiment, we also experimented by reducing the number of items being given the consistent ranks, $n^*$, as well as the level of precision in the data sampled from the Mallows, $\alpha$. Decreasing both $n^*$ and $\alpha$ makes the estimation via lowBMM more difficult, as less items show a detectable pattern, and the noise level in the data is higher. Figure \ref{fig:sim_rank_consistency} shows the results obtained with $\alpha=0.5$: as expected, we see a much larger uncertainty in the ranks with respect to what observed in Figure \ref{fig:toy_topk_vs_rankcons_alpha10} (right) for $\alpha=10$. It is interesting to note the performance of the method: despite the much larger uncertainty, still most weight in the marginal posterior of $\bm{\rho}$ is given to the correct items (looking at the bar plots at the top), and the ordering of the items is consistent with the true ordering (rainbow grid on top). This shows that the accuracy of lowBMM is unaffected by decreasing $\alpha$ or $n^*$, even in the case of a rank consistency data generating process. Note that in this experiment we have used the same $n^*$ both in the data generation and in the analysis, as the main point here was assessing the robustness of lowBMM to a less detectable pattern. In the next Section \ref{sec:sim_sens_study} we will instead verify the method robustness to the misspecification of $n^*$.

\subsection{Sensitivity study on tuning parameters}\label{sec:sim_sens_study}

\emilie{A sensitivity study was carried out to verify the effect of the tuning parameters $L$ and $l$ on lowBMM, as well as the effect of misspecifying the number of relevant items for the selection, $n^*$. We first focused on studying the effect of $L$ and $l$ and therefore we set: $n=100$, $N=10$, $n^*=10$ and $\alpha=5$. The data was generated according to the top-rank data generating process described in Section \ref{sec:sim_data_processes}: $n^*$ items were given ranks sampled from the Mallows model, while the rest of the items were ranked randomly lower. We experimented by varying the two tuning parameters introduced in Section \ref{sec:mh}: $L$, i.e., the number of items perturbed in $\mathcal{A}^*$ for proposing a new set $\mathcal{A}^*_{\textnormal{prop}}$, and $l$, i.e., the number of items perturbed in the current $\bm{\rho}$ for proposing a new consensus $\bm{\rho}^\prime$. Values for $L$ and $l$ were explored over a grid $(L,l) \in [1,2,3,4,5] \times [1,2,3,4,5]$.} Results were collected after running the method on 20 simulated datasets, when using $M=5000$ MCMC iterations, which always showed to be more than enough for convergence. 

\emilie{To evaluate the results we computed the footrule distance between the true and the estimated consensus ranking, normalized by the number of correctly selected items $n_\textnormal{corr}=|\mathcal{A}^*\bigcap\hat{\mathcal{A}}^*|$: $d_{\textnormal{norm}}(\bm{\rho}_{\mathcal{A}^*}, \bm{\hat{\rho}}_{\mathcal{A}^*}) = d(\bm{\rho}_{\mathcal{A}^*}, \bm{\hat{\rho}}_{\mathcal{A}^*})/n_\textnormal{corr}$ (see Section \ref{sec:mh_postprocess} for details on the computation of the posterior summaries $ \bm{\hat{\rho}}_{\mathcal{A}^*}$ and $\hat{\mathcal{A}}^*$). For brevity, we will sometimes indicate $d_{\textnormal{norm}}(\bm{\rho}_{\mathcal{A}^*}, \bm{\hat{\rho}}_{\mathcal{A}^*})$ simply with $d_{\textnormal{norm}}$.}

\begin{figure}[!htb]
\minipage{0.33\textwidth}
  \includegraphics[width=\linewidth]{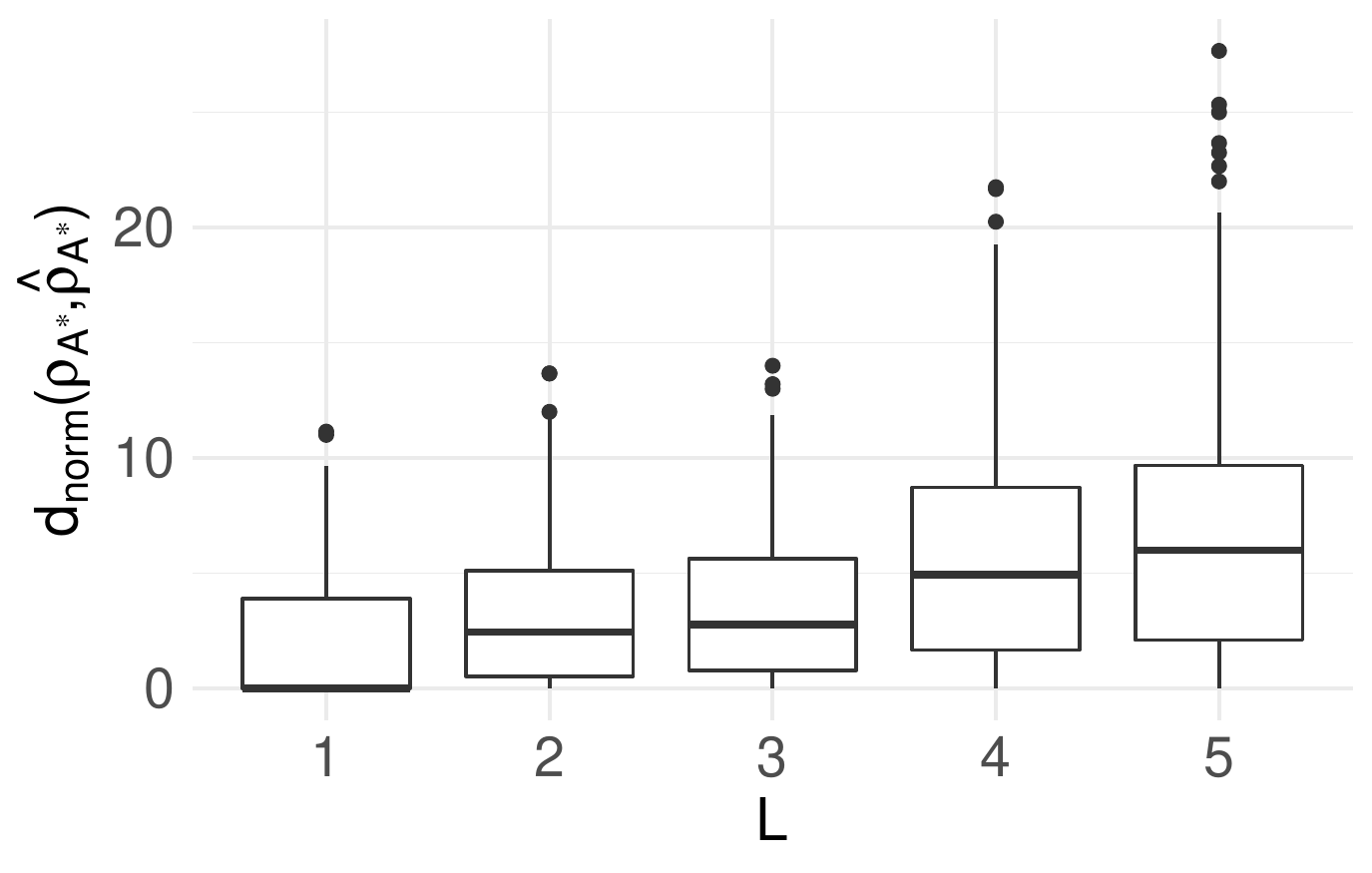}
\endminipage\hfill
\minipage{0.33\textwidth}
  \includegraphics[width=\linewidth]{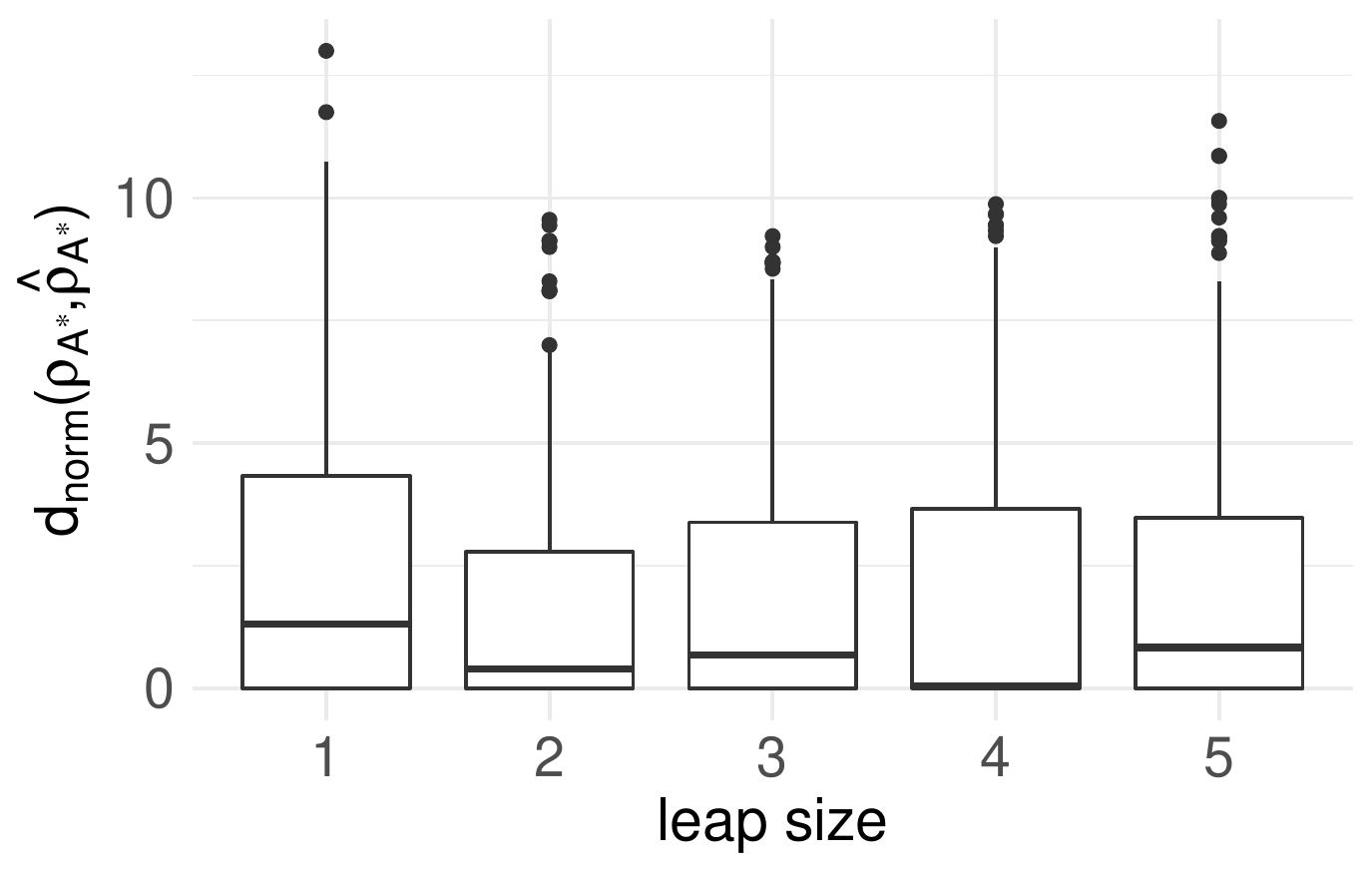}
\endminipage
\minipage{0.33\textwidth}
  \includegraphics[width=\linewidth]{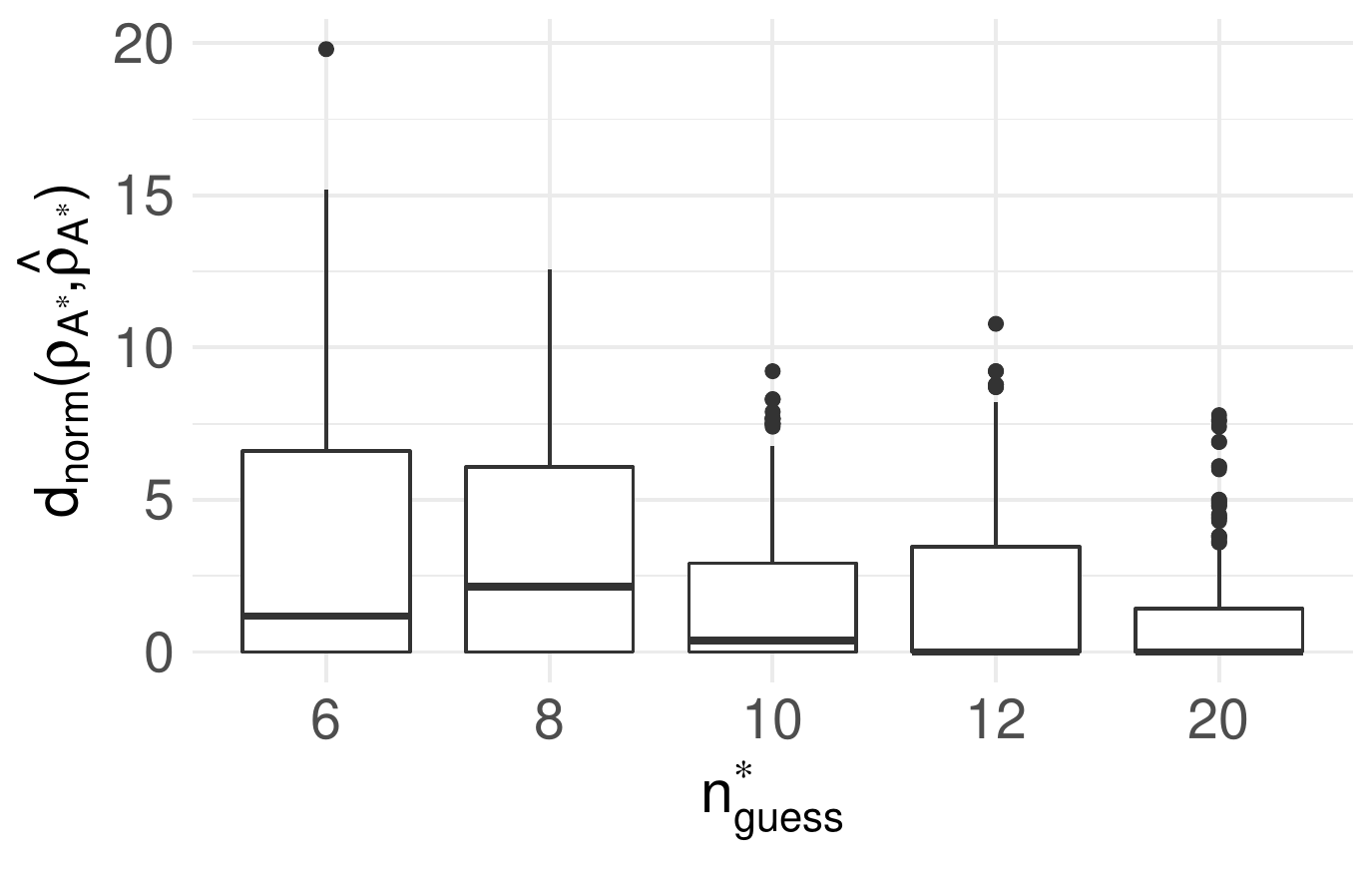}
\endminipage
\caption{Results from the sensitivity study on the tuning parameters described in Section \ref{sec:sim_sens_study}: boxplots of $d_{\textnormal{norm}}$, on 20 different datasets. From left to right: for varying $L$, leap size $l$ and $n^*_\textnormal{guess}$. $n=100$, $N=10$, $n^*=10$, $\alpha=5$, $M=5000$.}
\label{fig:sens_study_l1_dist}
\end{figure}

\emilie{The left panel in Figure \ref{fig:sens_study_l1_dist} shows $d_{\textnormal{norm}}(\bm{\rho}_{\mathcal{A}^*}, \bm{\hat{\rho}}_{\mathcal{A}^*})$ over the 20 runs and for varying $L$ and the left panel in Figure S2 in the supplementary material displays the proportion of correctly selected items in $\mathcal{A}^*$ for each $L$. The results showed a clear indication that a smaller $L$ results in a lower error overall. Similarly, the middle panel in Figure \ref{fig:sens_study_l1_dist} (and also the middle panel in Figure S2 in the supplementary material) indicates that the leap size $l$ has a smaller effect on the error, even if $l=\textnormal{round}(n^*/5)$ yields a slightly better result, both in terms of the footrule distance ($d_\textnormal{norm}$) and in terms of the selection of items. Similar conclusions can be made by inspecting the heatplots of the marginal posterior distributions of $\bm{\rho}$ for varying values of the tuning parameters (see Figure S3 in the supplementary material for $\alpha=3$, and Figure S4 in the supplementary material for $\alpha=10$), where it is evident that varying $l$ does not affect the error (but does contribute slightly to the degree of uncertainty in the ranks), while varying $L$ does seem to affect the error to a slight higher degree compared to $l$. Furthermore, as noticeable in the barplots in Figures S3 and S4, increasing $L$ contributes to worsening the degree of exploration of the items space when estimating the posterior distribution for $\mathcal{A}^*$, thus supporting the decision to keep $L$ small.}

\begin{figure}[!htb]
\centerline{
  	\includegraphics[width=0.4\linewidth]{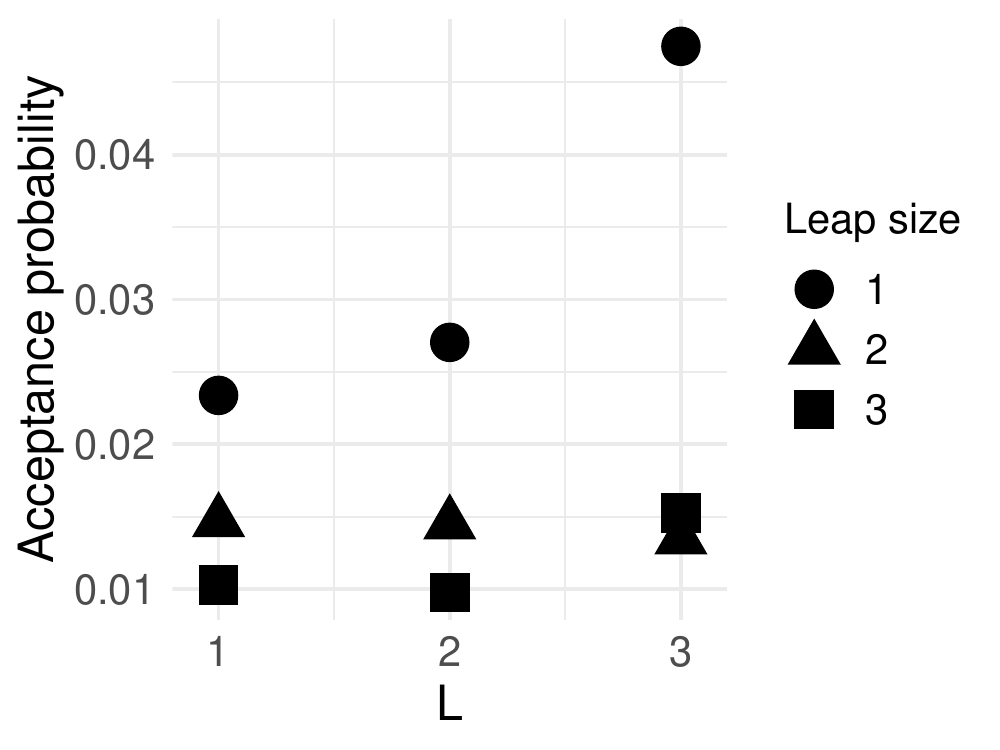}
  	\hspace{0.5cm}
    	\includegraphics[width=0.4\linewidth]{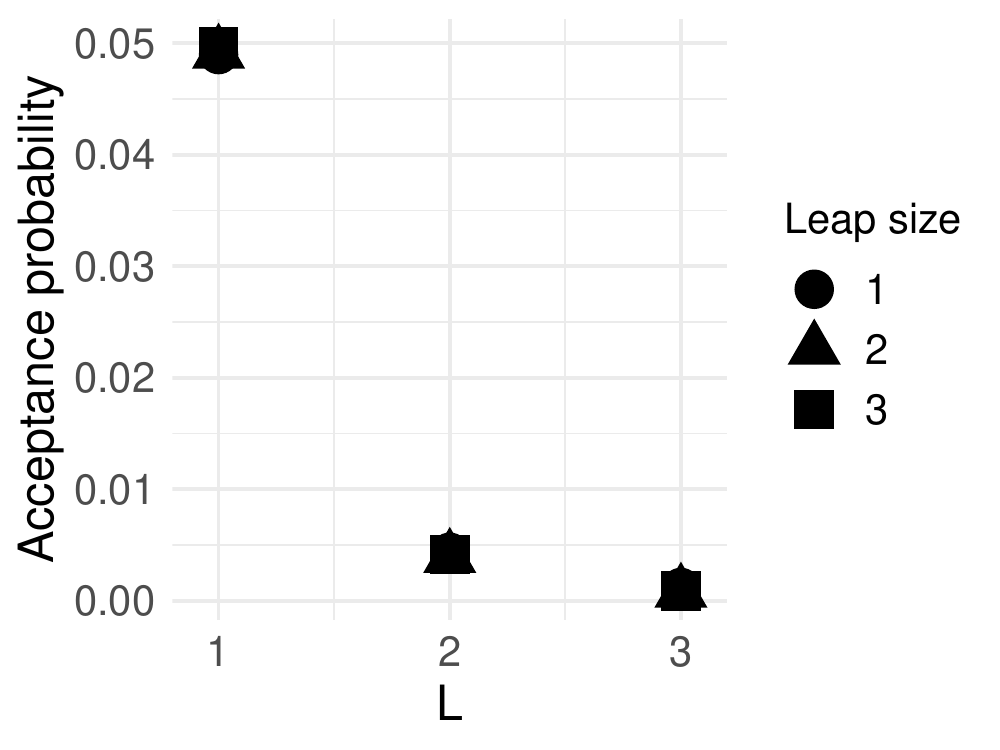}
}
\caption{Results from the sensitivity study on the tuning parameters as described in Section \ref{sec:sim_sens_study}: acceptance probabilities of the consensus ranking $\bm{\rho}$ (left), and of the set of selected items $\mathcal{A}^*$ (right). The x-axis corresponds to varying $L=[1,2,3]$, while the points shapes correspond to varying the leap size $l=[1,2,3]$. Simulated scenario with $n=20$, $N=50$, $n^*=8$, $\alpha=3$.}
\label{fig:alpha3_nstar8_sim_small_accept}
\end{figure}

It is also worth exploring variations in the acceptance probability of the two parameters estimated in the MCMC, $\bm{\rho}$ and $\mathcal{A}^*$, when varying the tuning parameters $L$ and $l$. From inspection of Figure \ref{fig:alpha3_nstar8_sim_small_accept} ($\alpha=3$, $n^*=8$, $n=20$ and $N=50$ in this scenario), it seems that the acceptance probability for $\bm{\rho}$ increases with increasing $L$, while it decreases with increasing $l$; the latter behaviour is expected, while the former is less trivial to understand. For what concerns $\mathcal{A}^*$, the acceptance probability decreases greatly with increasing $L$, and it seems to not be much affected by the leap size $l$, which are both quite natural behaviours. Therefore, we generally recommend to keep $L=1$ to explore the space appropriately when estimating $\mathcal{A}^*$, and to tune $l$ by inspecting the acceptance probability of $\bm{\rho}$. The tuning of $l$ in the BMM has been extensively studied, resulting in a recommendation to use $l=\textnormal{round}(n/5)$ as a default value to ensure a good balance between space exploration and accuracy (see \cite{vitelli2018} for further discussions). However, the respective values of $n^*$ and $n,$ which constitute the novelty of lowBMM with respect to the original BMM, might make it challenging to choose a combination of tuning parameters that results in a chain that explores the space well. Therefore, even if $l=\textnormal{round}(n^*/5)$ will generally be a solid choice, we do recommend to try some values of $l$ before setting it.

\begin{figure}[!htb]
\minipage{0.49\textwidth}
  \includegraphics[width=\linewidth]{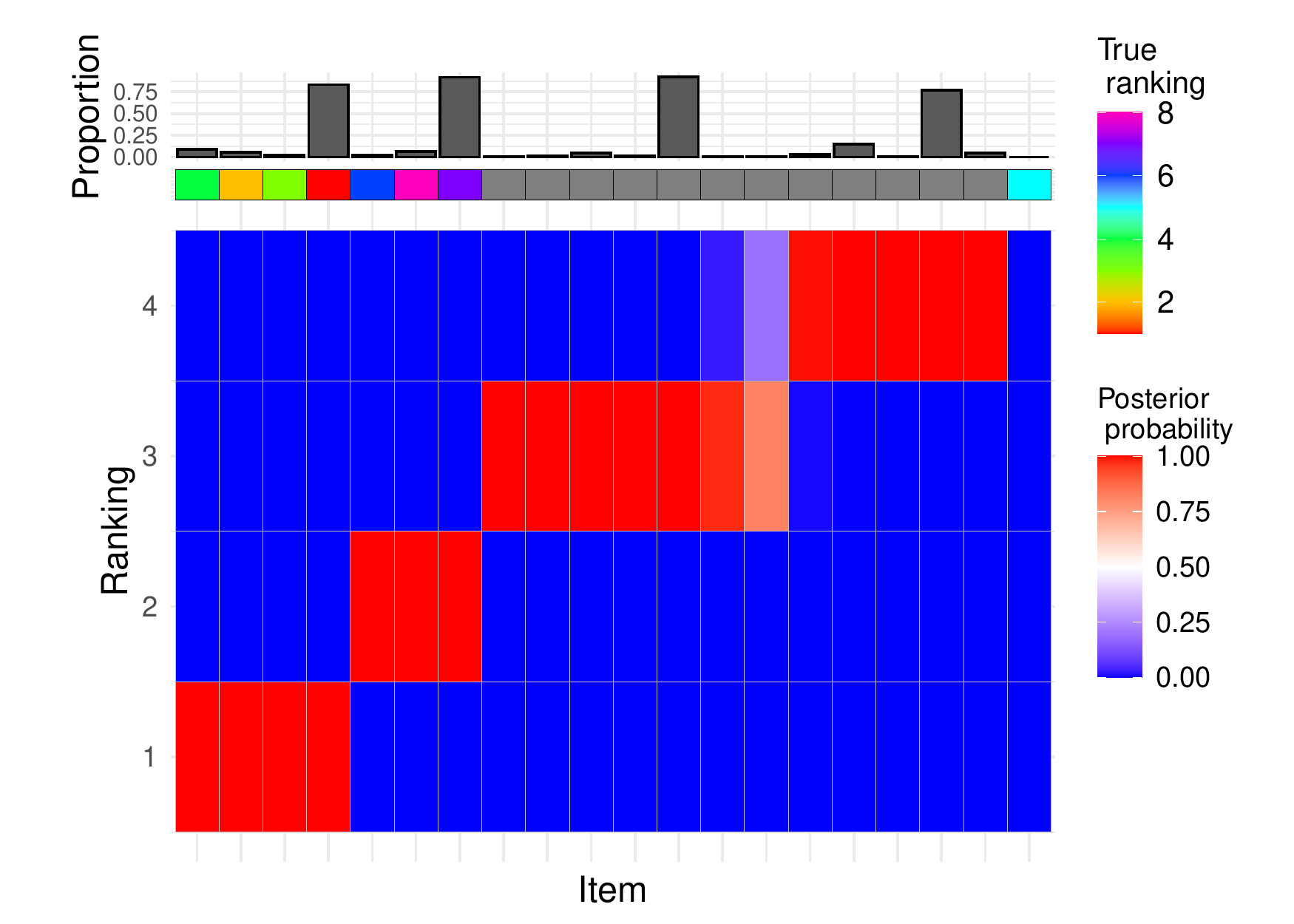}
\endminipage\hfill
\minipage{0.49\textwidth}
  \includegraphics[width=\linewidth]{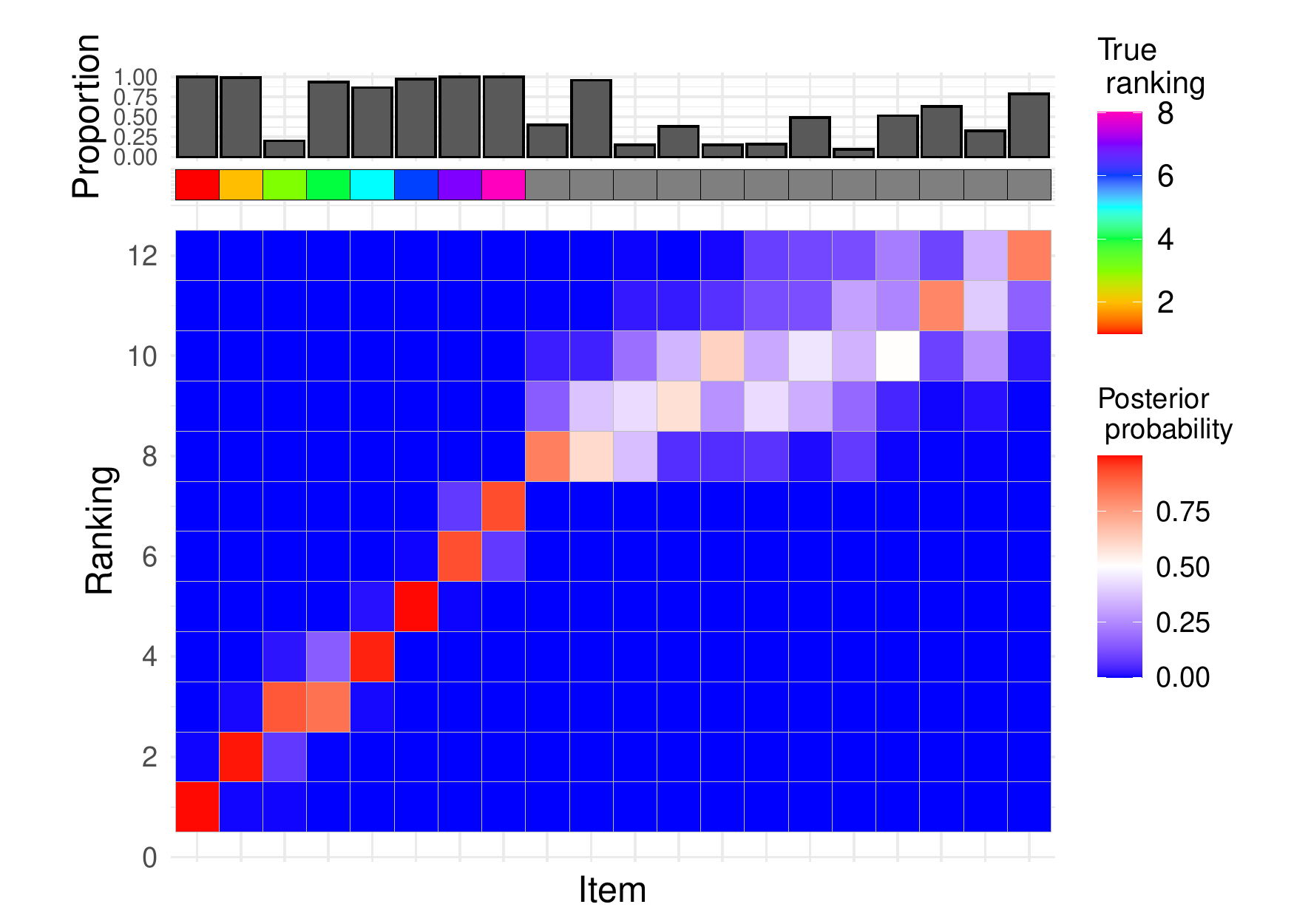}
\endminipage
\caption{Results from the sensitivity study on the tuning parameters as described in Section \ref{sec:sim_sens_study}, here when setting  $n^*_{\textnormal{guess}}$ differently from the true $n^*=8$. Both panels display heatplots of the marginal posterior distribution of $\bm{\rho}$ with the items ordered according to $\bm{\Hat{\rho}}_{\mathcal{A}^*}$ on the x-axis; left: $n^*_{\textnormal{guess}}=4$, right: $n^*_{\textnormal{guess}}=12$. The rainbow grid indicates the true ranking $\bm{\rho}_{\mathcal{A}^*}$, and the bar plot indicates the proportion of times the items were selected in $\mathcal{A}^*$ over all MCMC iterations. $n=20$, $N=50$, $\alpha=3$, $L=1$ and $l=\textnormal{round}(n^*_{\textnormal{guess}}/5)$.}
\label{fig:alpha3_nstar4_vs_nstar12}
\end{figure}

\emilie{Since the number of relevant items $n^*$ is in general unknown, we also tested the performance of lowBMM when assuming a different $n^*$ than what was used when generating the data. We did a systematic study with $n^*_\textnormal{guess} \in [6,8,10,12,20]$, and with the truth set to $n^*=10$. The results from the study are displayed in the right panels in Figure \ref{fig:sens_study_l1_dist} and in Figure S2 in the supplementary material, both suggesting that setting $n^*_{\textnormal{guess}}>n^*$ increases the accuracy. This is not surprising, as it allows the algorithm to search for a good selection in a larger permutations space. }
\emilie{It is also interesting to evaluate how the consensus ranking $\bm{\rho}$ changes when $n^*_\textnormal{guess}$ differs from the true value: the marginal posterior distribution of $\bm{\rho}$ obtained when $n^*_{\textnormal{guess}}<n^*$ is displayed in Figure \ref{fig:alpha3_nstar4_vs_nstar12} on the left, and when $n^*_{\textnormal{guess}}>n^*$ on the right. By inspecting the left panel we see that assuming $n^*_{\textnormal{guess}}<n^*$ yields a clustered structure, thus implying that the best lower-dimensional solution of ranked data in a higher-dimension is consistent with a partial ordering, also called ``bucket solution'' \cite{dambrosio2019}. On the other hand, assuming $n^*_{\textnormal{guess}}>n^*$ (Figure \ref{fig:alpha3_nstar4_vs_nstar12}, right panel) yields a top-rank solution, as the algorithm seems to be able to indirectly suggest the true $n^*$ by showing a much larger uncertainty around the items ranked with larger values than the true $n^*$. Overall, when $n^*_{\textnormal{guess}}>n^*$, the posterior distribution of $\bm{\rho}$ is anyway accurate, since the subset of relevant items has larger marginal posterior probability of being top-ranked. Moreover, the marginal posterior distribution of $\bm{\rho}$ implicitly suggests the most suitable value for $n^*_{\textnormal{guess}}$, by suddenly showing larger uncertainty around items unnecessarily selected in $\mathcal{A}^*$. Similar trends were observed in simulation studies when varying $\alpha$ and the tuning parameters $l$ and $L$, see Figures S5 and S7 in the supplementary material for $\alpha=3$, and Figures S6 and S8 in the supplementary material for $\alpha=10$, with $n^*_{\textnormal{guess}} \neq n^*$.}

\subsection{Robustness to noise experiment}\label{sec:sim_noise}
\begin{figure}[!htb]
\minipage{0.49\textwidth}
  \includegraphics[width=\linewidth]{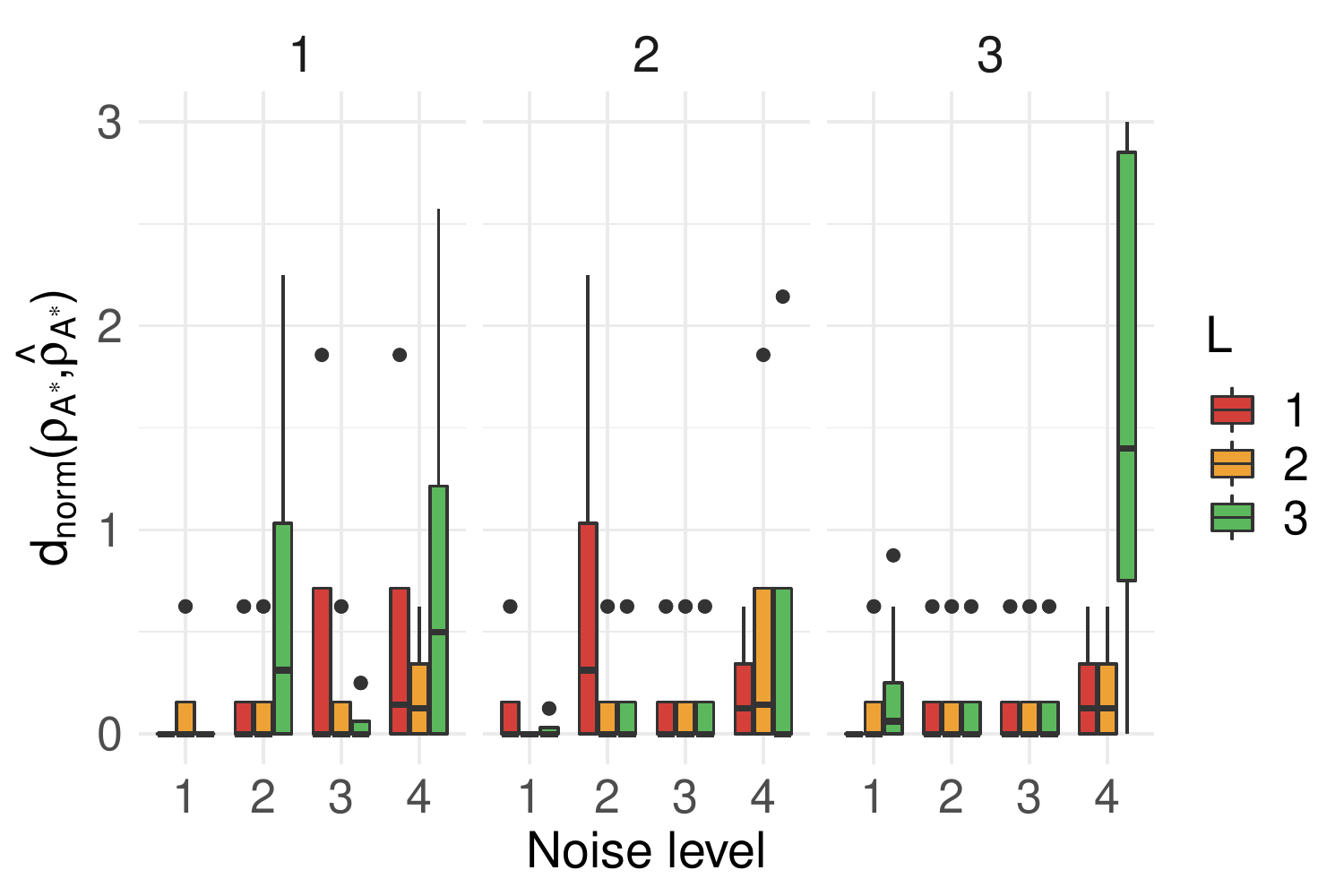}
\endminipage \hfill
\minipage{0.49\textwidth}
  \includegraphics[width=\linewidth]{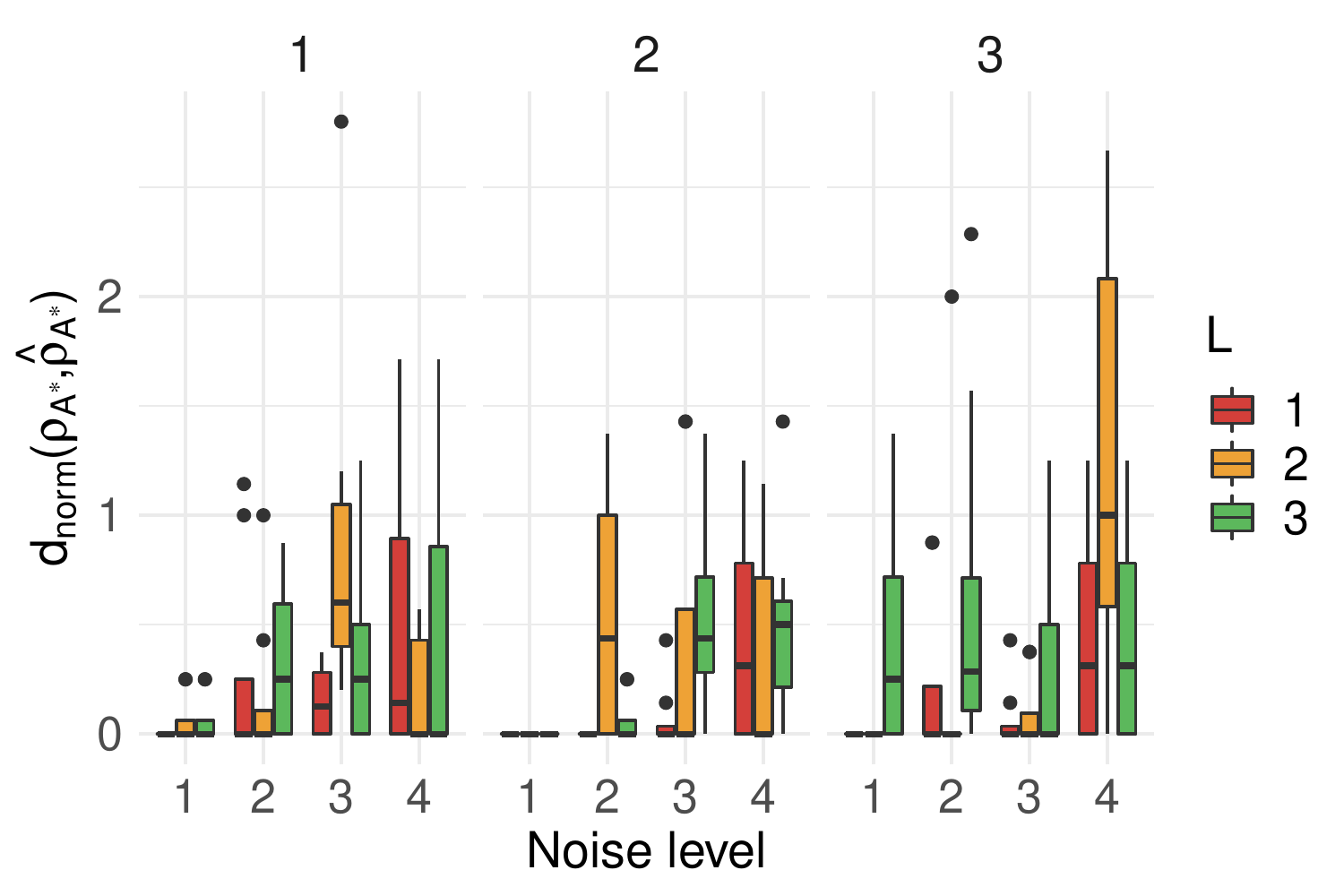}
\endminipage
\caption{Results from the simulation experiments described in Section \ref{sec:sim_noise} with $n=20$, $N=25$, $n^*=8$: boxplots of $d_{\textnormal{norm}}$ for each level of noise added to the data. Left: $\alpha=3;$ right: $\alpha=10$. Each subplot from left to right displays the results obtained when varying the leap size $l=1,2,3$ (leap size), and the color of the bars indicate $L=1,2,3$.}
\label{fig:noise_boxplot}
\end{figure}

In the simulated data experiments conducted so far, we only tested a varying degree of agreement among the rankings provided by the assessors by varying $\alpha$. It could then be argued that both data generating processes were somewhat easy to handle for lowBMM, since the noise structure was only induced by controlling a single model parameter.
To test the robustness of lowBMM to increasingly large noise levels, regardless the value of $\alpha$, we performed another simulation experiment: data were generated as previously, but an increasingly large random perturbation of the rankings was then enforced in the data. Precisely, we first simulated the data according to the top-rank data generating process described in Section \ref{sec:sim_data_processes} (with $n^*=8,$ $n=20$ and $N=25$), and then we iteratively swapped the rankings of items in the top $n^*$ (sampled from the Mallows) with those of items ranked below. In order to perform this iterative rank swapping, we followed the following procedure: at level 1, we take the ``bottom item'' in the true set $\mathcal{A}^*$ (the item whose rank is $n^*$ in the true $\bm{\rho}_{\mathcal{A}^*}$), and we swap the rank for that item with the rank of an item outside of the true set, for 90$\%$ of the assessors. Then, at any level $i>1$, we repeat the same procedure for the items in the true set $\mathcal{A}^*$ whose ranks in $\bm{\rho}_{\mathcal{A}^*}$ are $n^*-i$. So for instance, if we repeat the swapping procedure four times, at level 4 the ranks of four items in the true set have been swapped with the ranks of four items from the outside. The tuning parameters were distributed over the grid $(L,l) \in [1,2,3] \times [1,2,3]$. The results can be seen in Figure \ref{fig:noise_boxplot}, displaying boxplots of $d_{\textnormal{norm}}(\bm{\rho}_{\mathcal{A}^*}, \bm{\hat{\rho}}_{\mathcal{A}^*})$. The results generally show a very good performance of lowBMM even with an increasing level of noise, i.e., several levels in the iterative perturbation/swapping. The figure also suggests that a lower $L$ is more robust to noise, while the leap size $l$ has very little effect: therefore, studying lowBMM's robustness to noise also confirmed the conclusions about the tuning of $L$ and $l$ reached in Section \ref{sec:sim_sens_study}. 

\subsection{Comparison with other methods}\label{sec:sim_comparison_all_methods}

\emilie{To further assess the accuracy and solidity of the lowBMM method, it is also important to compare its performance with competitor approaches. In order to carry out a fair comparison, we set up simulation experiments with varying computational demands, taking inspiration from the simulation study in \cite{zhu2021}. The first experiment is a toy example with a small number of items, $n=20$, so that all compared approaches can manage to provide a solution in a small setting. Two additional experiments with increasing dimensions, instead, are used to test how the various methods are able to handle a larger number of items, specifically: $n=100$ and $n=1000$. The methods included in the comparison were: four Markov chain-based methods $\textnormal{MC}_1$, $\textnormal{MC}_2$, $\textnormal{MC}_3$ \cite{lin2010}, and Cross Entropy Monte Carlo (CEMC) \cite{linding2009}; three Mallows-based methods: the original Mallows Model (MM) \cite{mallows1957}, the Extended Mallows Model (EMM) \cite{li2020}, and the Partition Mallows model (PAMA) \cite{zhu2021}. Where possible, a comparison with the original BMM \cite{vitelli2018} was also included. Finally, we included BORDA \cite{borda} as a reference, as it is a simple method and quite commonly used as a ``baseline'' comparison. The \texttt{TopKLists} R package was used for the MC methods, CEMC and BORDA; \texttt{PerMallows} was used for MM;  \texttt{ExtMallows} was used for EMM; the \texttt{PAMA} package was used for PAMA; and finally the \texttt{BayesMallows} package was used for BMM. All methods were implemented for all simulation experiments, however some methods did not manage to complete/converge by the set time limit. }

\emilie{We computed three performance measures in order to compare and evaluate the methods. Let $\bm{\rho}_{\mathcal{A}^*}$ be the true consensus ranking of the relevant items, and $\bm{\hat{\rho}}_{\mathcal{A}^*}$ be its estimate (the point estimate for the frequentist methods, and the default posterior summary for the Bayesian ones). Similarly, let $\mathcal{A}^*$ be the true set of relevant items, and $\hat{\mathcal{A}}^*$ be its estimate, with $|\hat{\mathcal{A}}^*|=|\mathcal{A}^*|=n^*$. We then consider the following measures of performance:}

\begin{enumerate}
    \itemsep0em
    \item the footrule distance between the true and the estimated consensus ranking, normalized by the number of correctly selected items $n_\textnormal{corr}$, denoted as $d_{\textnormal{norm}}$ (this is the same measure as described in Section \ref{sec:sim_sens_study}). 
    \item the coverage, i.e. the proportion of correctly selected items $\hat{p} = n_\textnormal{corr}/n^*$. 
    \item the recovery distance $d_{R} = d_\tau(\bm{\rho}_{\mathcal{A}^*}, \bm{\hat{\rho}}_{\mathcal{A}^*}) + (n^* - n_\textnormal{corr}) \times (n + n^* +1)/2$ as defined in \cite{zhu2021}, where $d_\tau$ is the Kendall distance between two rankings, and $n^* - n_\textnormal{corr}$ is the number of wrongly selected items.
\end{enumerate}

\begin{table}[!htb]
\centering
\begin{tabular}{|l|rrrrrrrrr|}
\hline
 & BORDA & CEMC & EMM & lowBMM & MC1 & MC2 & MC3 & MM & PAMA \\ 
   \hline
 & \multicolumn{9}{c|}{\it{Top-rank, $n = 20$}}  \\ 

  \hline
  $d_R$ & \bf{12.24} & 80.71 & \bf{11.02} & \bf{12.06} & 72.52 & 42.78 & 27.26 & 80.53 & 38.14 \\ 
  $\hat{p}$ & \bf{1.00} & 0.43 & \bf{1.00} & \bf{1.00} & 0.50 & 0.75 & 0.88 & 0.42 & 0.78 \\ 
  $d_\textnormal{norm}$ & 5.29 & 19.03 & \bf{4.87} & 5.17 & 14.07 & 8.35 & 6.64 & 18.60 & 7.87 \\ 
  time (sec) & 0.02 & 58.42 & 0.05 & 0.54 & 4.42 & 4.42 & 4.42 & 0.0003 & 17.58 \\ 
   \hline
 & \multicolumn{9}{c|}{\it{Top-rank, $n = 100$}}  \\ 
 \hline
  $d_R$ & \bf{19.88} & $\cdot$ & \bf{19.40} & 21.70 & $\cdot$ & $\cdot$ & $\cdot$ & 525.35 & 249.33 \\ 
  $\hat{p}$ & \bf{1.00} & $\cdot$ & \bf{1.00} & \bf{1.00} & $\cdot$ & $\cdot$ & $\cdot$ & 0.09 & 0.59 \\ 
  $d_\textnormal{norm}$ & \bf{25.79} & $\cdot$ & \bf{24.97} & \bf{25.10} & $\cdot$ & $\cdot$ & $\cdot$ & Inf & 93.28 \\
  time (sec) & 0.64 & $\cdot$ & 3.19 & 1.48 & $\cdot$ & $\cdot$ & $\cdot$ & 0.01 & 876.17 \\ 
  \hline
\end{tabular}
\caption{\label{tab:compare_methods_toy_mid} Average results over 50 repetitions of the top-rank simulation experiments described in Section \ref{sec:sim_comparison_all_methods}. Upper portion of the table: $N=5$, $n=20$, $n^*=8$, $\alpha=2$, $M=1000$; lower portion of the table: $N=10$, $n=100$, $n^*=10$, $\alpha=2$, $M=2000$.}
\end{table}

\begin{figure}[!htb]
\minipage{0.49\textwidth}
  \includegraphics[width=\linewidth]{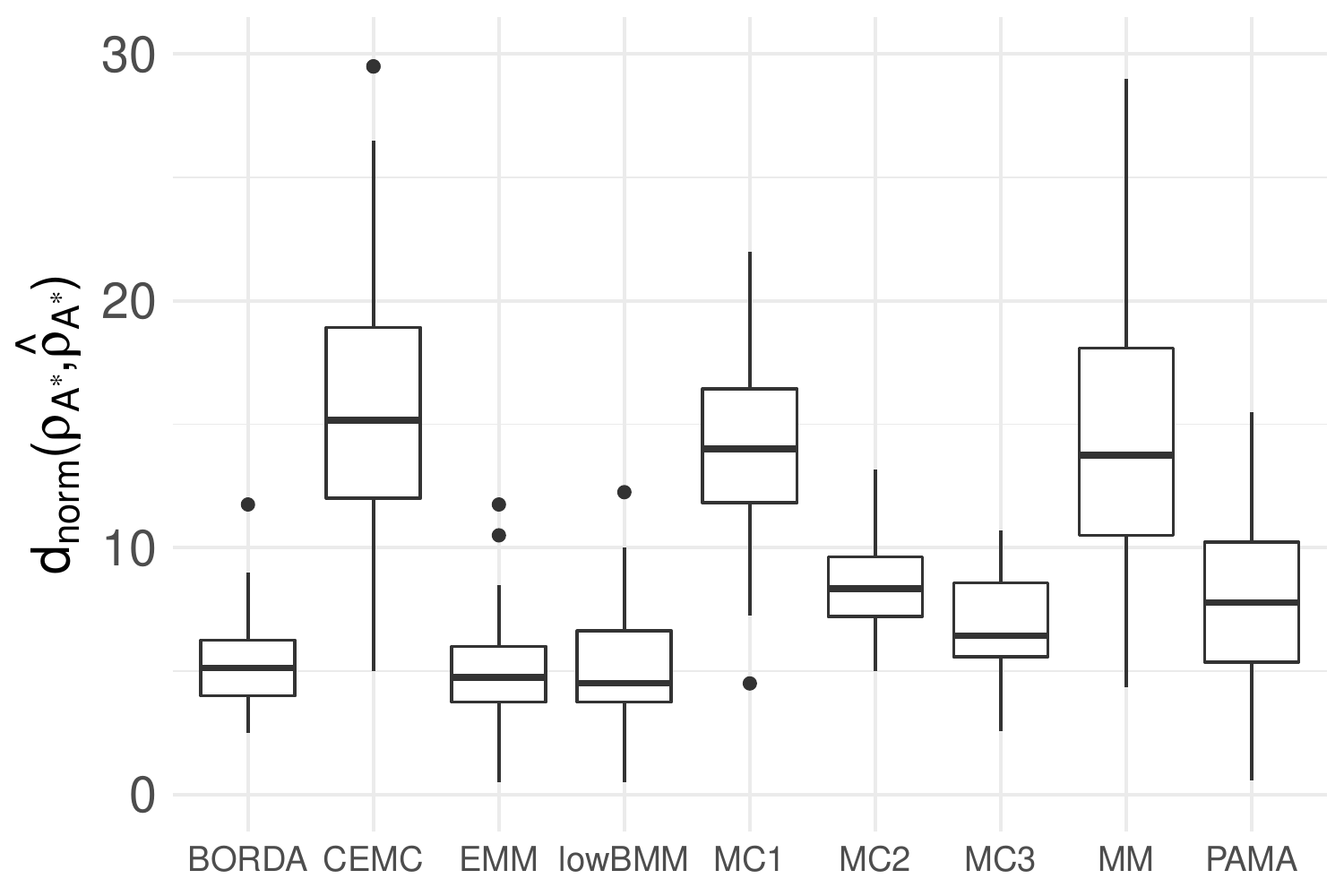}
\endminipage\hfill
\minipage{0.49\textwidth}
  \includegraphics[width=\linewidth]{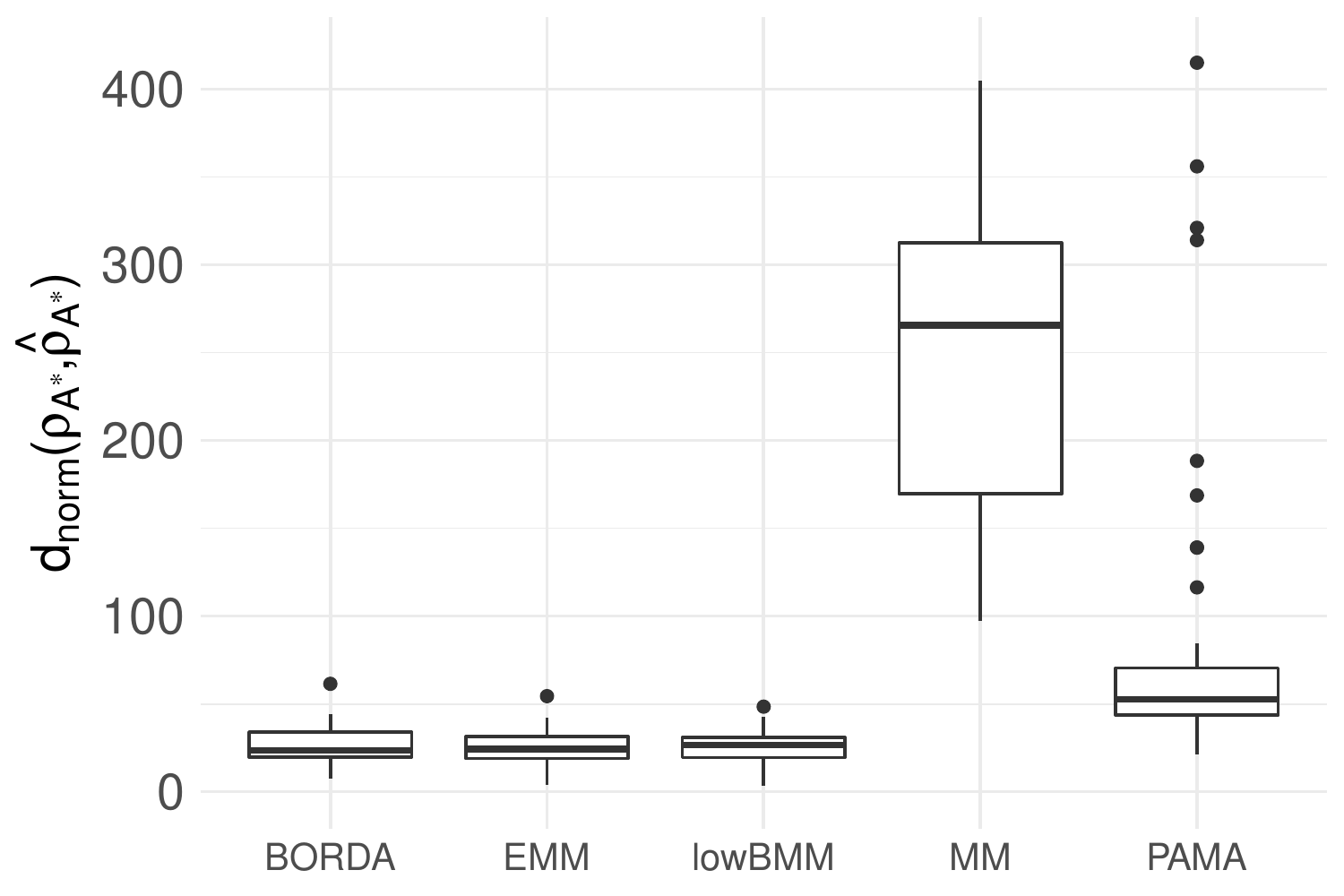}
\endminipage
\caption{Results of the top-rank simulation experiments described in Section \ref{sec:sim_comparison_all_methods}: boxplots of the values of $d_\textnormal{norm}$ over 50 repetitions obtained with several alternative methods. Left panel: $\alpha=2$, $n=20$, $N=5$, $n^*=8$, $M=1000$; right panel: $\alpha=2$, $n=100$, $N=10$, $n^*=10$, $M=2000$.}
\label{fig:sim_compare_methods_toy_mid}
\end{figure}

\begin{figure}[!htb]
\minipage{0.49\textwidth}
  \includegraphics[width=\linewidth]{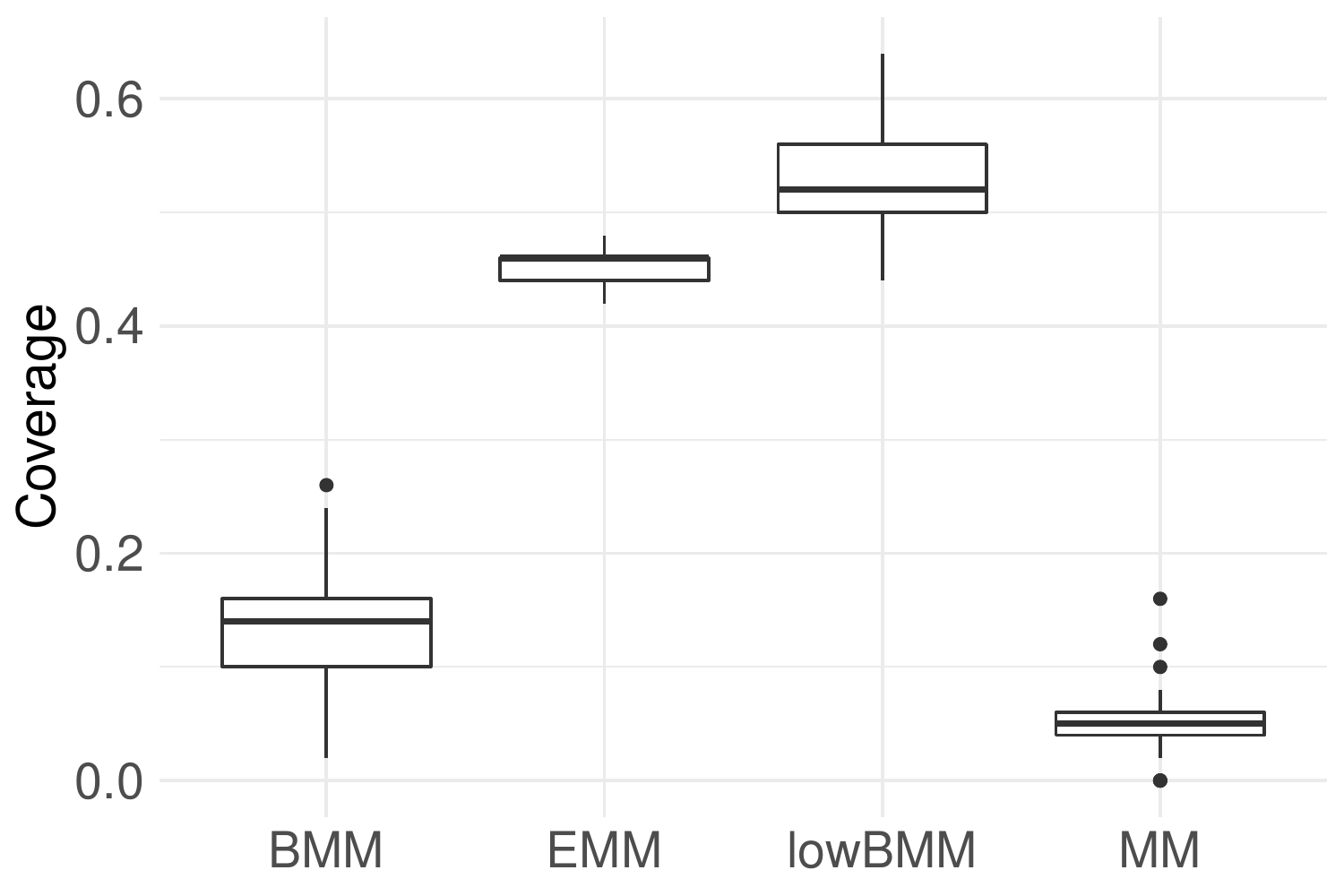}
  \endminipage\hfill
  \minipage{0.49\textwidth}
  \includegraphics[width=\linewidth]{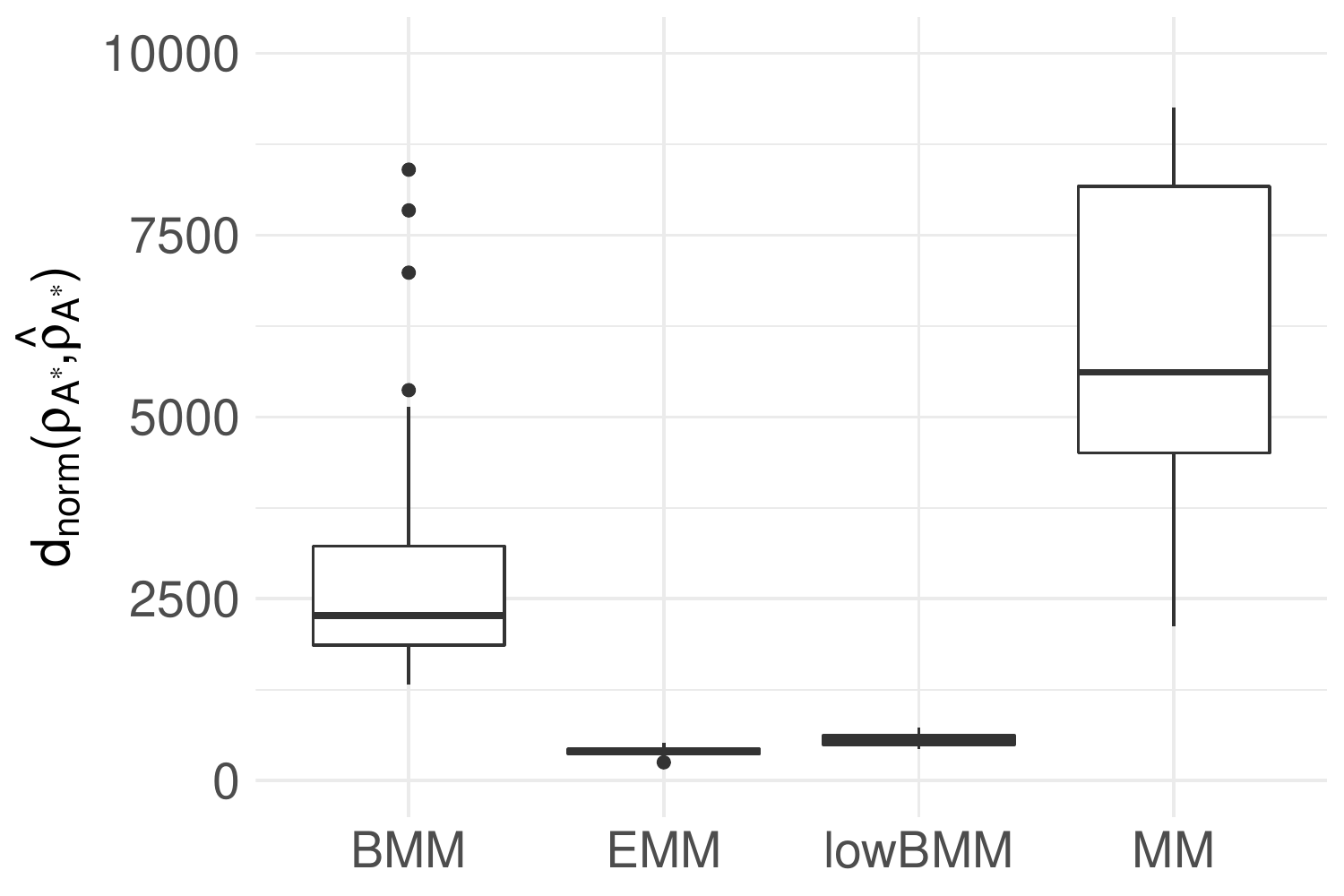}
\endminipage
\caption{Results of the rank consistency simulation experiment described in Section \ref{sec:sim_comparison_all_methods} with $n=1000$, $N=50$, $n^*=50$, $M=7.5\cdot 10^4$, $\alpha=5$: boxplots of the values of $\hat{p}$ (left) and $d_\textnormal{norm}$ (right) over 50 repetitions obtained with several alternative methods (BMM, EMM, lowBMM and MM). }
\label{fig:sim_compare_methods_big_rank_cons}
\end{figure}

\emilie{The results from the first experiment with $n=20$ items are shown in the upper portion of Table \ref{tab:compare_methods_toy_mid}, and in the left panel of Figure \ref{fig:sim_compare_methods_toy_mid}. BORDA, EMM and lowBMM all perform similarly, outperforming the other methods in terms of accuracy, both in the selection of the items, and also in their respective rankings. Similar results were obtained for the experiment with $n=100$ items, where however we were only able to run the methods displayed in the right panel of Figure \ref{fig:sim_compare_methods_toy_mid} in the set time limit of 72hrs (see also the lower portion of Table \ref{tab:compare_methods_toy_mid}). Finally, concerning the large experiment with $n=1000$ items, we simulated data according to the two data generating processes (top-rank and rank consistency) described in Section \ref{sec:sim_data_processes}. Results are reported in Table \ref{tab:compare_methods_big}: only the Mallows-based methods, and not the PAMA, were able to converge in these scenarios. The results for the top-rank simulation experiment are similar to those obtained for the smaller set of items, however with EMM performing slightly better in terms of $d_R$ and $\hat{p}$. On the other hand, EMM was by far the slowest method in this computationally demanding scenario, as the algorithm is based on an iterative optimization procedure and not on MCMC-sampling. In the rank consistency experiment, lowBMM is performing slightly better than EMM for $d_R$ and $\hat{p}$ (see Figure \ref{fig:sim_compare_methods_big_rank_cons}, and the right portion of Table \ref{tab:compare_methods_big}). These results also show that BMM and MM are not even close to converging to the correct solution in the rank consistency scenario. It is also worth noting that lowBMM and BMM are Bayesian methods that provide uncertainty quantification, while EMM and MM are methods that result in point estimates. }

\begin{table}[!htb]
\centering
\begin{tabular}{|l|rrrr|rrrr|}
\hline
\multicolumn{1}{|l|}{} & \multicolumn{4}{c|}{\it{Top-rank}} & \multicolumn{4}{c|}{\it{Rank consistency}}\\ \hline
\multicolumn{1}{|l|}{} & BMM & EMM & lowBMM & MM & BMM & EMM & lowBMM & MM \\
  \hline
  $d_R$ & 9269.69 & \bf{611.10} & 1621.02 & 25550.13 & 23344.03 & 14776.84 & \bf{12933.85} & 25518.04 \\ 
  $\hat{p}$ & 0.67 & \bf{1.00} & 0.96 & 0.05 & 0.13 & 0.45 & \bf{0.53} & 0.05 \\ 
  $d_\textnormal{norm}$ & 594.76 & \bf{323.83} & 336.73 & Inf & 3105.60 & \bf{398.23} & 557.30 & Inf \\ 
    time (sec) & 116.36 & 2307.42 & 236.92 & 9.48 & 105.41 & 1586.66 & 227.79 & 8.76 \\
   \hline
\end{tabular}
\caption{\label{tab:compare_methods_big}
Average results over 50 repetitions of the large simulation experiment described in Section \ref{sec:sim_comparison_all_methods} with $N=50$, $n=1000$, $n^*=50$, $M=7.5\cdot 10^4$ for the two data generating processes: a top-rank simulation experiment with $\alpha=2$ (left portion of the table), and a rank-consistency simulation experiment with $\alpha=5$ (right portion of the table).}
\end{table}

\normalsize 

\subsubsection{Comparison with the Bayesian Mallows Model (BMM)}\label{sec:sim_comparison_bmm}

Among the performance comparisons of lowBMM with alternative methods, a special role is played by BMM: indeed BMM, as introduced in \cite{vitelli2018} and subsequently implemented in the \texttt{BayesMallows} R package \cite{sorensen2019}, is the original full dimensional version of lowBMM, so we also would like to verify that lowBMM is consistent with the full model in a limiting situation. To this aim, we ran the full BMM as well as the variable selection lowBMM on the same simulated dataset with $n=20$, $N=50$, $n^*=8$. For a comparison of the marginal posterior distribution of $\bm{\rho}$ obtained on the reduced set of items with the lowBMM to the one obtained with BMM on the complete set of items, see Figure S10 in the supplementary material, right and left panel, respectively. Both models show consistent results for the top items, with the full model being slightly more certain in its top ranks. The noisy items seem to be randomly distributed on the bottom in the full model, while in lowBMM they are assigned the bottom ranks.

\begin{figure}[!htb]
\centering
\minipage{\textwidth}
  \includegraphics[width=\linewidth]{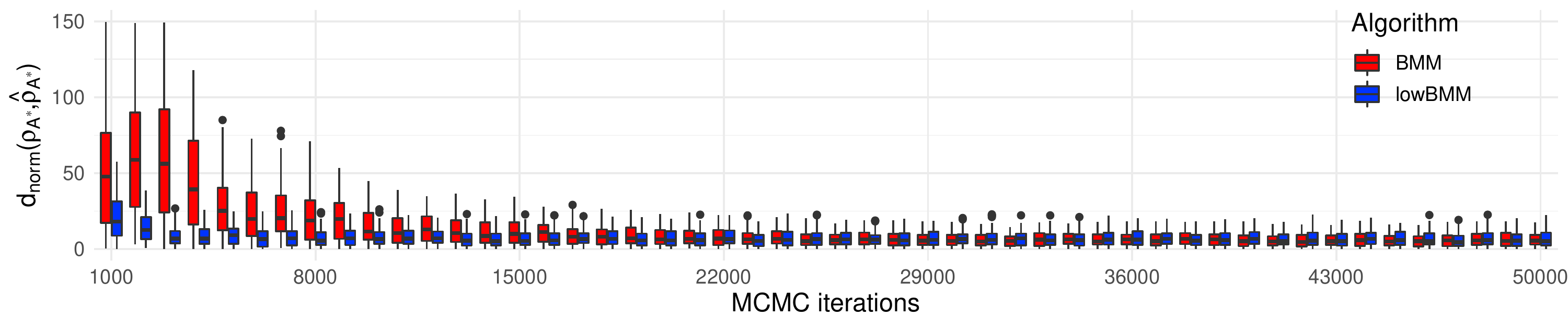}
\endminipage \hfill
\hfill
\minipage{\textwidth}
  \includegraphics[width=\linewidth]{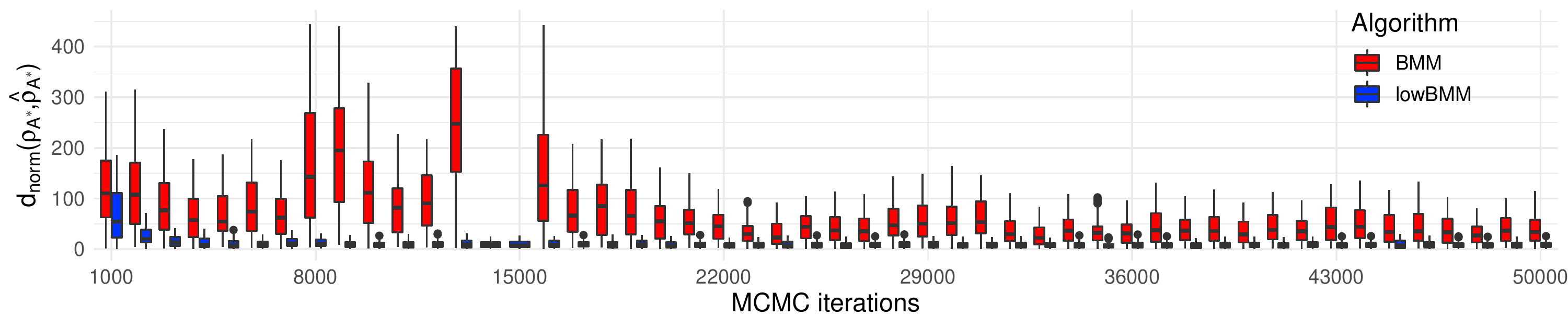}
\endminipage
\caption{Results from the simulation experiment described in Section \ref{sec:sim_comparison_bmm} with $n=1000$, $N=50$, $n^*=50$: boxplots of the posterior distribution of $d_\textnormal{norm}$ evolving along MCMC iterations for BMM (red) and lowBMM (blue). Top panel: top-rank simulation scenario with $\alpha=5$; bottom panel: rank-consistency simulation scenario with $\alpha=2$. }
\label{fig:bmm_comparison_sim_large}
\end{figure}

We then compared BMM and lowBMM on a larger simulated scenario ($n=1000$, $N=50$, $n^*=50$), when using both data generating processes: Figure \ref{fig:bmm_comparison_sim_large} demonstrates that lowBMM outperforms BMM in terms of accuracy and stability in both scenarios. In the top-rank simulation experiment, lowBMM converges to the correct solution with fewer iterations, while BMM reaches the solution at a later stage. This shows the consistency of lowBMM, as we obtain the same solution as for the full model, however in the correct reduced dimension and with lower computational demands. For the rank-consistency simulation experiment, lowBMM clearly outperforms BMM: while lowBMM converges to the solution after 3000 iterations, BMM exhibits an oscillating behaviour, and it never manages to reach convergence. This last simulation example further shows the flexibility in the variable selection procedure under different data generating procedures, even in a high-dimensional setting with basically no possible (Bayesian) competitor.

\subsection{Off-line estimation of $\alpha$} \label{sec:off_line_alpha_simulation}
Finally, we also aimed at testing the approach described in Section \ref{sec:off_line_alpha_notation} for the off-line estimation of $\alpha$, in a simulation setting with the same top-rank data generating process as previously used. We therefore generated datasets with varying dimensions $(N,n)$, where $n^*=\textnormal{round}(n/3)$ items were sampled from the Mallows model with a fixed $\alpha_{\textnormal{true}}$, and the rest of the items were sampled from a uniform distribution. For each dataset in a given dimension, we ran the estimation procedure of Section \ref{sec:off_line_alpha_notation} by generating datasets of equal dimension over a grid of $\alpha_0$ values, and computed the mean distance between assessors in each dataset. Finally, we obtained the final estimate $\hat{\alpha}_{n^*}$ by determining the intersection between the distance measure for the simulated datasets and the distance measure for the dataset generated from $\alpha_{\textnormal{true}}$, in each dimension. Performance of the estimation procedure can be inspected in Table \ref{tab:alpha_sim_est}, where the method stability is demonstrated even for increasing $n$. The estimated values $\hat{\alpha}_{n^*}$ are consistently slightly smaller than $\alpha_{\textnormal{true}}$, which however does not affect the method performance, but simply makes the posterior distributions for $\bm{\rho}$ and $\mathcal{A}^*$ slightly more peaked to absorb the unnecessary uncertainty. We remark that a smaller $\alpha$ than necessary is in theory better for estimation, as it allows for more model flexibility and for better space exploration in the MCMC.

\begin{table}[!htb]
\centering
\begin{tabular}{|l|l|r|r|r|}
\hline
\multicolumn{2}{|l|}{$N$} & \multicolumn{1}{c|}{$10$} & \multicolumn{1}{c|}{$20$} & \multicolumn{1}{c|}{$50$}\\ \hline
\multicolumn{1}{|l|}{$n$} & \multicolumn{1}{|l|}{$n^*$} & \multicolumn{3}{c|}{$\hat{\alpha}_{n^*}$} \\ 
  \hline
     20   & 7 &  2.37 & 2.22 & 2.18\\ 
     150  & 50 & 2.15 & 2.18 & 2.23\\
     500  & 167 & 2.14 & 2.15 & 2.24\\
     1000 & 333 & 2.05 & 2.09 & 2.21\\
   \hline
\end{tabular}
\caption{Results from simulation experiments described in Section \ref{sec:off_line_alpha_simulation}, for testing the off-line estimation of $\alpha$ for varying $(N,n)$ as described in Section \ref{sec:off_line_alpha_notation}. $\alpha_\textnormal{true}=3$ for all scenarios.}
\label{tab:alpha_sim_est}
\end{table}

\section{Application to RNA-seq data from ovarian cancer patients}\label{sec:application}
In this section we describe the application of the lowBMM method on RNAseq data from TCGA ovarian cancer patients. The gene level expression data was available at the TCGA Data Portal (https://tcga-data.nci.nih.gov/tcga/)\footnote{The BAM files were downloaded from CGHub (www.cghub.ucsc.edu), and converted to FASTQs, where all samples were processed similarly aligning reads to the hg19 genome assembly using MapSplice \cite{mapsplice2010}. Gene expression was quantified for the transcript models corresponding to the TCGA GAF 2.13, using RSEM4 and normalized within-sample to a fixed upper quartile \cite{rsem2011}.}, and the selected ovarian cancer patients were the same as previously analysed \cite{hoadley2014}. 

Serous ovarian tumors share similar genetic origin and features with basal-like breast cancers \cite{begg2017examining}: both are difficult-to-treat cancer types, and they are highly commonly characterized at the molecular level, especially for what concerns mutations (with TP53 being the most frequent example). Massive whole-genome data analyses are therefore key to understanding individual tumor biology for these cancer types. This is the main motivation for focusing on RNAseq data from serous ovarian tumors, as selecting basal-like breast cancer patients would require an additional subtyping preprocessing step that we would like to avoid.
Additionally, since our method assumes complete data, preprocessing steps included keeping only the items (genes) with less than 50$\%$ missing values: all of these missing values were imputed using k-nearest neighbor averaging \cite{troyanska2001}. This resulted in a final dataset of ultra-high dimension with $n=15348$ genes and $N=265$ patients.

For running our lowBMM method, we set $n^*=500$, $l=20$ and $L=1$ (tuning parameters were set based on the conclusions from the simulation studies in Section \ref{sec:sim_sens_study}, and from some trial-and-error). We also performed an off-line estimation of $\alpha$ on the experimental dataset, \emilie{on a search grid $\alpha \in [10^{-15}, 10^{-5},1,10,50,100]$ that was set as such to ensure adequate coverage over several orders of magnitude}, as described in Section \ref{sec:off_line_alpha_notation}, resulting in $\alpha=10$. Due to the high-dimensionality of the data, we ran two chains, each with $M=5\times 10^6$ MCMC iterations on a big memory HPC server. We discarded the first $5\times 10^4$ iterations as burn-in in both chains before merging them for post-processing (see Section \ref{sec:mh_postprocess} for details on the post-processing of the results). As $\mathcal{A}^*$ is a probabilistic selection, we will get slightly different posterior summaries from different runs of lowBMM. Moreover, $\mathcal{A}^*$ is a discrete parameter defined in a space whose dimension grows exponentially with $n$, and in this particular case the space will be exceedingly large. This will in turn result in small variations in each run of the algorithm, as we are limited by computational capacities in how much we can realistically explore this space. Furthermore, this behaviour can also be affected by aspects related to the current problem setting: particularly as the $n^*$ value used in the analysis might be too low compared to $n$. We remark that the \emilie{sensitivity study on the tuning parameters} in Section \ref{sec:sim_sens_study} suggested that running lowBMM with a smaller $n^*$ compared to the ``truth'' might make it more difficult for the method to converge to the true $\bm{\rho}_{\mathcal{A}^*}$. Moreover, during the off-line estimation of $\alpha$, we estimated the $n^*$ that best fits the experimental data to $n^* \approx 2400$. However, we kept the final $n^* = 500$ value as it is more suited to allow results interpretability, and as it also allows the MCMC to perform a good exploration of the state space given the current computational capacity. The computations were performed on resources provided by UNINETT Sigma2, the National Infrastructure for High-Performance Computing and Data Storage in Norway. 

\begin{figure}[!htb]
\minipage{0.48\textwidth}
  \includegraphics[width=\linewidth]{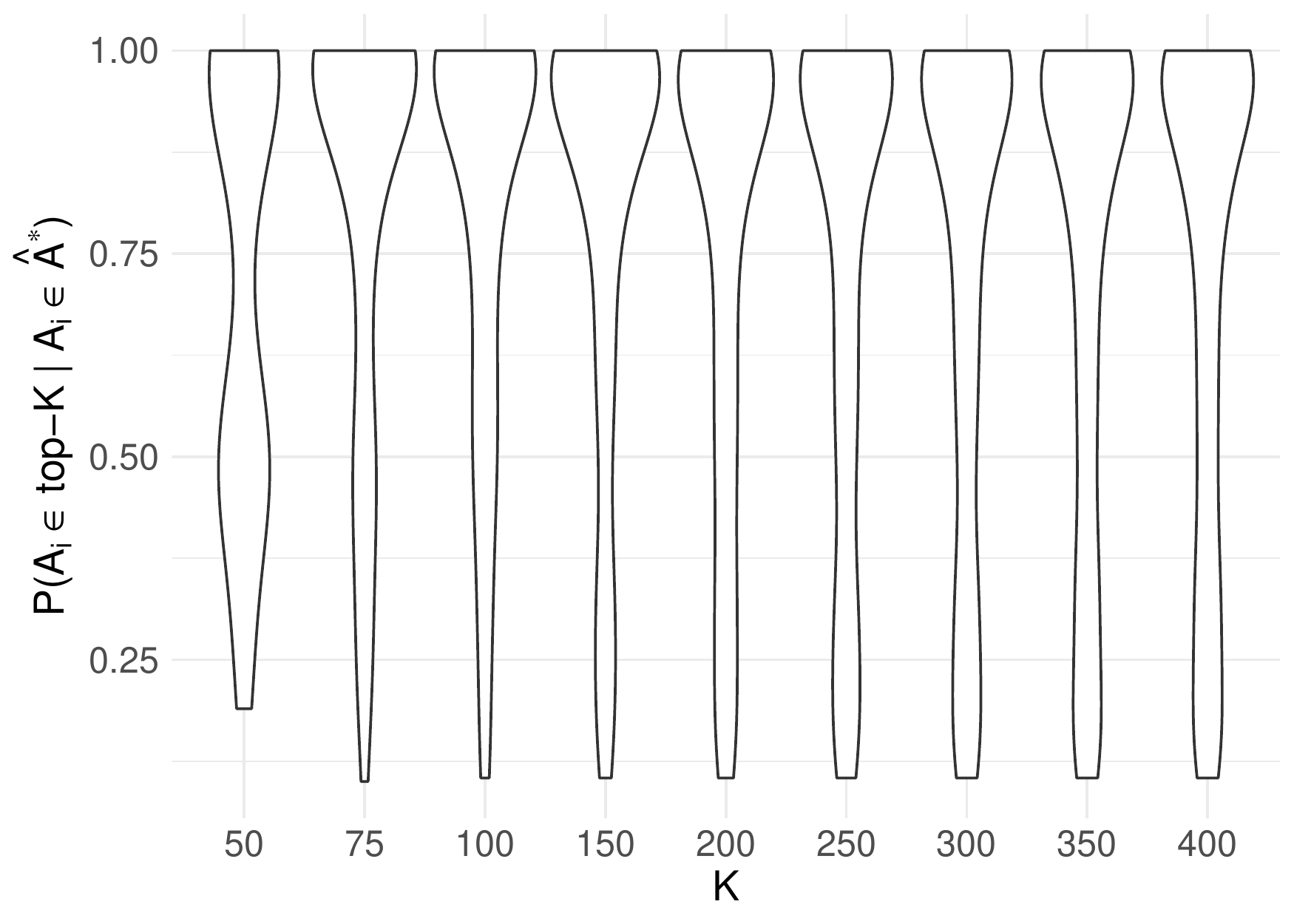}
\endminipage\hfill
\minipage{0.48\textwidth}
  \includegraphics[width=\linewidth]{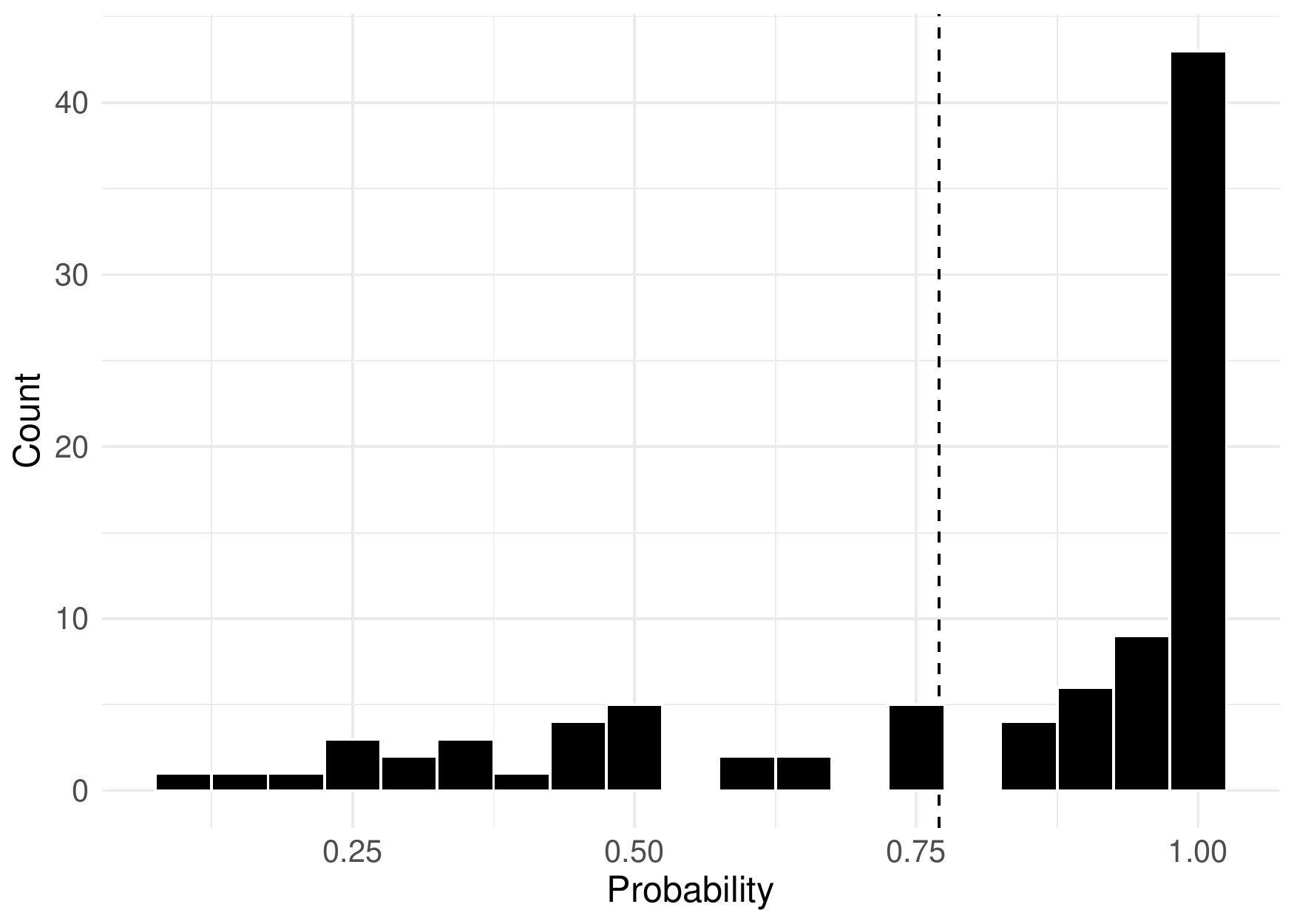}
\endminipage
\caption{Results from the analysis of the RNAseq data described in Section \ref{sec:application}. Left: violin plots of $P(A_i \in \textnormal{top-}K \,  | \, A_i  \in \mathcal{A^*})$ for all items $A_i$, $i=1,...,n$ and for varying $K$. Right: histogram of $P(A_i \in \textnormal{top-}75 \,  | \, A_i  \in \mathcal{A^*})$ for all items $A_i$, $i=1,...,n$, with dashed line at $c=0.77$. $n^*=500$, $\alpha=10$, $L=1$, $l=10$, $M=10^7$ and burn-in=$10^5$.}
\label{fig:ovarian_violin_hist}
\end{figure}

To gain insight on the results provided by lowBMM on the experimental dataset, we computed a \emilie{``top probability selection'' as described in Section \ref{sec:mh_postprocess}}, to be used in a Gene Set Enrichment Analysis (GSEA) \cite{subramanian1005, mootha2003}. This analysis was done to verify whether the method could identify common genes related to cancer. To this aim, we computed \emilie{various top-$K$ sets, with a threshold for the lower items: genes that had probability less than $0.1$ of being in the different top-$K$ sets were dropped from the plot.} This resulted in the violin plots shown in Figure \ref{fig:ovarian_violin_hist}, left panel: this plot indicates that $K=75$ is a good choice for allowing many genes to have a quite large posterior probability of being top-ranked. We therefore further inspected the probability distribution of the top-75 probabilities across the genes, by drawing a histogram of the same probabilities (Figure \ref{fig:ovarian_violin_hist}, right panel): this second plot allowed us to select a cut-off point of $c=0.77$ for the top-75 probability set, thus resulting in a final selection of $|\mathcal{\hat{A}}^*_{\textnormal{top}}|=63$ genes where $\mathcal{\hat{A}}^*_{\textnormal{top}} = \{ A_i \in \mathcal{A}^* \, \textnormal{s.t.} \, P(A_i  \in \textnormal{top-}75, i=1,...,n \, | \, A_i  \in \mathcal{\hat{A}^*}) > 0.77 \}$.

\begin{figure}[!htb]
\centering
\minipage{\textwidth}
  \includegraphics[width=\linewidth]{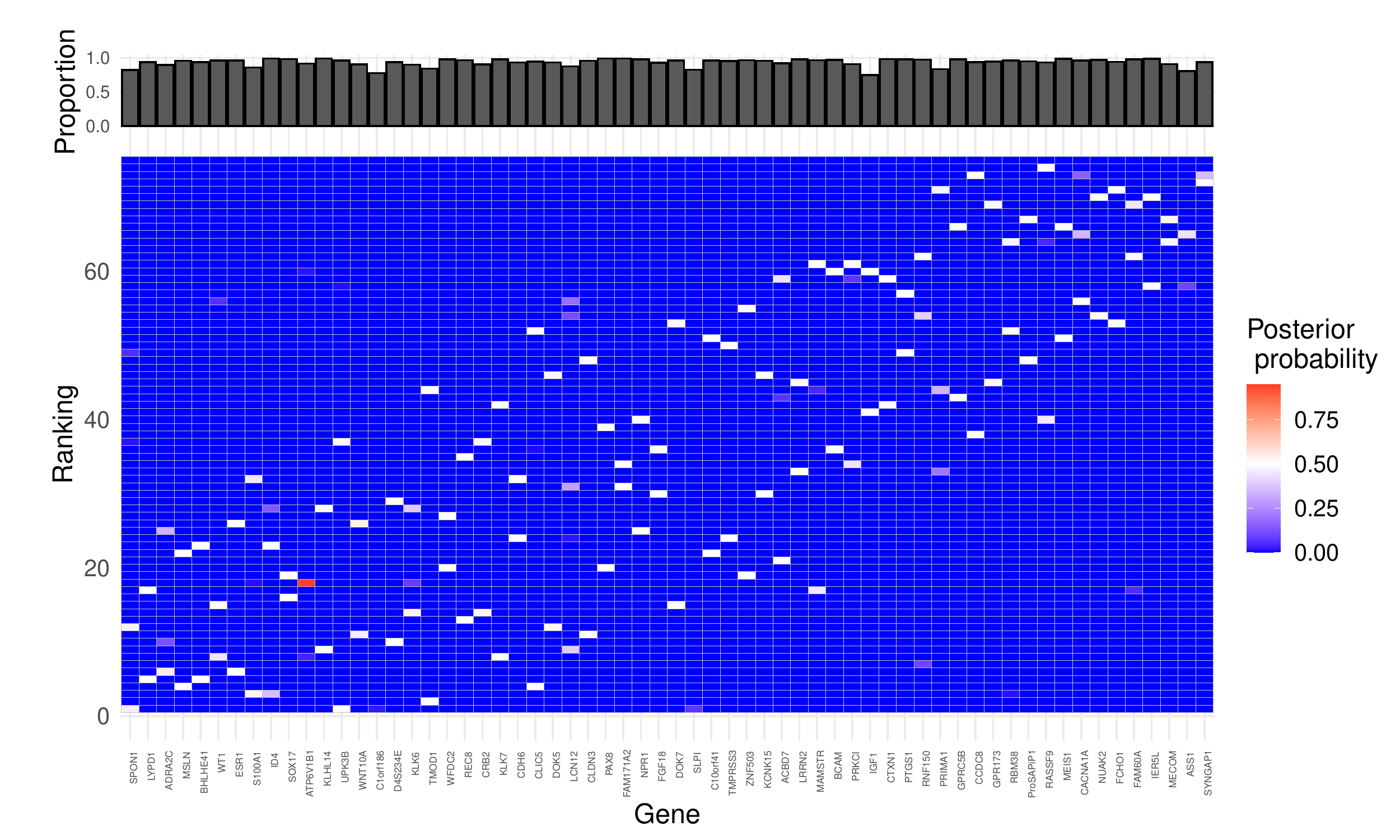}
\endminipage
\caption{Results from the analysis of the RNAseq data described in Section \ref{sec:application}: heatplot of the marginal posterior of $\bm{\rho}$ depicted only for the items in the ``top probability selection'' $\mathcal{\hat{A}}^*_{\textnormal{top}}$, and with the items ordered as in $\bm{\Hat{\rho}}_{\mathcal{A}^*}$ on the x-axis. The bars on top indicate the proportion the item was selected in $\mathcal{A}^*$ over all MCMC iterations, with $\alpha=10$, $L=1$, $l=10$, $M=10^7$ and burn-in=$10^5$.}
\label{fig:ovarian_refined_sel}
\end{figure}

Figure \ref{fig:ovarian_refined_sel} shows that the items in the refined selection $\mathcal{\hat{A}}^*_{\textnormal{top}}$ are selected with very large posterior probability, even if they sometimes ``compete'' for some of the available ranks: this effect had already been observed in the context of the simulation studies in Section \ref{sec:sim_sens_study}, and it might be due to $n^*$ being smaller than $n^*_\textnormal{true}$. There is overall more variability in the rankings of the items as compared to the simulation studies, which is to some degree expected in experimental data with a higher degree of noise. Higher variability in the ranks was also typical for the simulation studies where $n^*_\textnormal{guess}<< n^*_\textnormal{true}$ (see Figure S5 in the supplementary material).

The GSEA indicated that the lowBMM identified known genes involved in cell differentiation and cancer development. The top-ranked genes selected in $\mathcal{\hat{A}}^*_{\textnormal{top}}$ were enriched in pathways such as cell differentiation, morphogenesis and developmental processes (Table \ref{tab:gsea}), suggesting a role in the deregulation of cellular identity essential for cancer development.


\begin{table}[!htb]
\scriptsize
\centering
\begin{tabular}{l|r|r|r}
Gene Set Name & \# Genes in Gene set & \# Genes in overlap & p-value \\\hline
Regulation of cell differentiation (GOBP) & 1618 &	16 & 2.59E-9 \\
Regulation of multicellular organimsal development (GOBP) & 1397 & 13 &	2.05E-7 \\
Regulation of anatomical structure morphogenesis (GOBP) & 1006 & 11 &	4.17E-7 \\
Positive regulation of developmental process (GOBP) & 1284 & 12 &	6.17E-7 \\
Positive regulation of cell differentiation (GOBP) & 844 & 10 &	7.2E-7 \\
Response to endogenous stimulus (GOBP) & 1624 &	13 &	1.12E-6 \\
Cellular response to nitrogen compound (GOBP) & 	698	 & 9 &	1.37E-6 \\
Sensory organ development (GOBP) &	534	& 8	 & 1.84E-6 \\
Animal organ morphogenesis (GOBP) & 1025 & 10 & 4.08E-6 \\
Striated muscle cell differentiation (GOBP) & 269 & 6 & 4.12E-6
\end{tabular}
\caption{Overview of top-10 gene sets from GSEA for the selection $\mathcal{\hat{A}}^*_{\textnormal{top}}$, $|\mathcal{\hat{A}}^*_{\textnormal{top}}| = 63$.}
\label{tab:gsea}
\end{table}

\section{Discussion}\label{sec:discussion}
In this paper, we have developed a novel rank-based unsupervised variable selection method for high-dimensional data. The method extends the applicability of the Bayesian Mallows ranking model to unprecedented data dimensions and provides a much better modeling approach. Indeed, an important advantage of the variable selection procedure is the ability to distinguish relevant genes from the background ones, and to provide an estimate of the relative importance of the selected relevant genes. Furthermore, the variable selection procedure is able to work in high-dimensional settings where the number of items is $n >> 10^3,$ with no competitor model for ranking data capable of scaling to such setting. 
The simulation studies described in Section \ref{sec:simulations} showed that the method performs well on datasets of varying sizes and under varying data generating procedures. \emilie{ Additionally, lowBMM is superior in terms of accuracy and computational time compared to existing methods in high dimensions, and has no competitor in \emph{ultra}-high dimensions.} We applied the proposed model to RNAseq data of ovarian cancer samples from TCGA (Section \ref{sec:application}). Although we consider a cancer genomics application, the methodological contribution given in this paper is very general, and can thus be applied to any high-dimensional setting with heterogeneous rank data.

Throughout this work, we have assumed that the number of relevant items to be selected, $n^*$, is known. This is reasonable in many practical problems where a specific $n^*$ can be tuned according to the research questions/demands or has to be decided according to computational capacity. Simulation studies showed that assuming $n^*_\textnormal{guess}<< n^*_\textnormal{true}$ might pose problems in high-dimensional settings, and therefore the joint estimation of this parameter as part of the existing hierarchical framework constitutes a very interesting direction for future research.

\emilie{Nonetheless, as mentioned in the simulation studies while commenting upon the marginal posterior distribution of $\bm{\rho}$ obtained when $n^*_\textnormal{guess} < n^*_\textnormal{true}$, our method can be seen as an alternative to the constrained median bucket order technique introduced in \cite{dambrosio2019}. It would then be interesting to fully compare the respective estimates obtained via lowBMM and the median bucket order, in the light of the relationship between the choice of permutation space (strong vs weak rankings), distance (any distance vs the Kemeny only), and estimation procedure (Bayesian vs frequentist). This comparison is clearly out of the scopes of the present paper, but makes up a very interesting future research direction.}

The current version of lowBMM works under several assumptions: no handling of missing data, no clustering, fixed scale parameter $\alpha$, all aspects which we plan to implement in future versions of the method. The somewhat challenging procedure of choosing the tuning parameters $l$ and $L$ might be solved with an adaptive MCMC and would allow the algorithm to explore the space more thoroughly. We leave this task for future consideration.

To our knowledge, no rank-based unsupervised variable selection procedure that is capable of scaling to the common data dimensions in -omics applications currently exists.

\bibliographystyle{plain}
\bibliography{ms_review}

\end{document}


\maketitle

\section{}\label{supp:alpha3_vs_alpha10_nstar_true}
\begin{figure}[H]
\minipage{0.49\textwidth}
  \includegraphics[width=\linewidth]{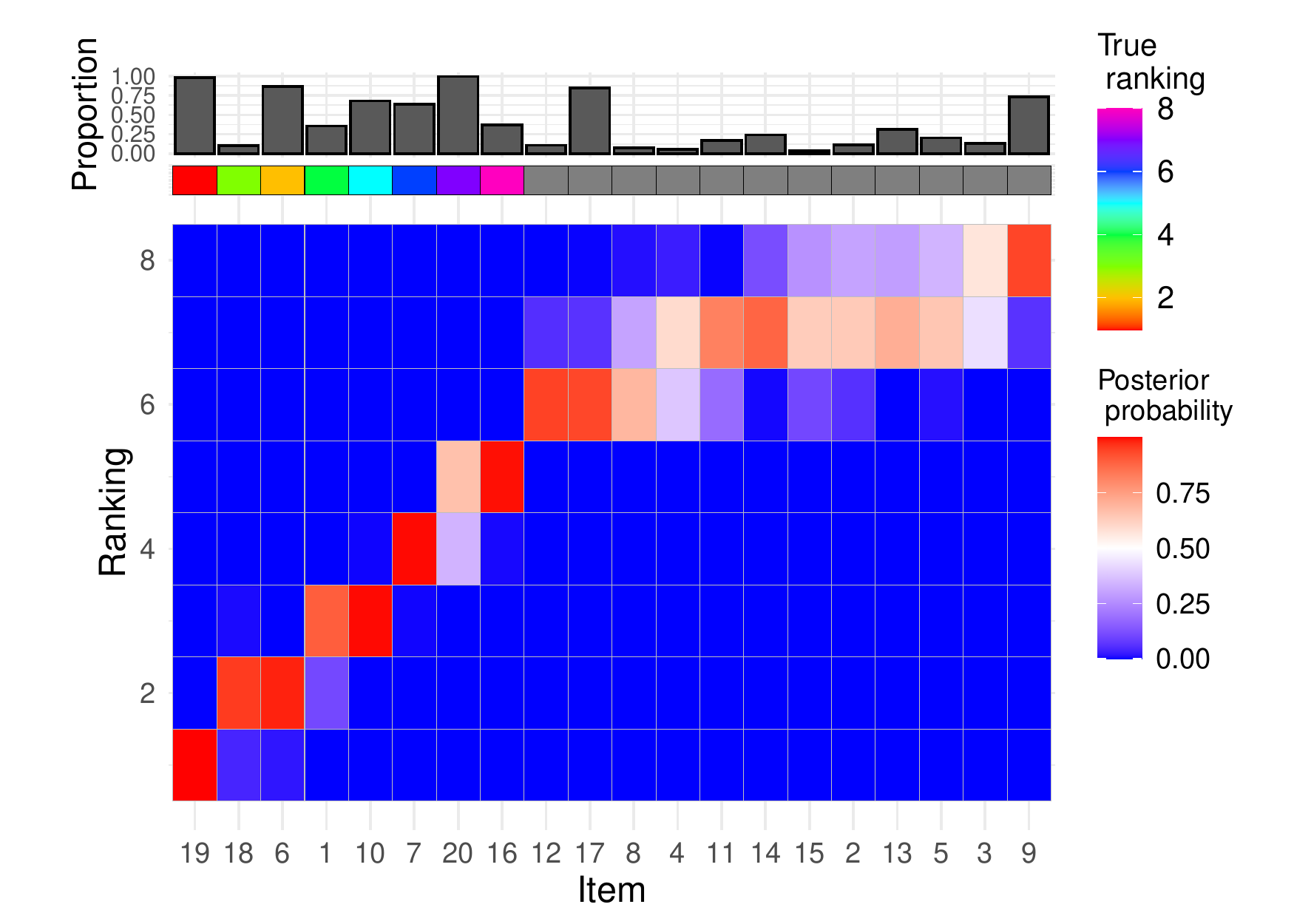}
\endminipage\hfill
\minipage{0.49\textwidth}
  \includegraphics[width=\linewidth]{fig/probitems_top20_item_selection_alphafixed10_simulation_top_items_nstartrue8_nstar8_L1_leap2_acceptance_v1_with_itemlabels.pdf}
\endminipage
\caption{Results from the top-rank simulation experiments described in Section 3.1: heatplots of the marginal posterior distribution of $\bm{\rho}$, where the items have been ordered according to $\bm{\Hat{\rho}}_{\mathcal{A}^*}$ on the x-axis. Left $\alpha=3$, right $\alpha=10$. The rainbow grid indicates the true $\bm{\rho}_{\mathcal{A}^*}$, and the bar plot indicates the proportion of times the items were selected in $\mathcal{A}^*$ over all MCMC iterations. $n=20$, $N=50$, $n^*=8$, $L=1$ and $l=2$. }
\end{figure}

\section{}\label{supp:sens_study_prop_correct}
\begin{figure}[H]
\minipage{0.33\textwidth}
  \includegraphics[width=\linewidth]{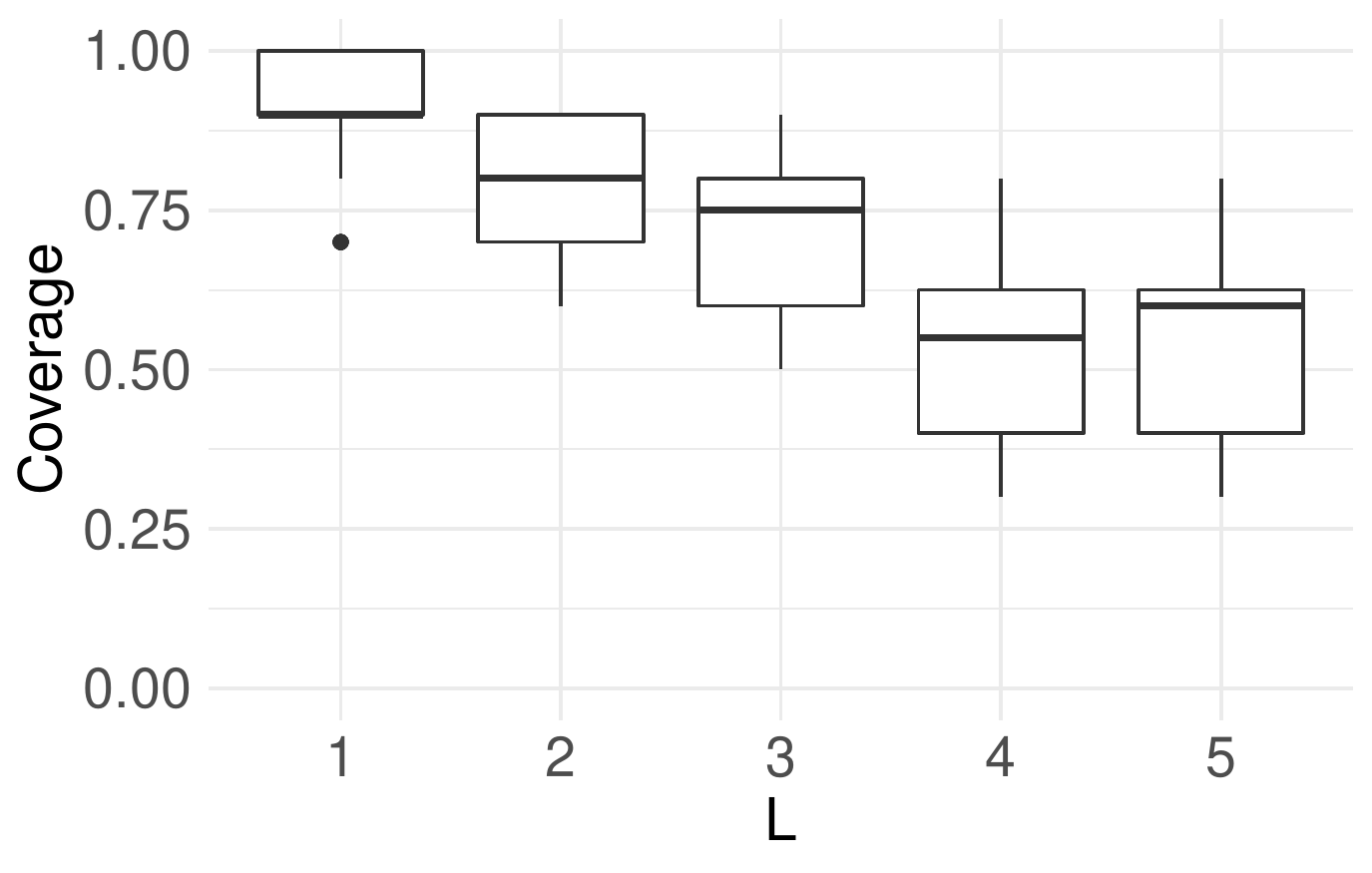}
\endminipage\hfill
\minipage{0.33\textwidth}
  \includegraphics[width=\linewidth]{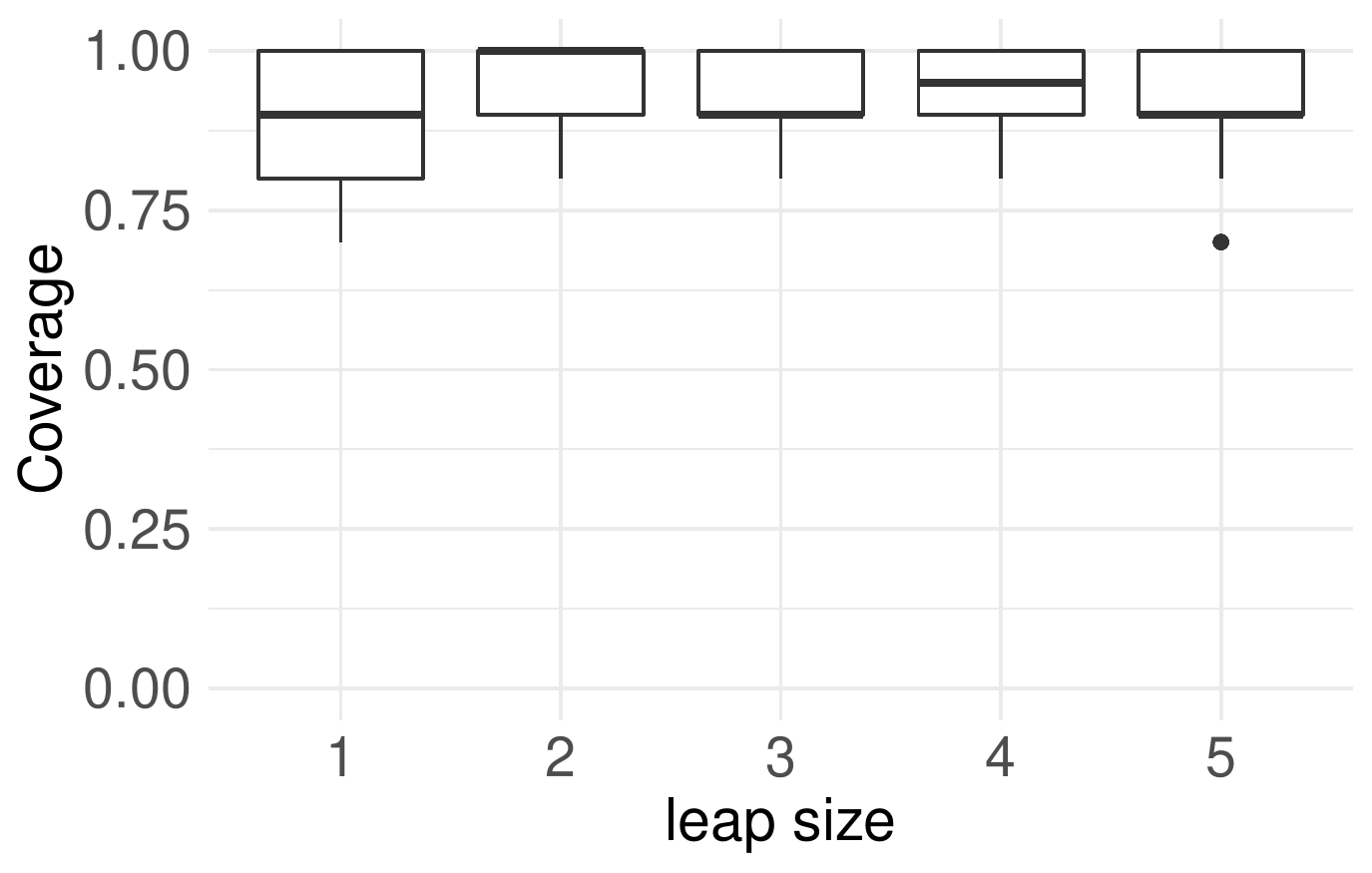}
\endminipage
\minipage{0.33\textwidth}
  \includegraphics[width=\linewidth]{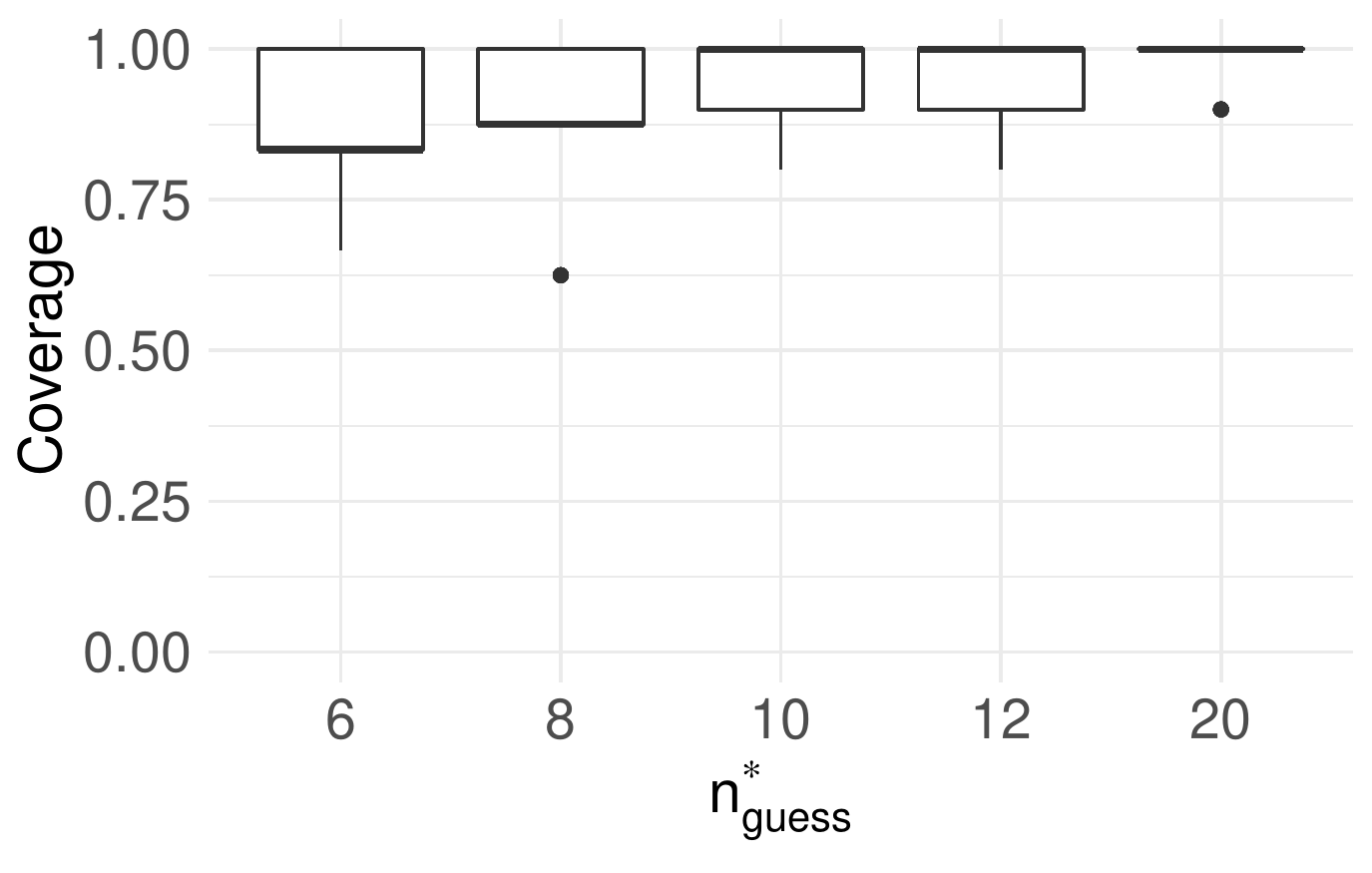}
\endminipage
\caption{Results from the sensitivity study described in Section 3.2: each panel displays boxplots of the proportion of correct items selected, $\Hat{p},$ over 20 runs on different datasets, for varying values of the tuning parameters on the x-axis. From left to right: varying $L$, $l$ and $n^*_\textnormal{guess}$, respectively. $n=100$, $N=10$, $n^*=10$, $\alpha=5$, $M=5000$.}
\end{figure}

\section{}\label{supp:alpha3_nstar_true_3by3plot}
\begin{figure}[H]
\minipage{0.33\textwidth}
  \includegraphics[width=\linewidth]{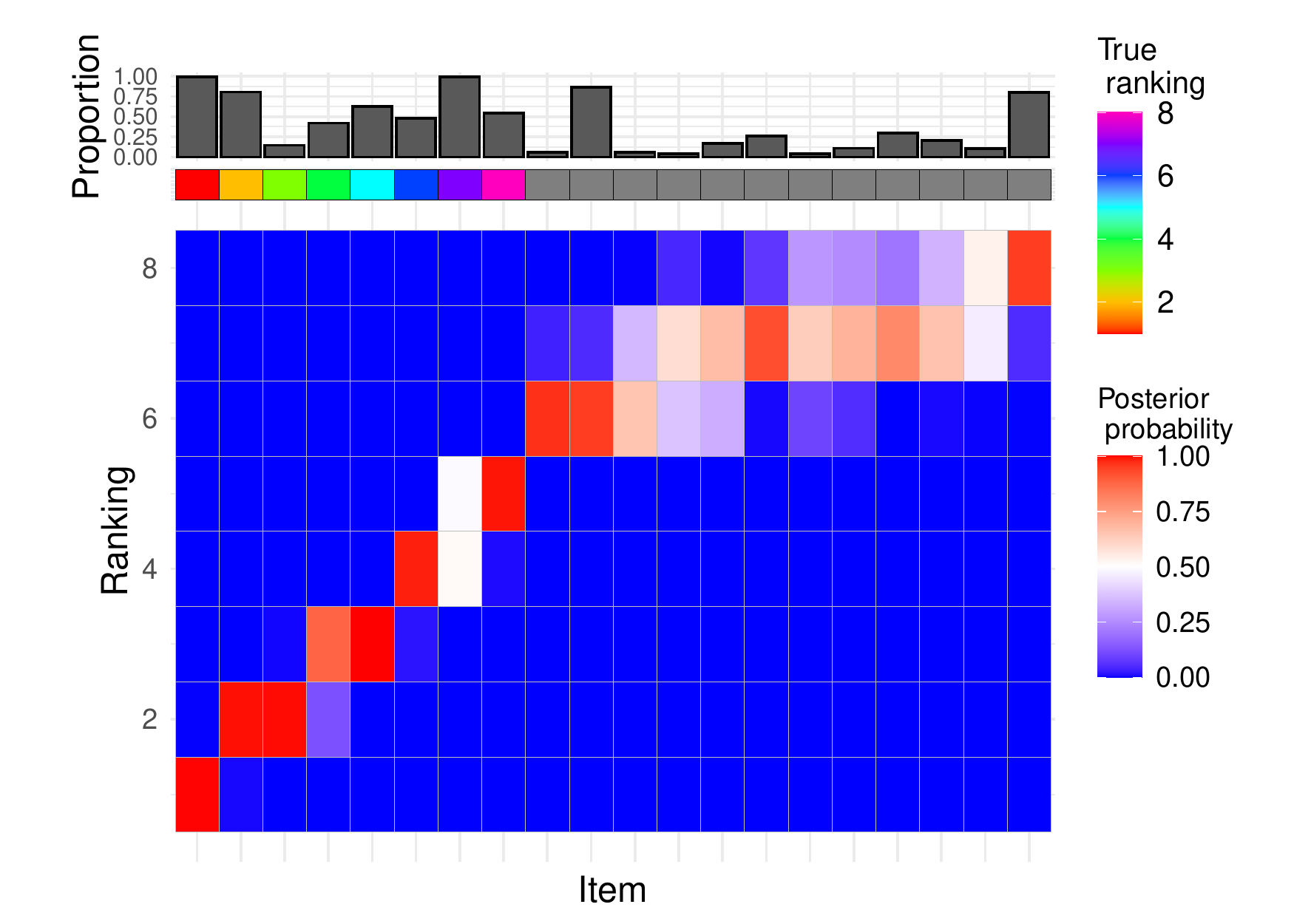}
\endminipage\hfill
\minipage{0.33\textwidth}
  \includegraphics[width=\linewidth]{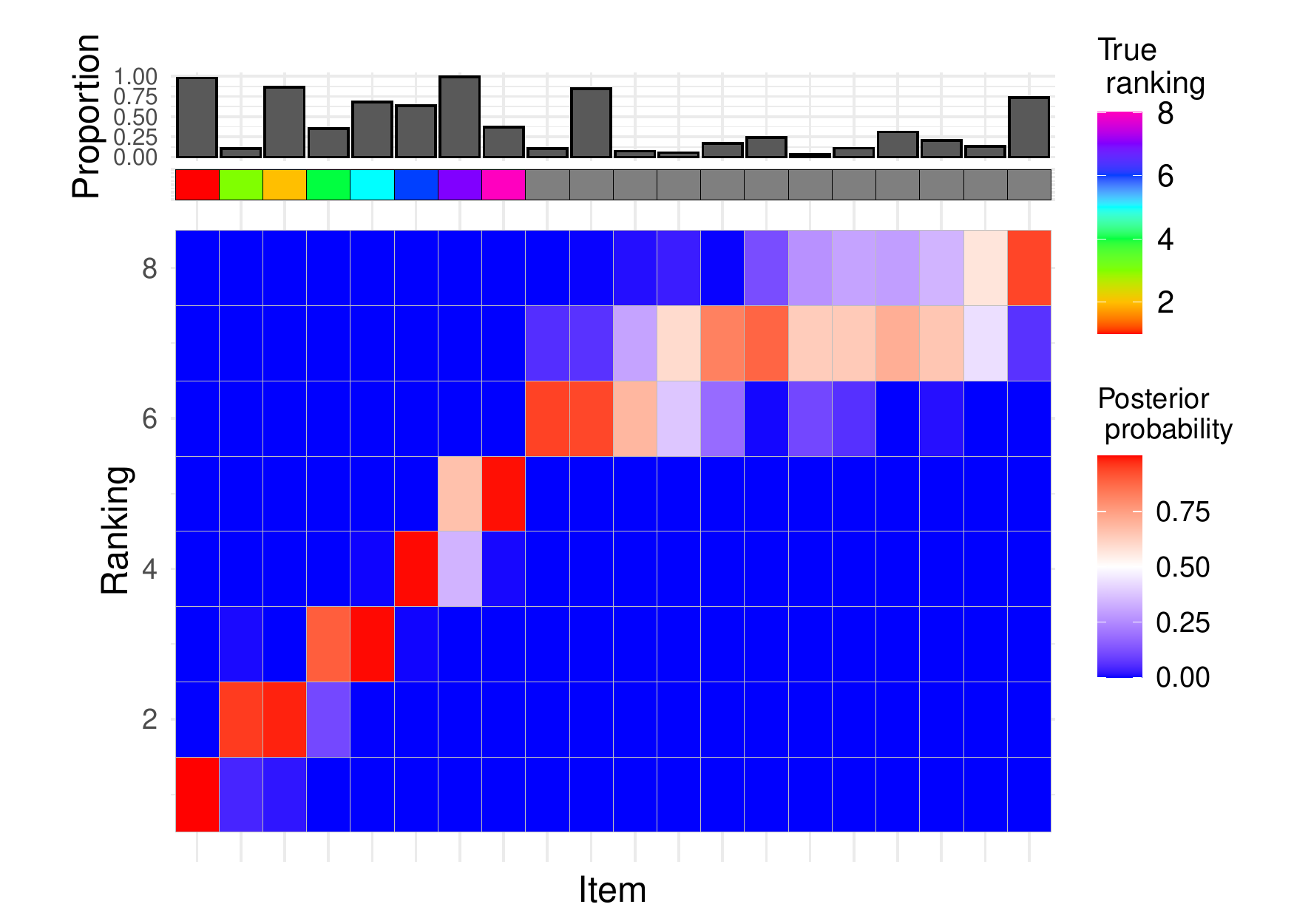}
\endminipage\hfill
\minipage{0.33\textwidth}%
  \includegraphics[width=\linewidth]{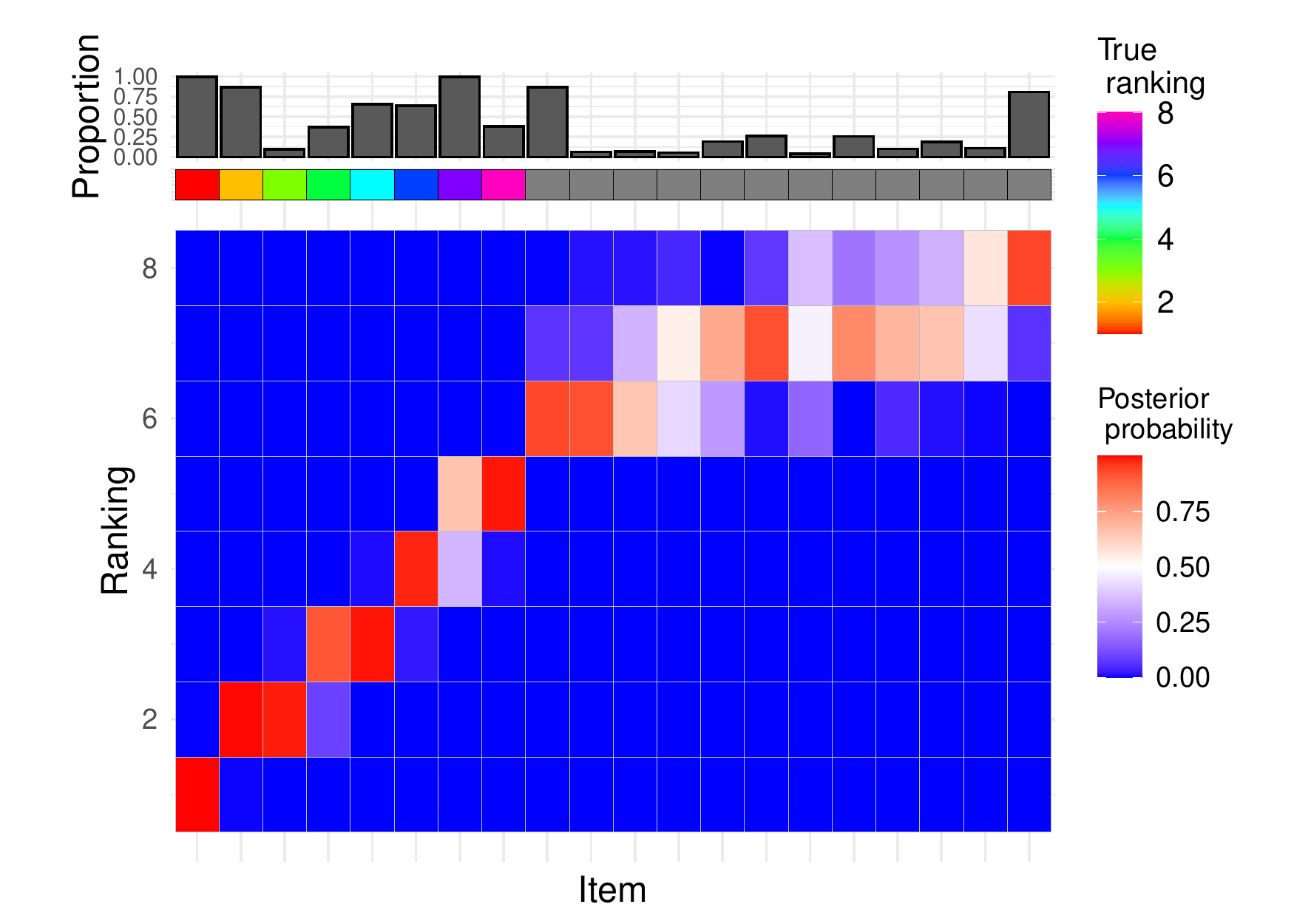}
\endminipage

\minipage{0.33\textwidth}
  \includegraphics[width=\linewidth]{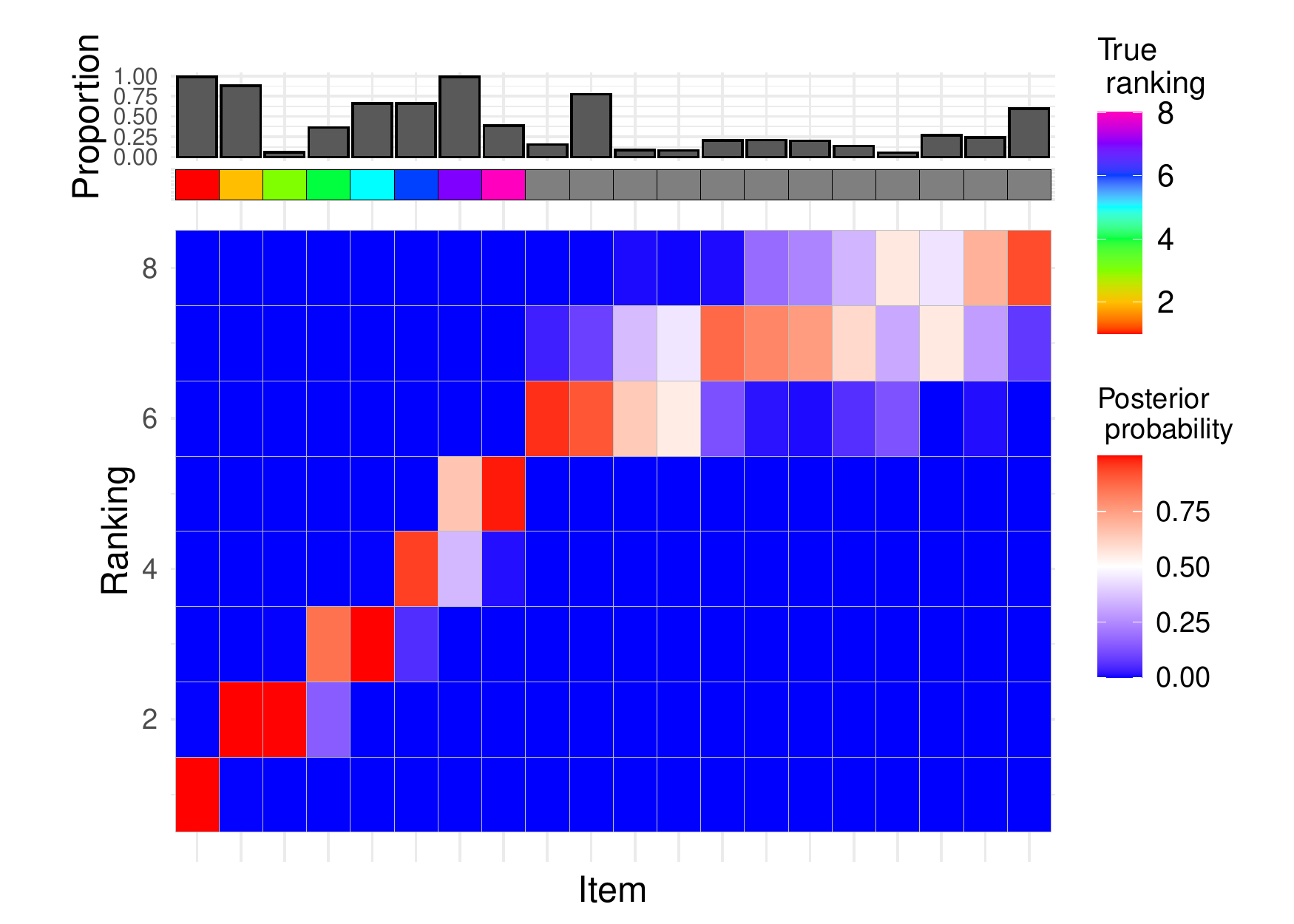}
\endminipage\hfill
\minipage{0.33\textwidth}
  \includegraphics[width=\linewidth]{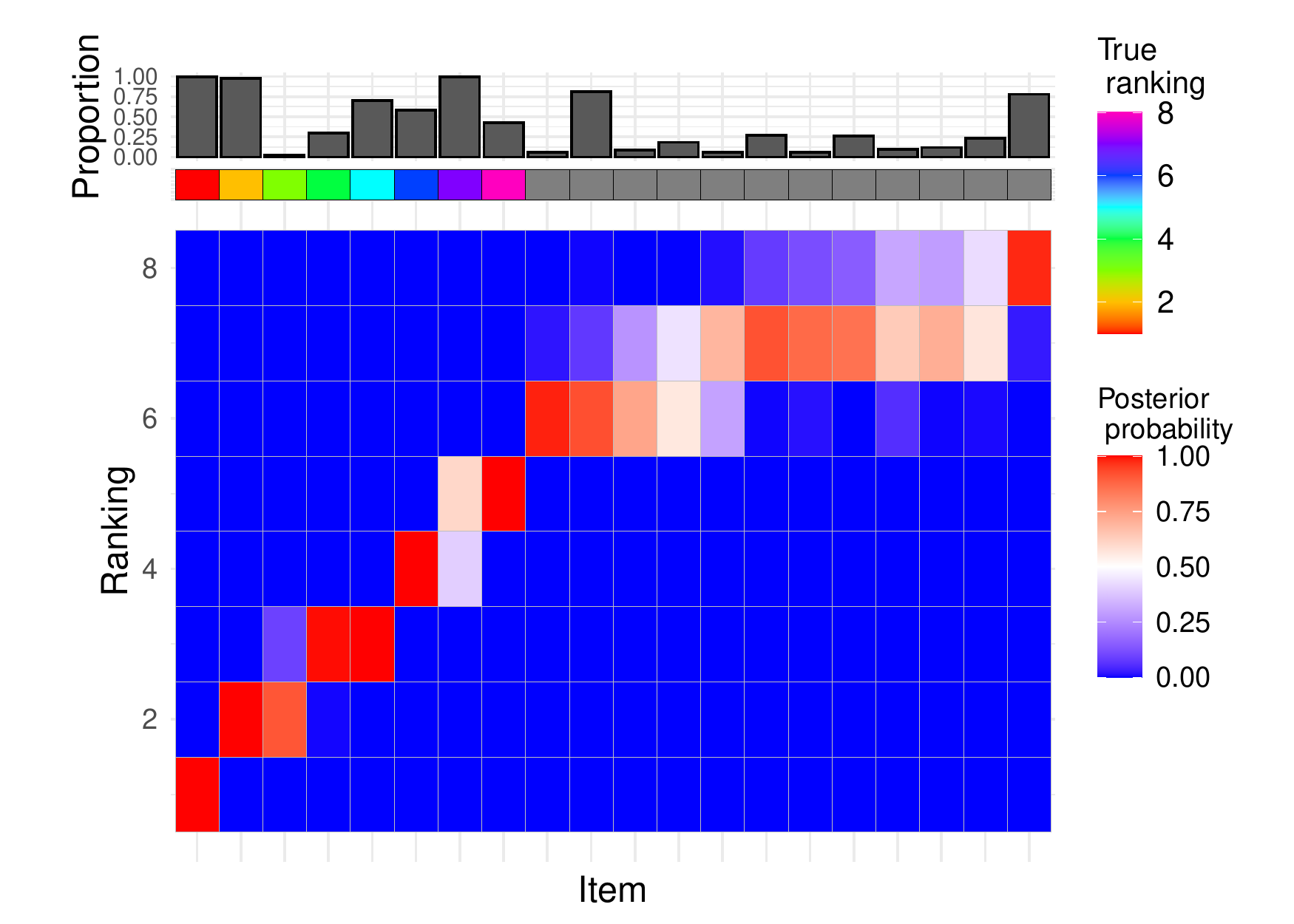}
\endminipage\hfill
\minipage{0.33\textwidth}%
  \includegraphics[width=\linewidth]{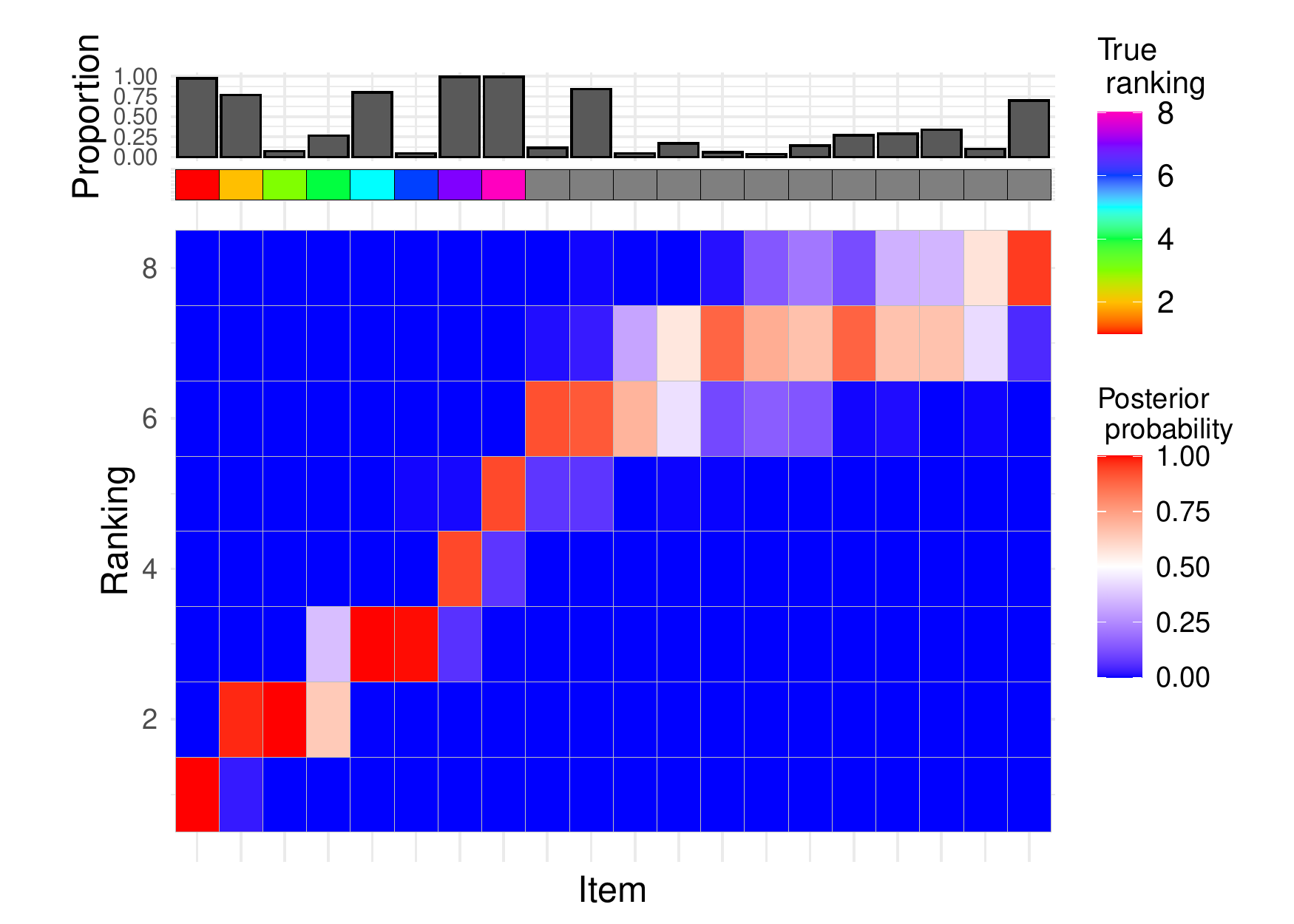}
\endminipage

\minipage{0.33\textwidth}
  \includegraphics[width=\linewidth]{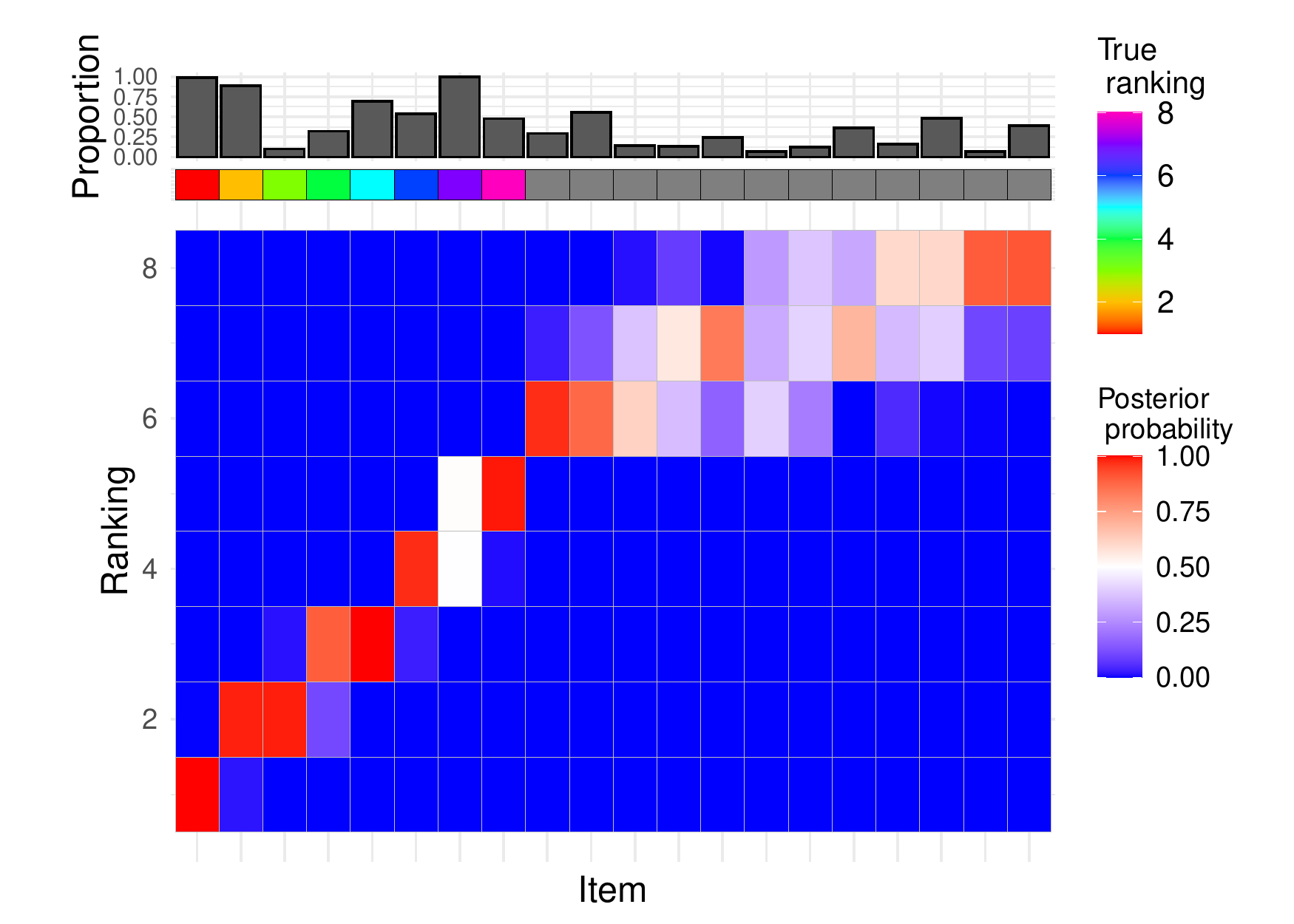}
\endminipage\hfill
\minipage{0.33\textwidth}
  \includegraphics[width=\linewidth]{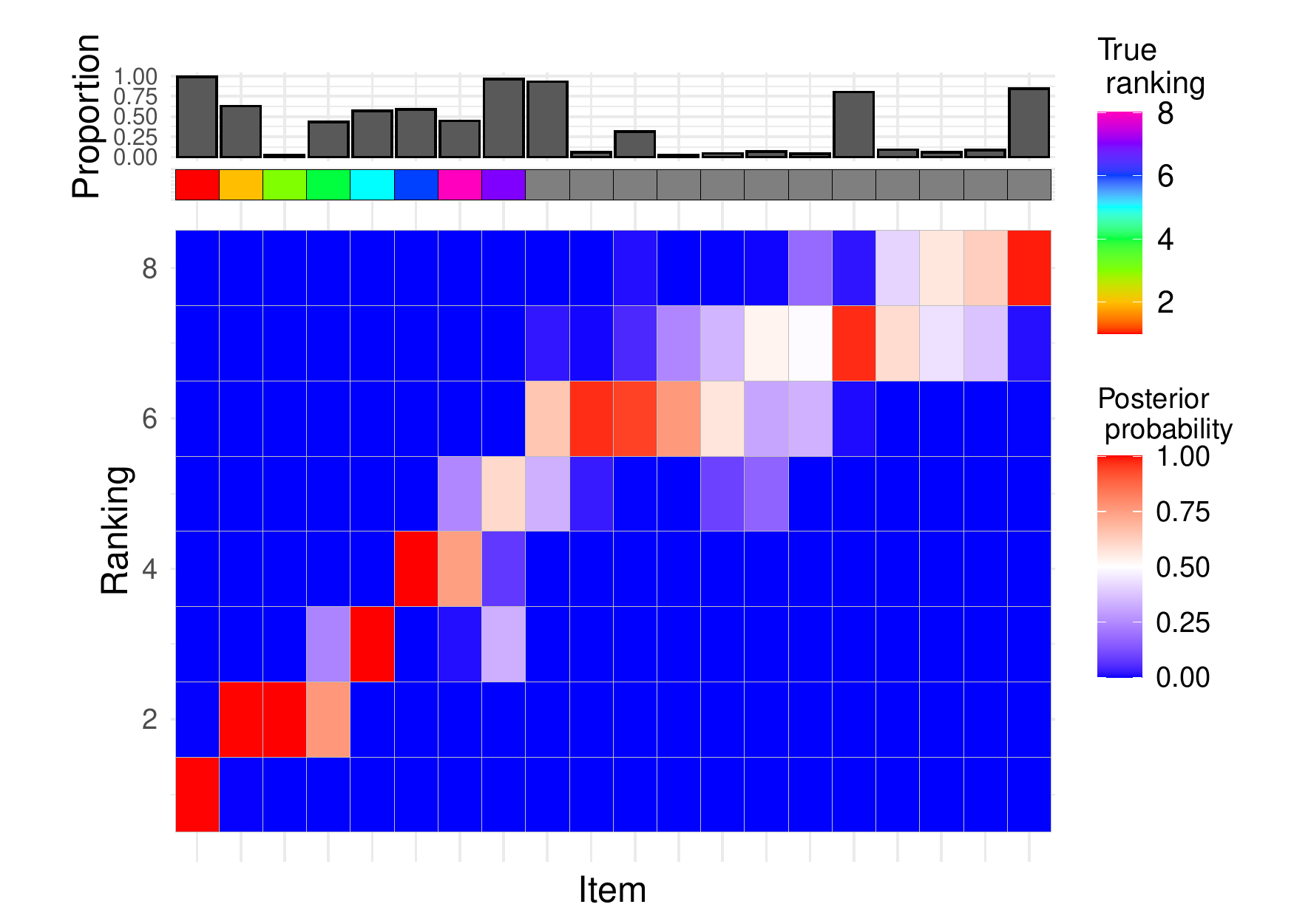}
\endminipage\hfill
\minipage{0.33\textwidth}%
  \includegraphics[width=\linewidth]{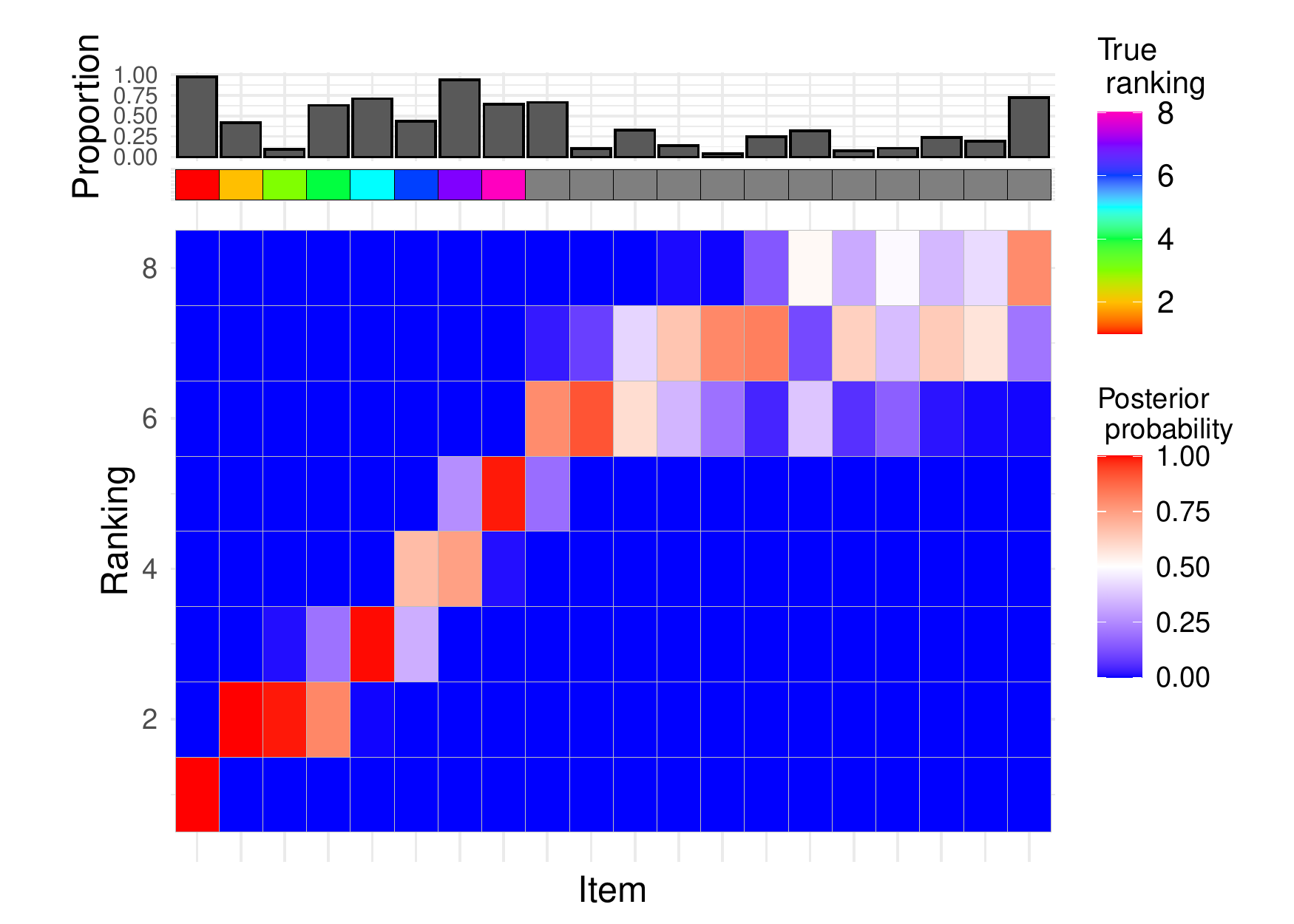}
\endminipage

\caption{Results from the top-rank simulation experiments described in Section 3.2: each panel displays the marginal posterior distribution of $\bm{\rho}$, where the items have been ordered according to $\bm{\Hat{\rho}}_{\mathcal{A}^*}$ on the x-axis. From left to right $l=1,2,3$, and from top to bottom $L=1,2,3$. The rainbow grid indicates the true $\bm{\rho}_{\mathcal{A}^*}$, and the bar plot indicates the proportion of times the items were selected in $\mathcal{A}^*$ over all MCMC iterations. $n^*=8$, $n=20$, $N=50$, $\alpha=3$.}
\end{figure}

\section{}\label{supp:alpha10_nstar_true}
\begin{figure}[H]
\minipage{0.33\textwidth}
  \includegraphics[width=\linewidth]{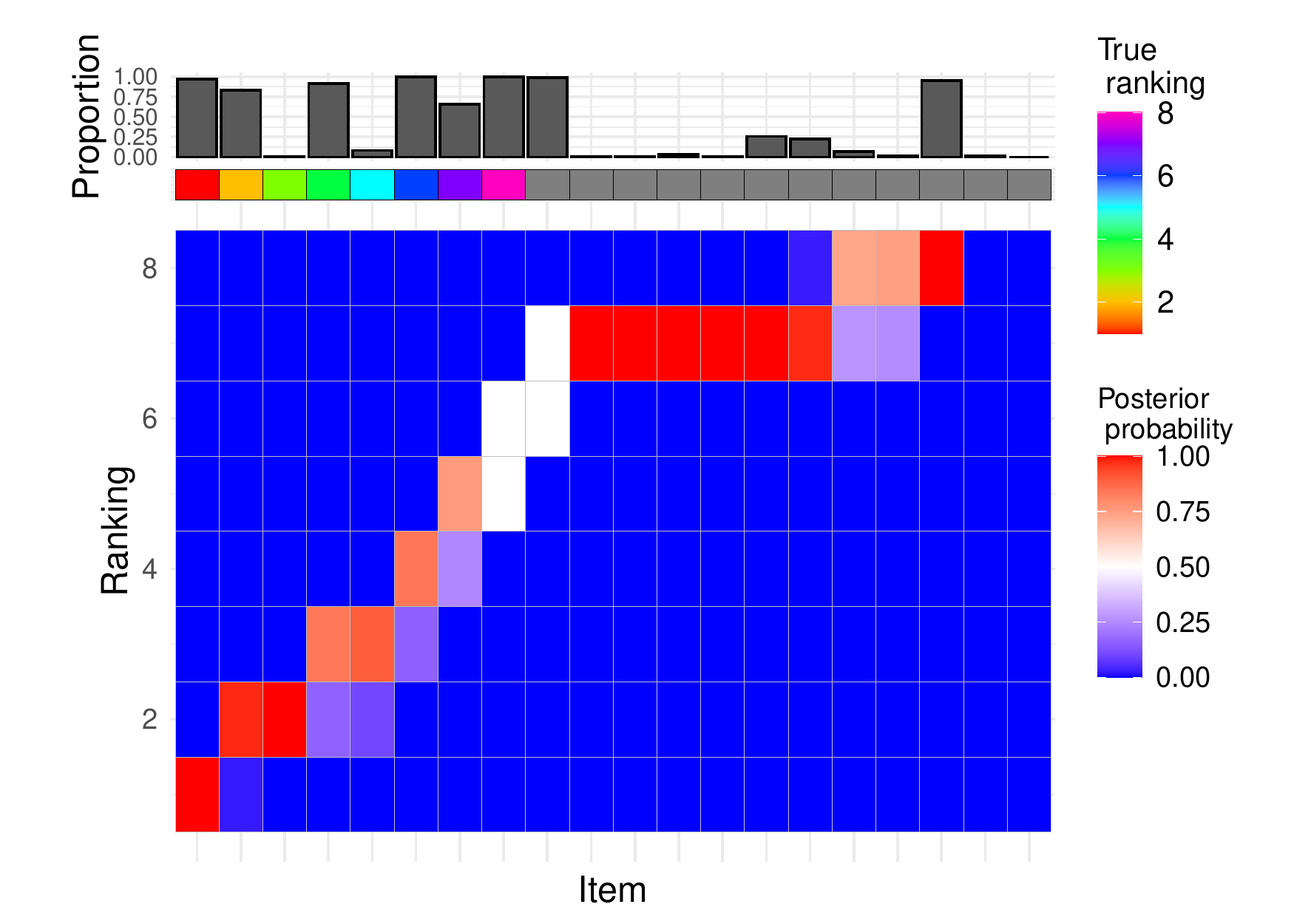}
\endminipage\hfill
\minipage{0.33\textwidth}
  \includegraphics[width=\linewidth]{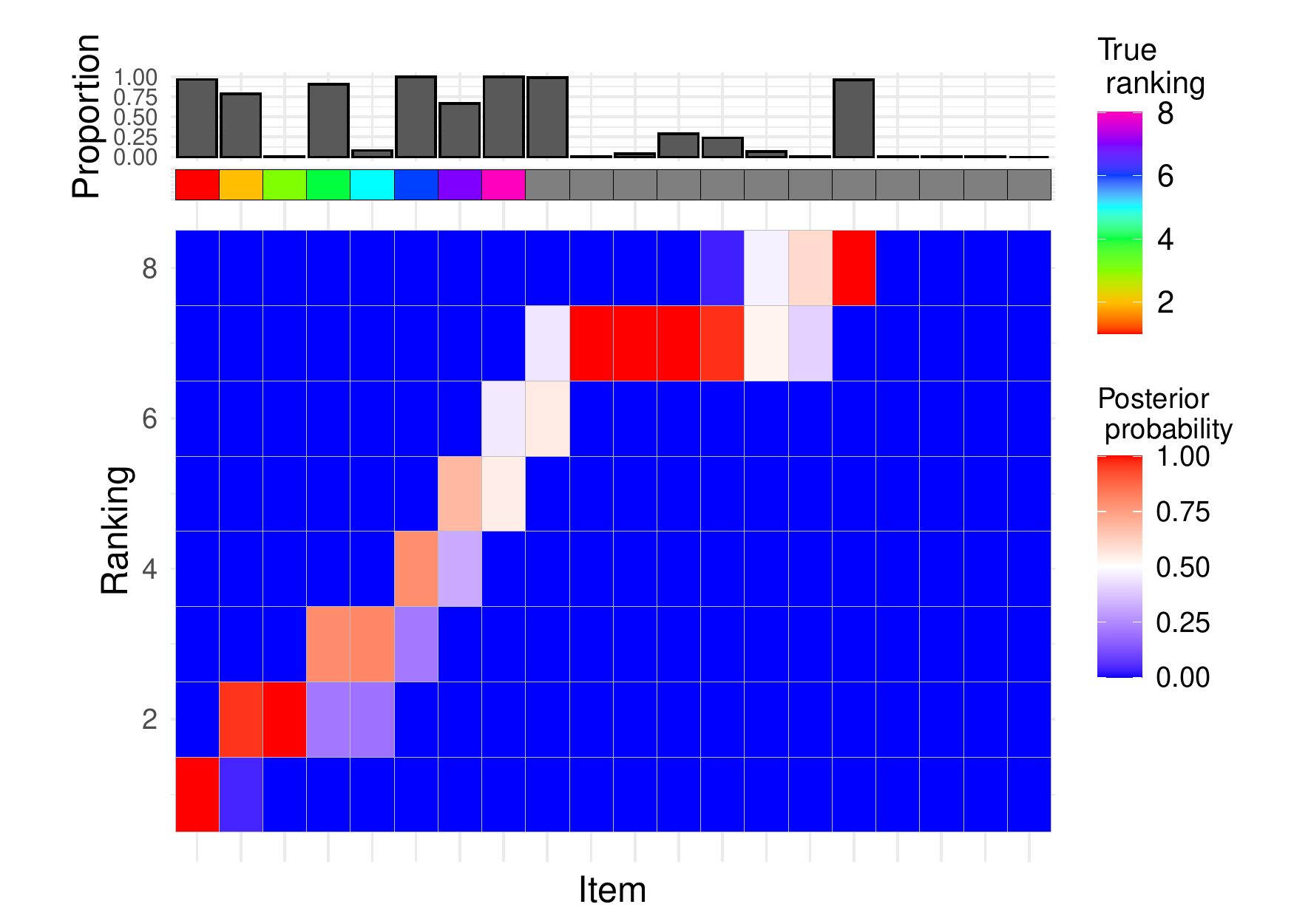}
\endminipage\hfill
\minipage{0.33\textwidth}%
  \includegraphics[width=\linewidth]{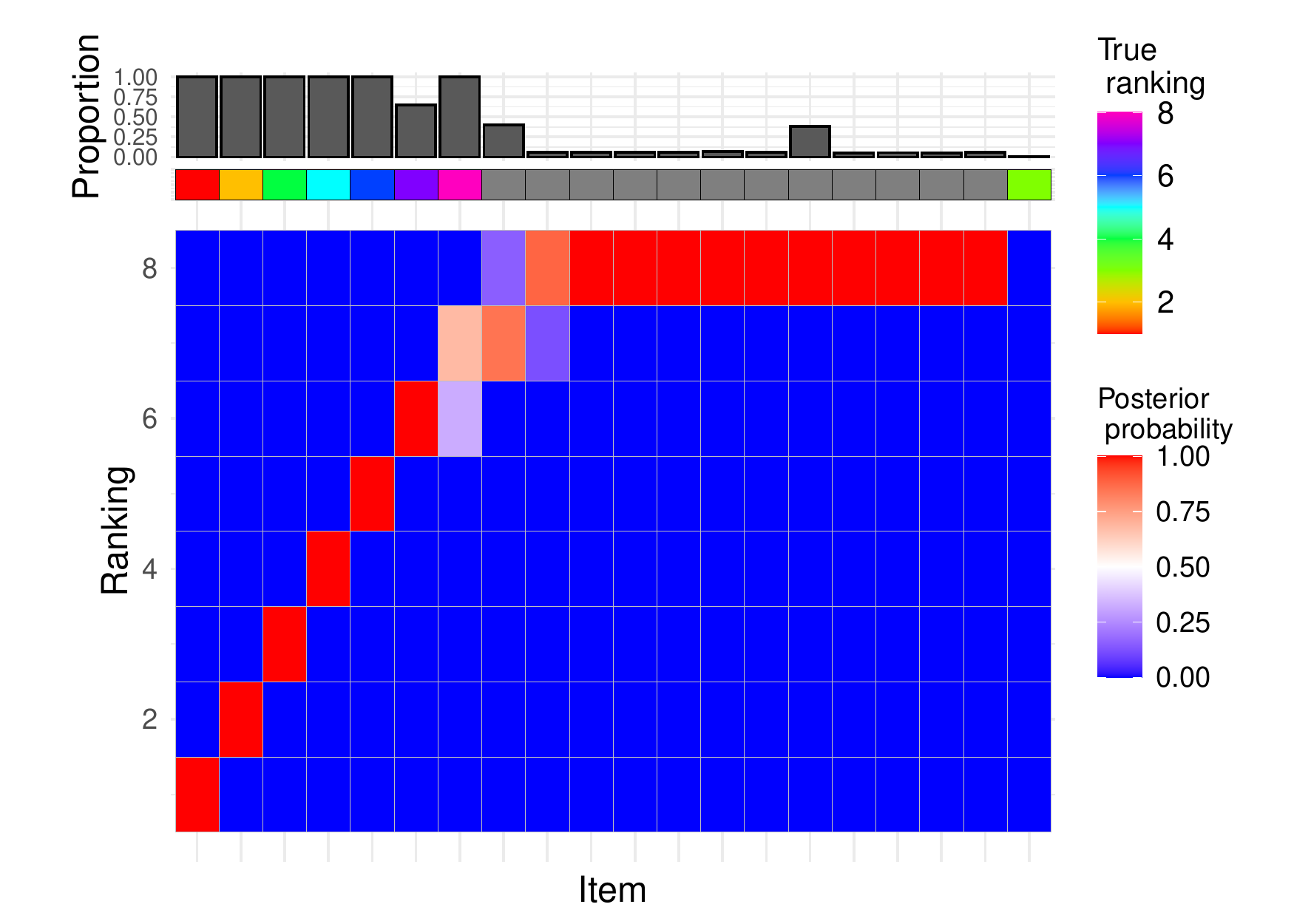}
\endminipage

\minipage{0.33\textwidth}
  \includegraphics[width=\linewidth]{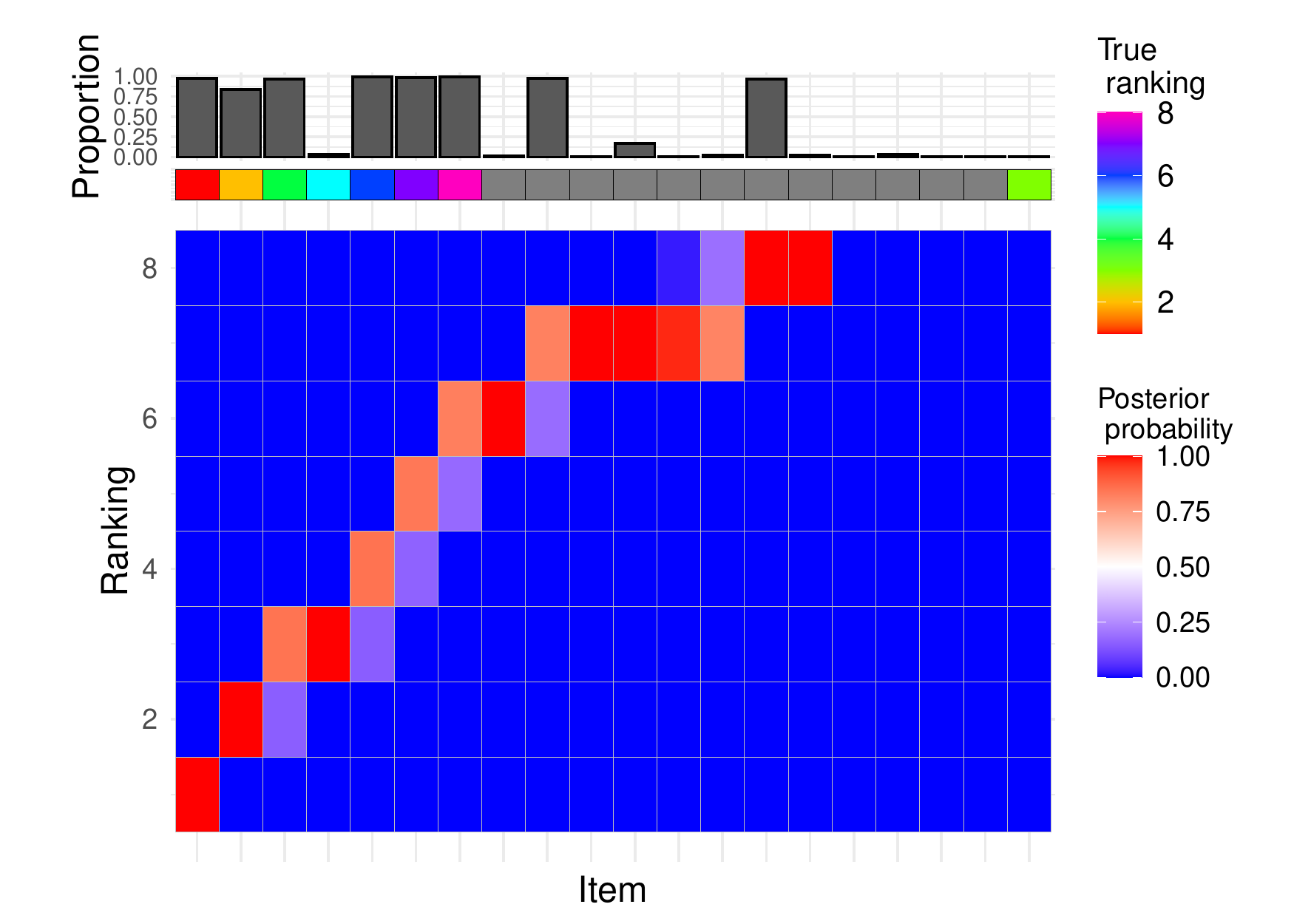}
\endminipage\hfill
\minipage{0.33\textwidth}
  \includegraphics[width=\linewidth]{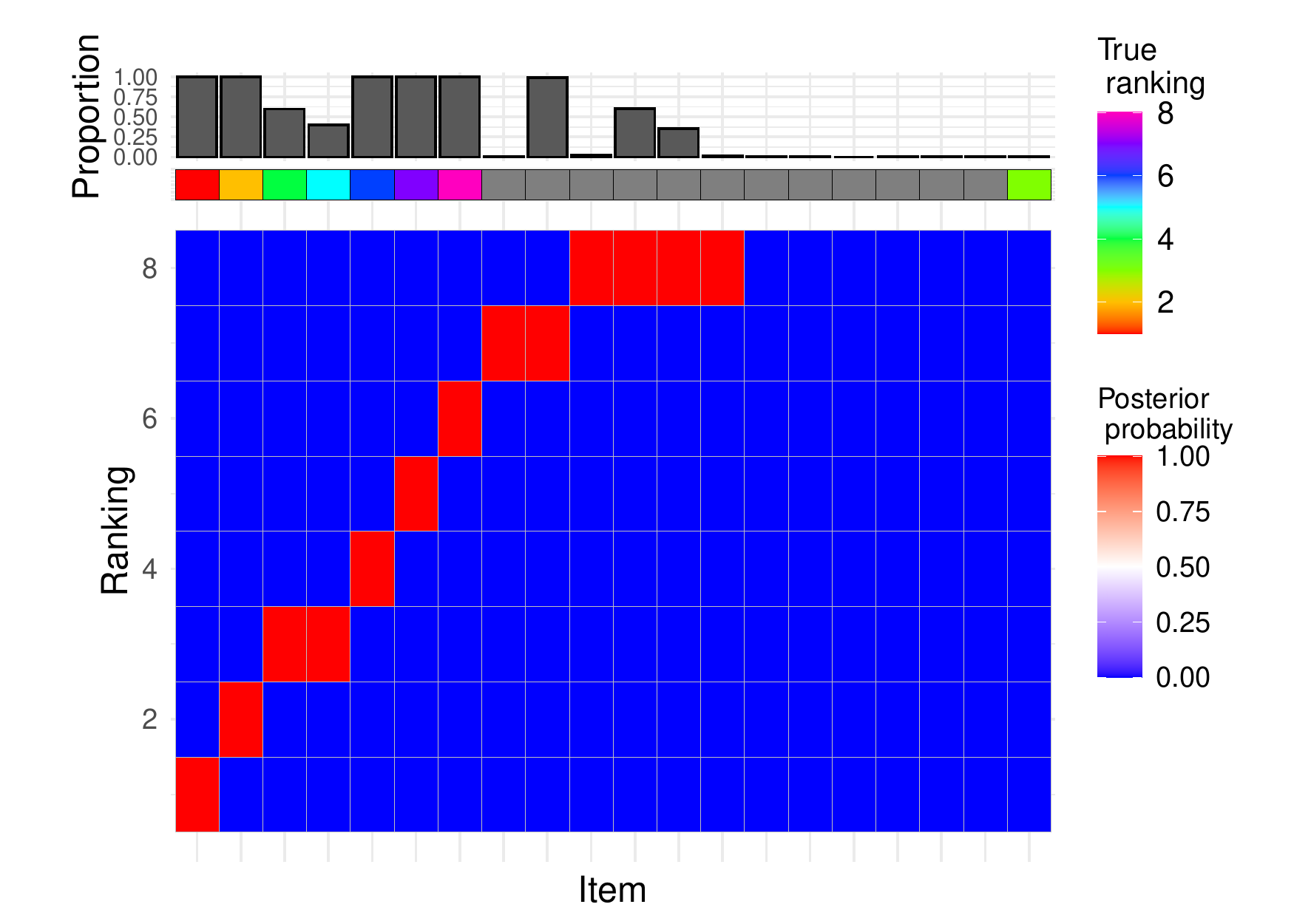}
\endminipage\hfill
\minipage{0.33\textwidth}%
  \includegraphics[width=\linewidth]{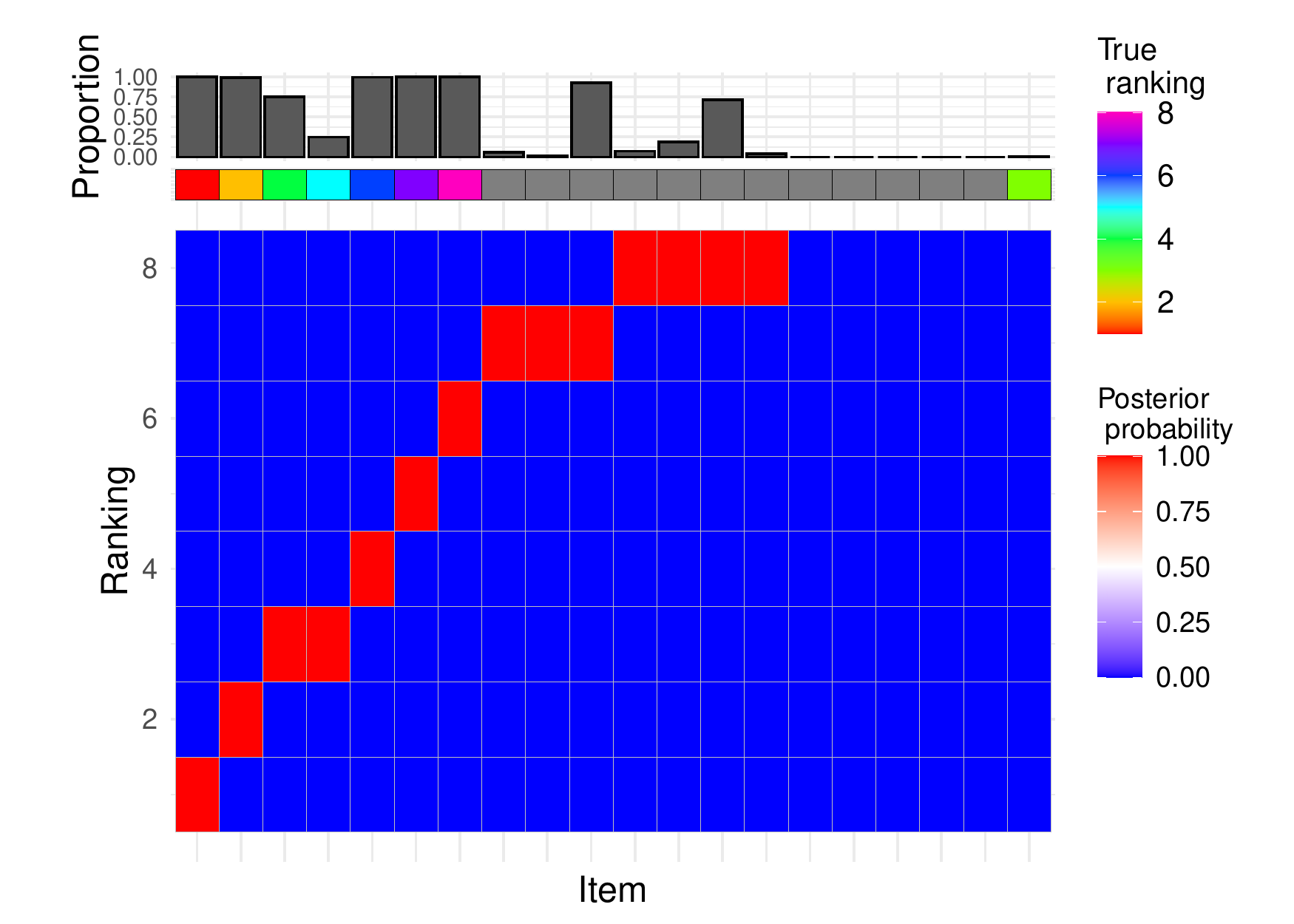}
\endminipage

\minipage{0.33\textwidth}
  \includegraphics[width=\linewidth]{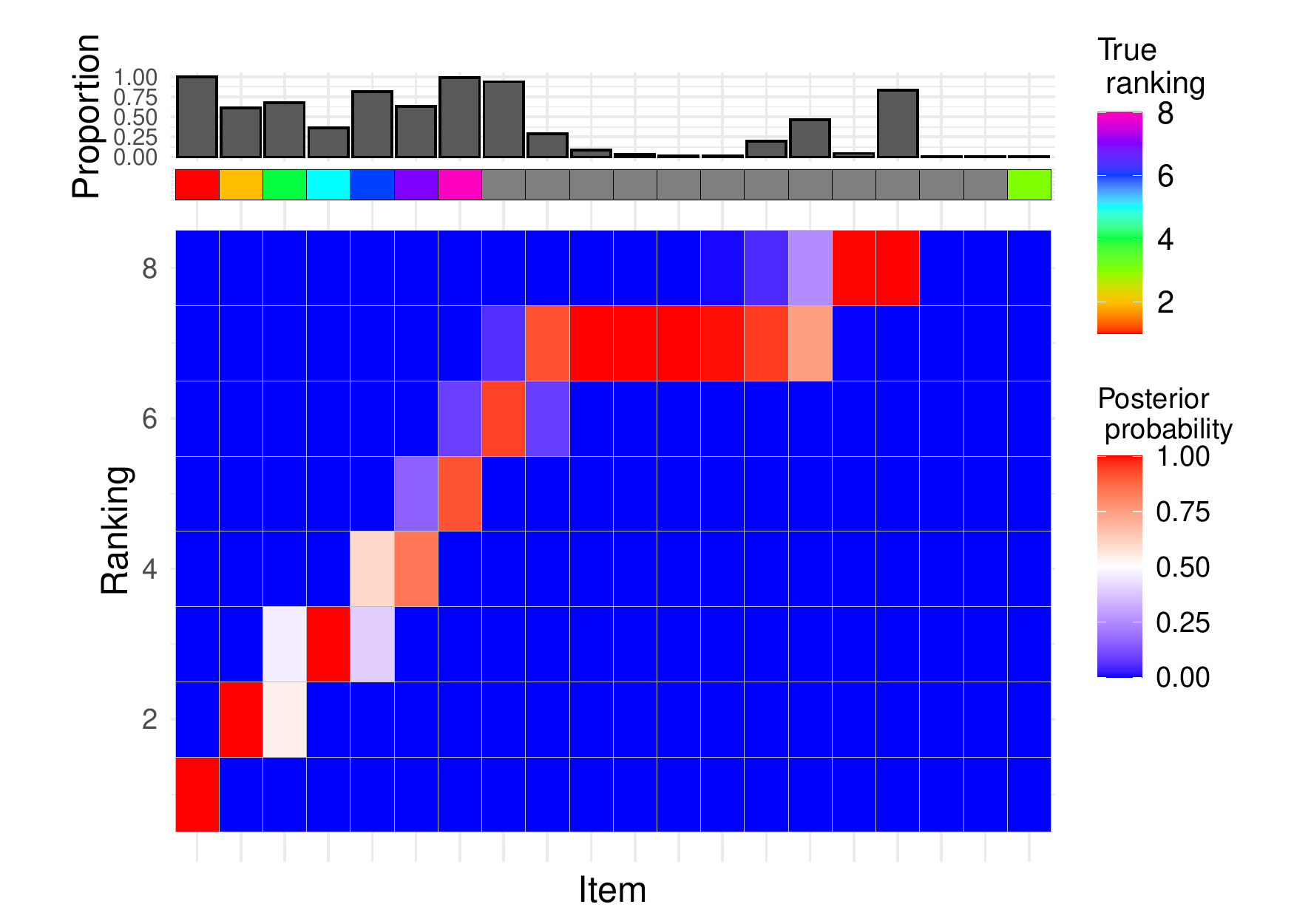}
\endminipage\hfill
\minipage{0.33\textwidth}
  \includegraphics[width=\linewidth]{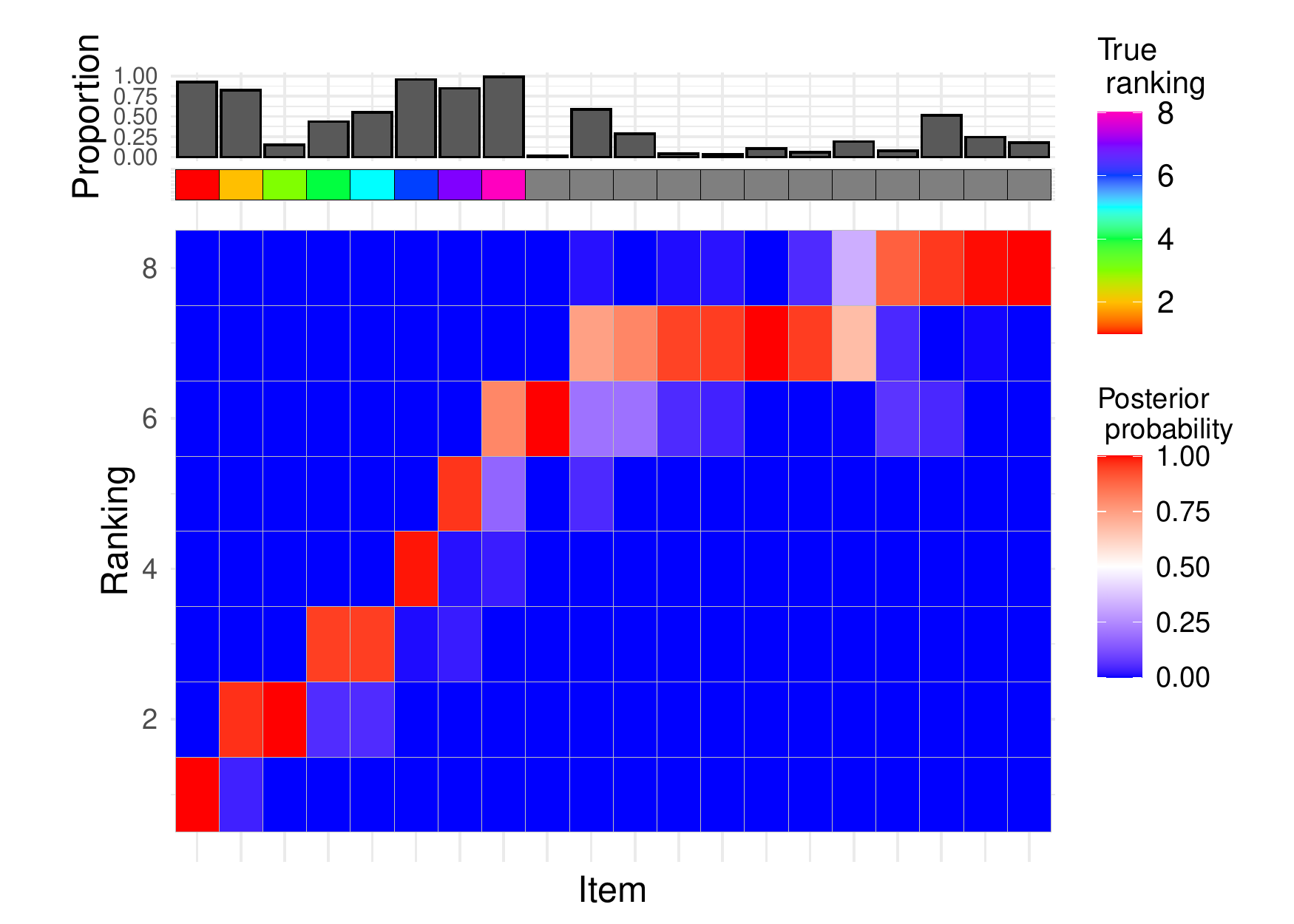}
\endminipage\hfill
\minipage{0.33\textwidth}%
  \includegraphics[width=\linewidth]{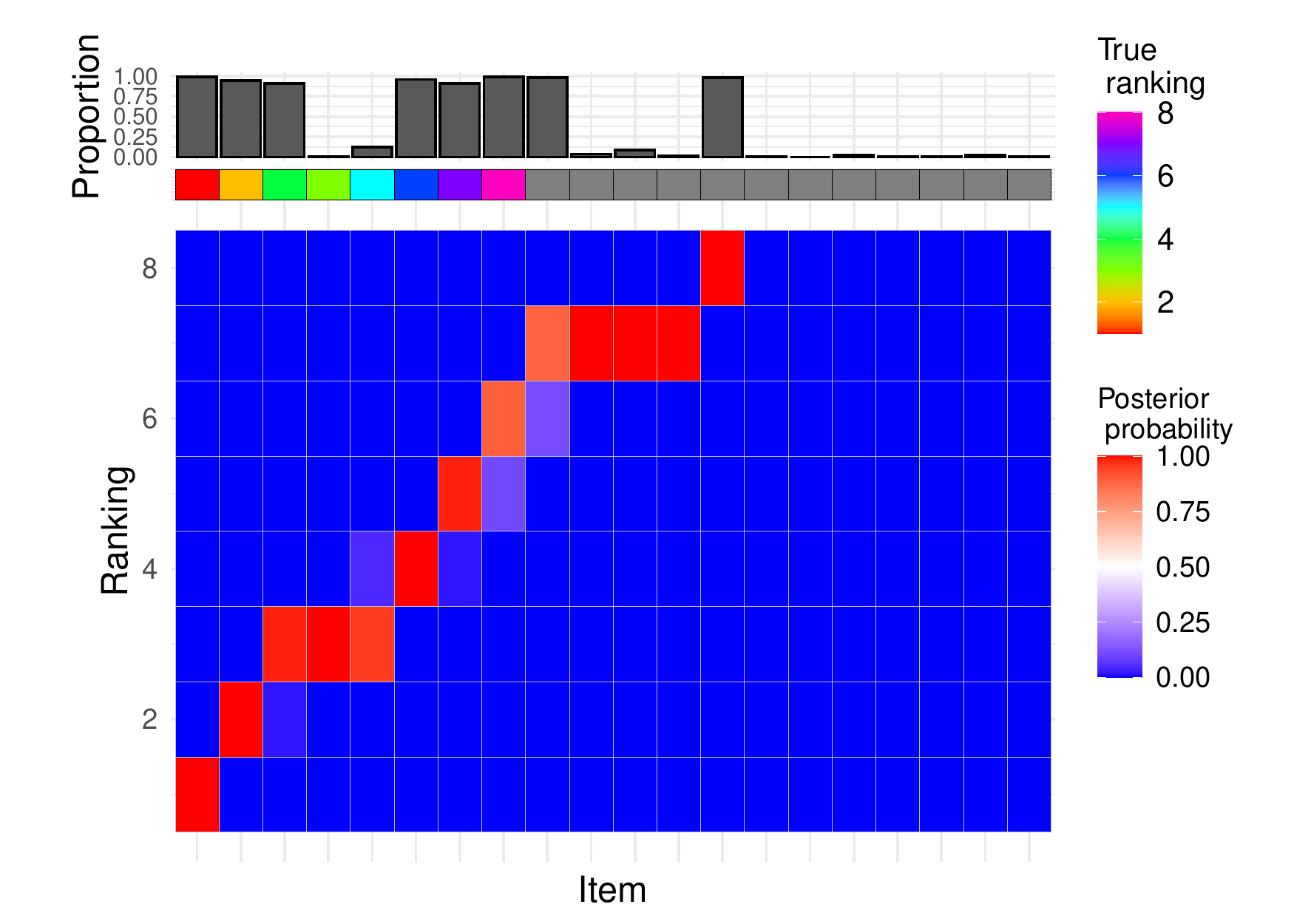}
\endminipage
\caption{Results from the top-rank simulation experiments described in Section 3.2: each panel displays the marginal posterior distribution of $\bm{\rho}$, where the items have been ordered according to $\bm{\Hat{\rho}}_{\mathcal{A}^*}$ on the x-axis. From left to right $l=1,2,3$, and from top to bottom $L=1,2,3$. The rainbow grid indicates the true $\bm{\rho}_{\mathcal{A}^*}$, and the bar plot indicates the proportion of times the items were selected in $\mathcal{A}^*$ over all MCMC iterations. $n^*=8$, $n=20$, $N=50$, $\alpha=10$.}
\end{figure}

\section{}\label{supp:alpha3_nstar_sub_four}
\begin{figure}[H]
\minipage{0.33\textwidth}
  \includegraphics[width=\linewidth]{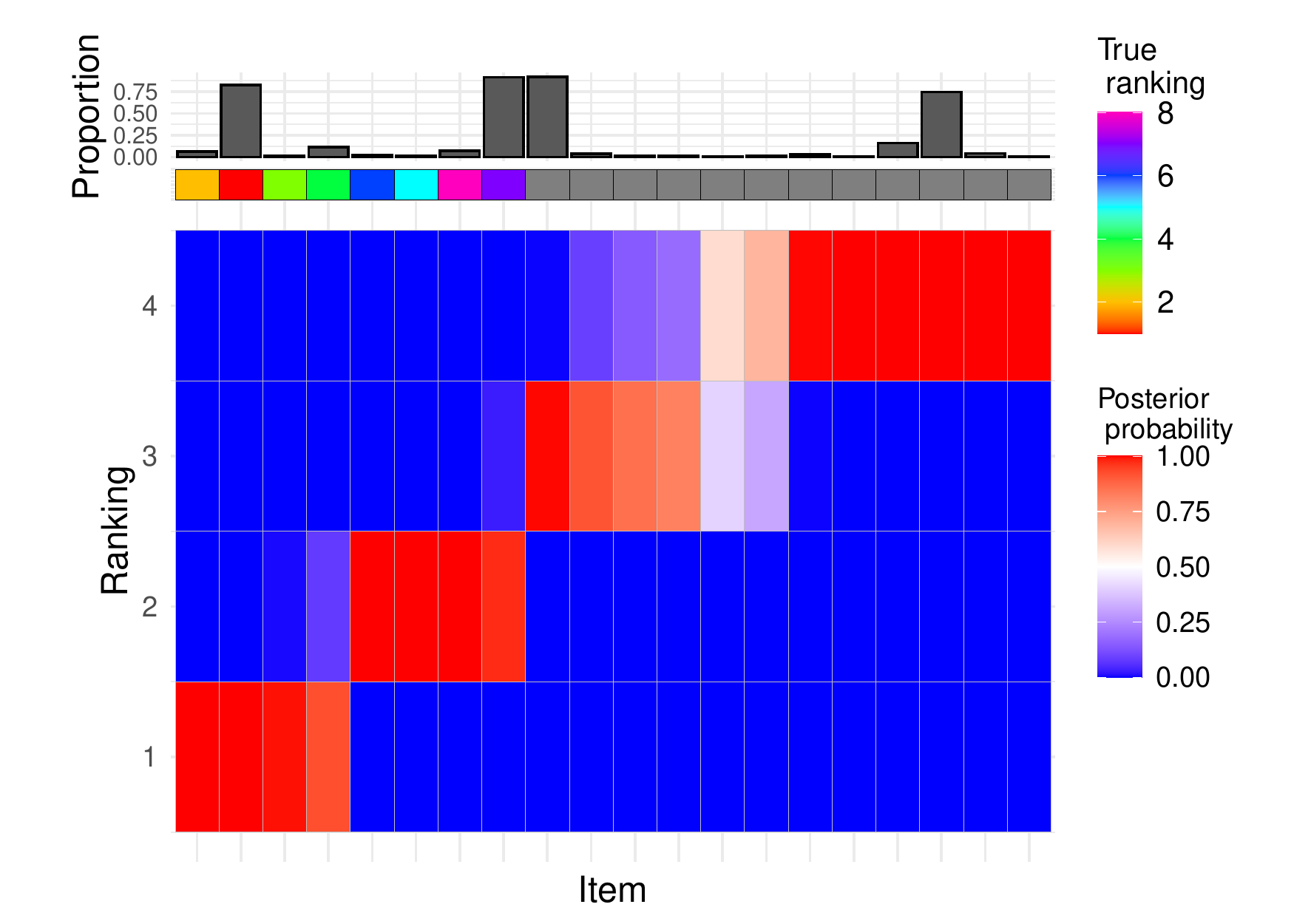}
\endminipage\hfill
\minipage{0.33\textwidth}
  \includegraphics[width=\linewidth]{fig/probitems_top20_item_selection_alphafixed3_simulation_top_items_nstartrue8_nstar4_L1_leap2_acceptance_v1.pdf}
\endminipage\hfill
\minipage{0.33\textwidth}%
  \includegraphics[width=\linewidth]{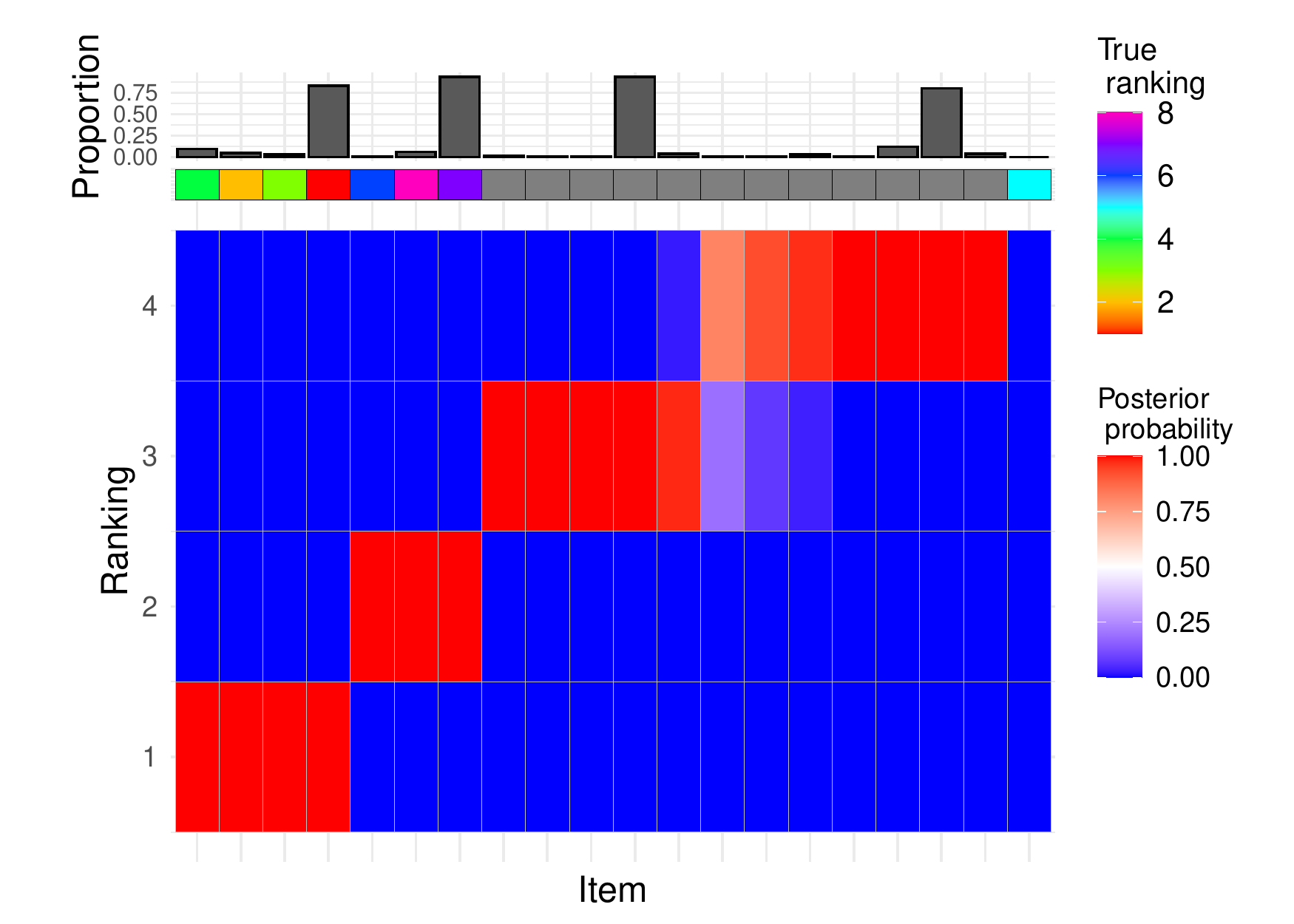}
\endminipage

\minipage{0.33\textwidth}
  \includegraphics[width=\linewidth]{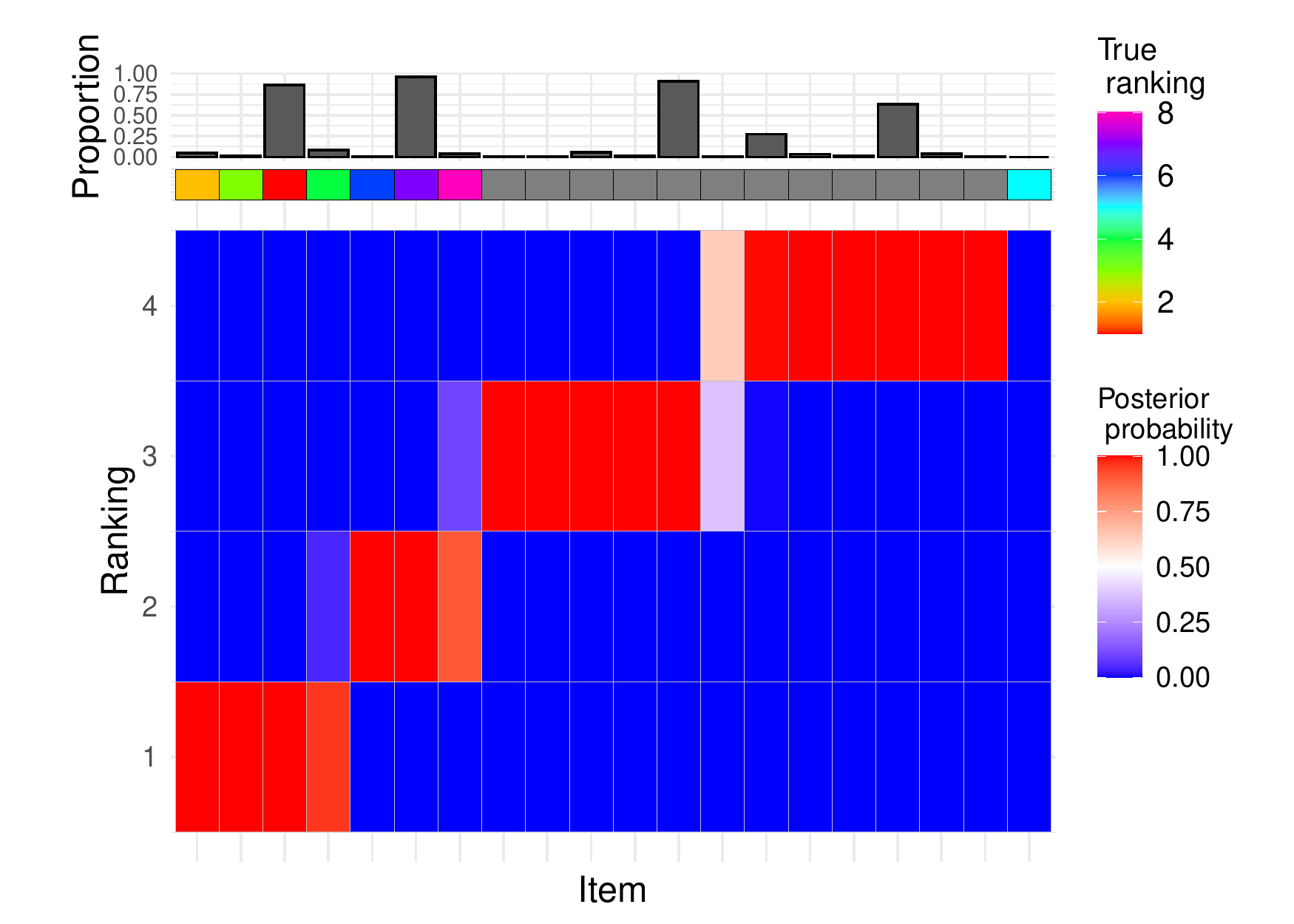}
\endminipage\hfill
\minipage{0.33\textwidth}
  \includegraphics[width=\linewidth]{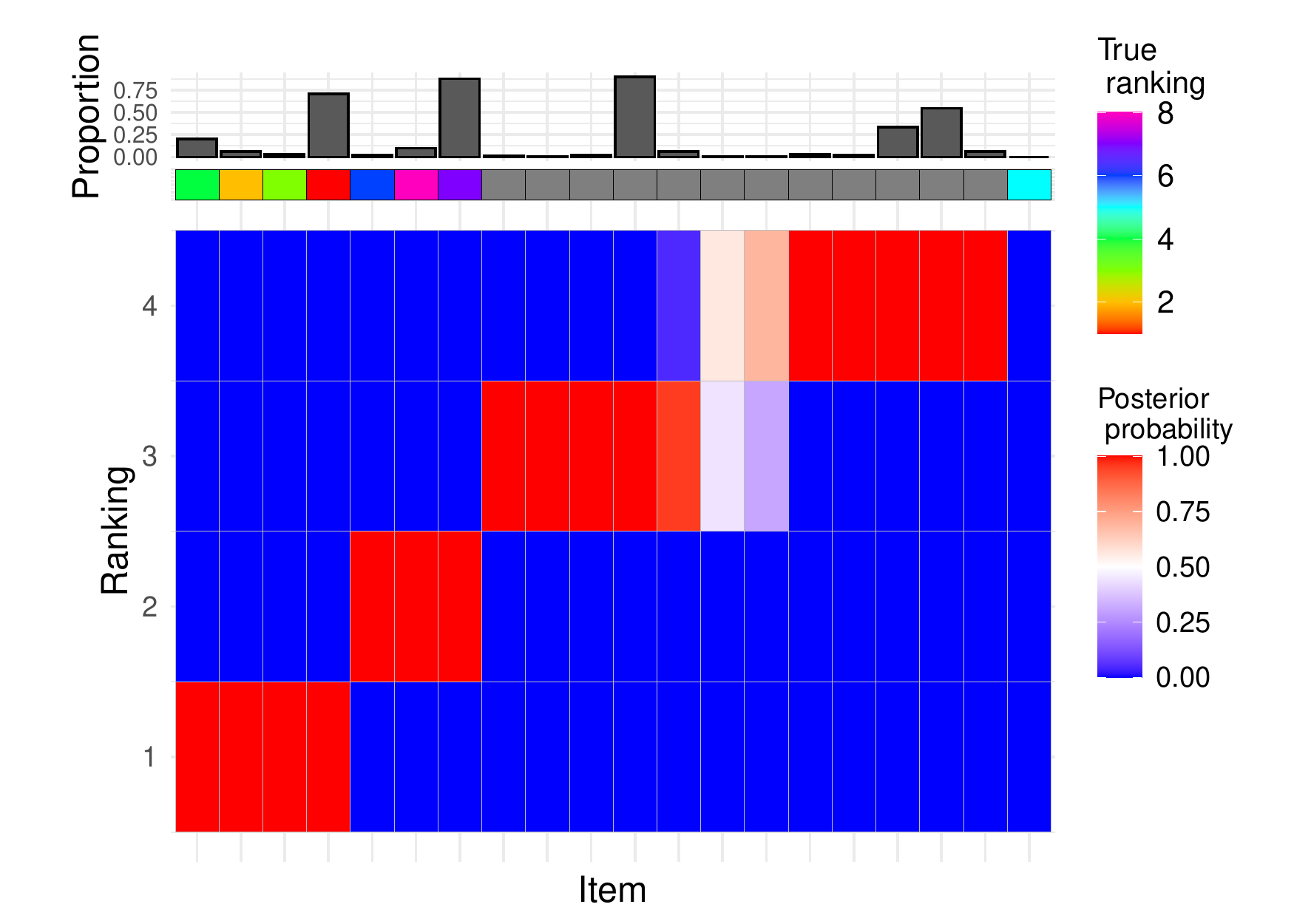}
\endminipage\hfill
\minipage{0.33\textwidth}%
  \includegraphics[width=\linewidth]{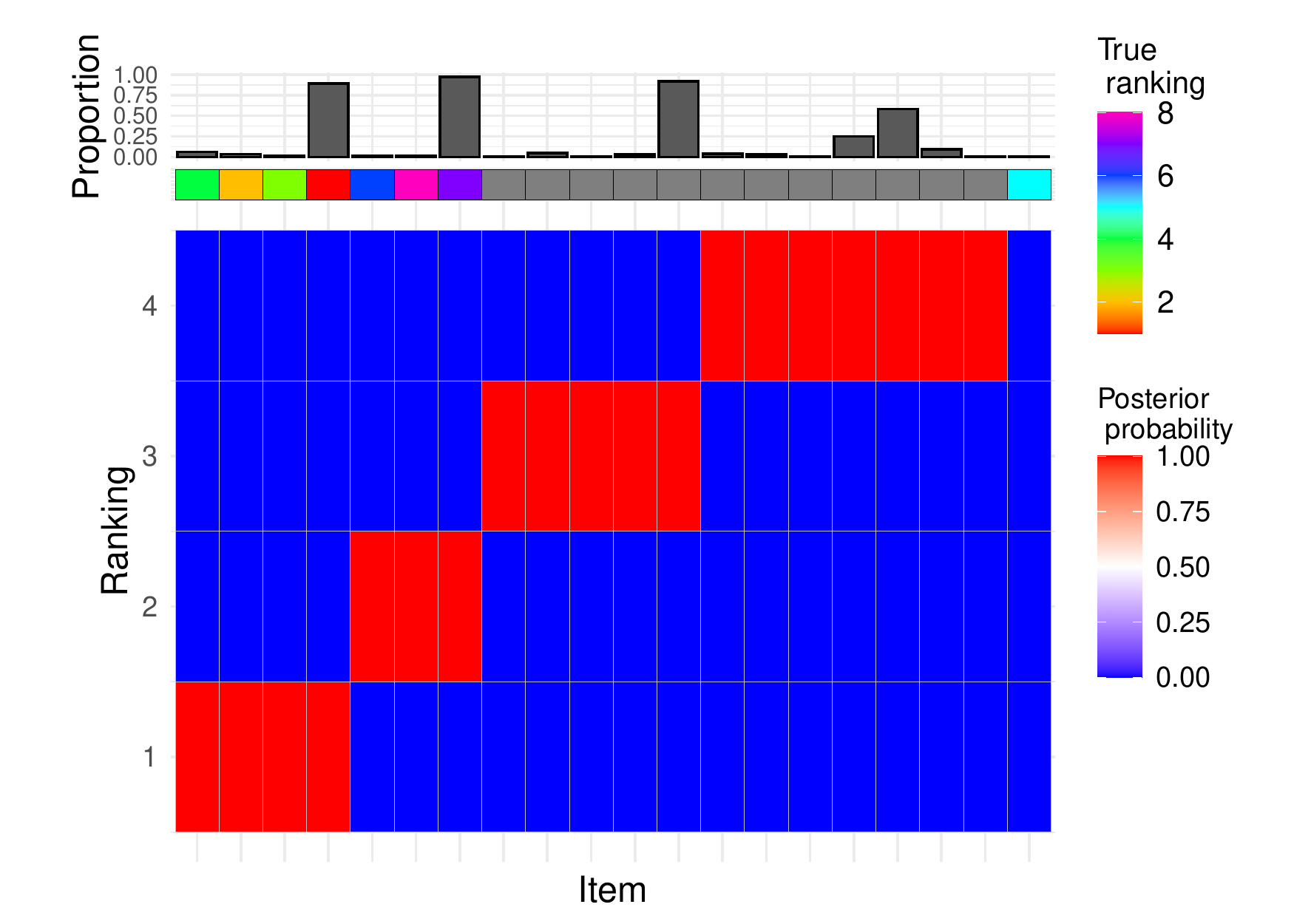}
\endminipage

\minipage{0.33\textwidth}
  \includegraphics[width=\linewidth]{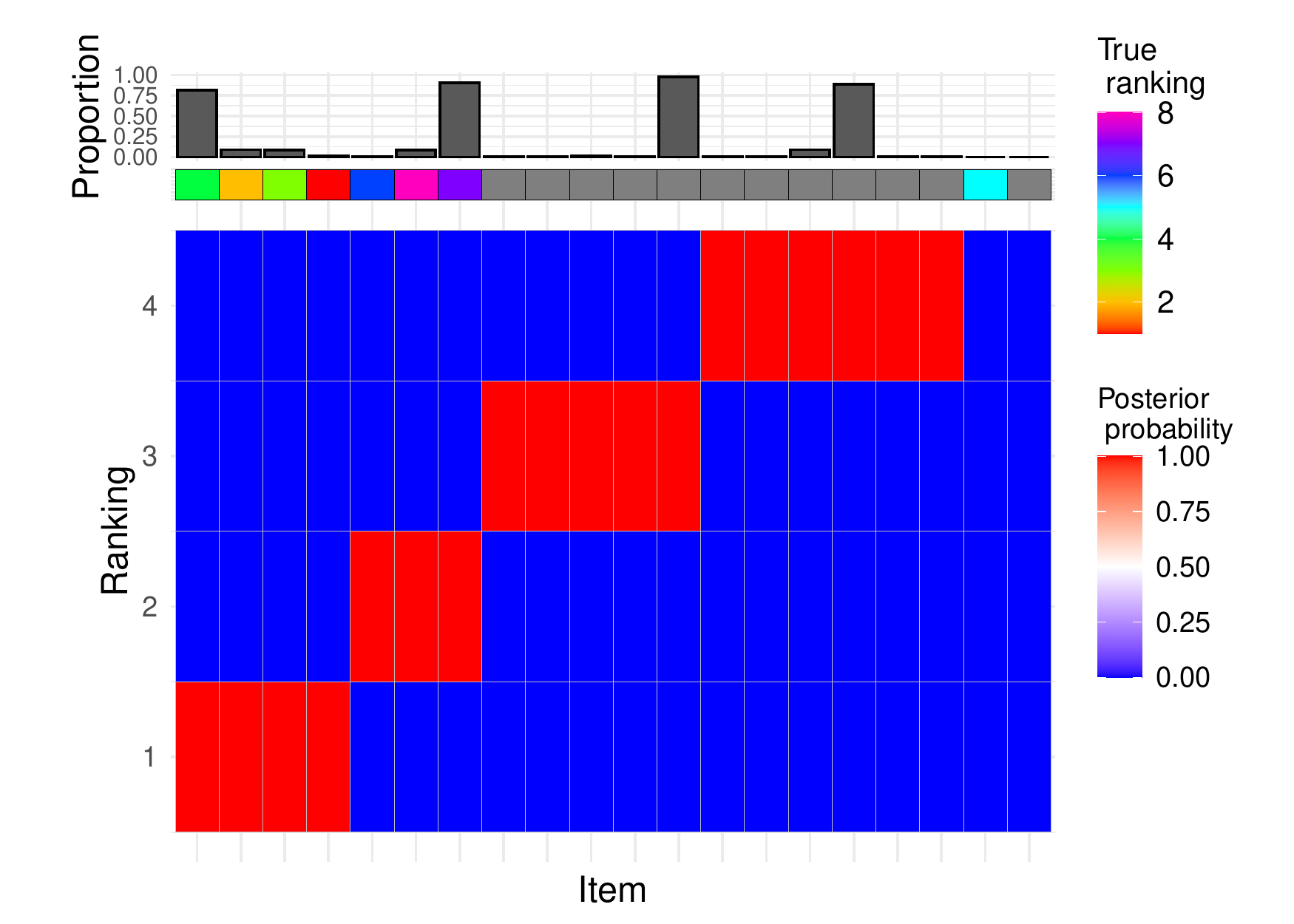}
\endminipage\hfill
\minipage{0.33\textwidth}
  \includegraphics[width=\linewidth]{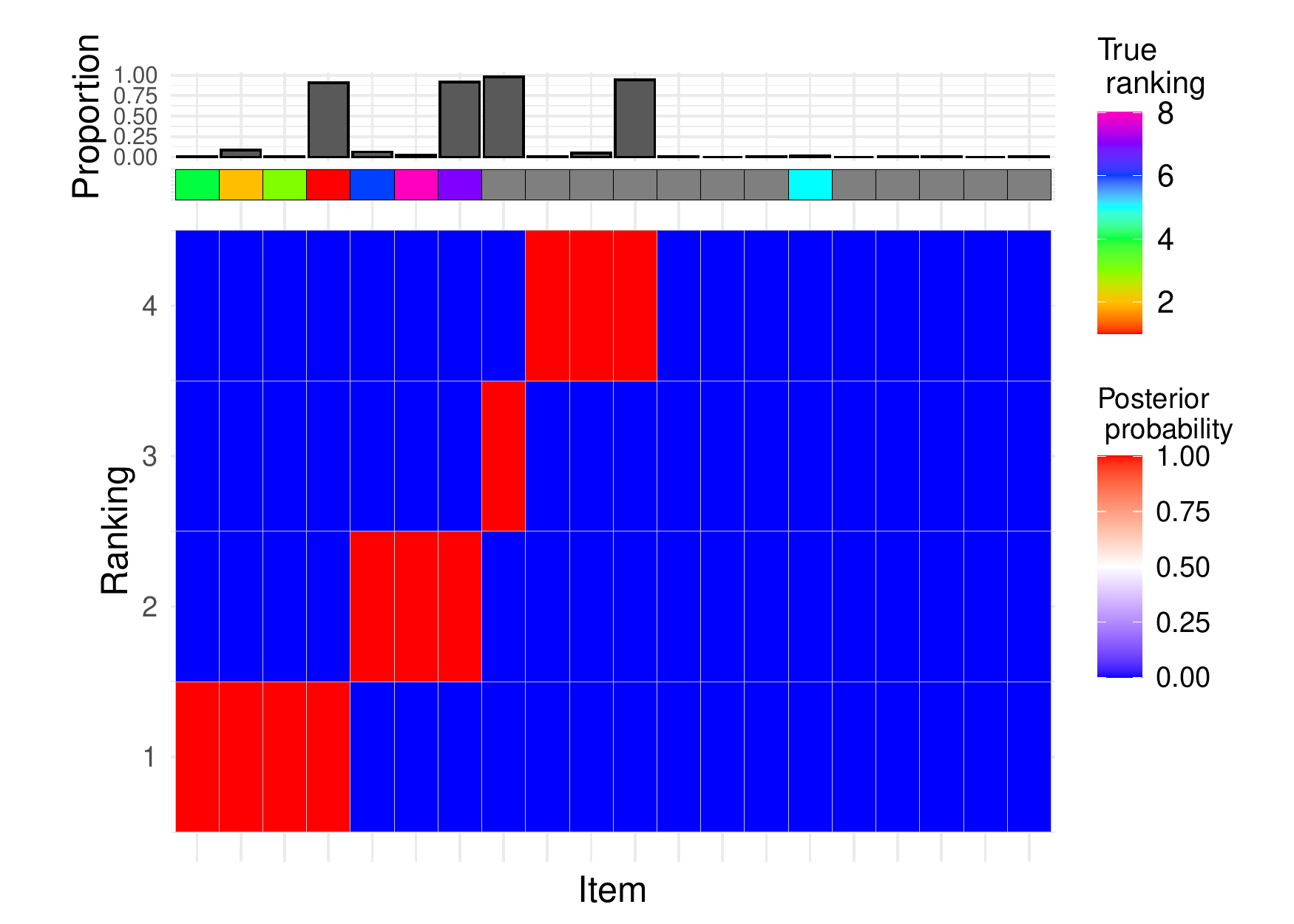}
\endminipage\hfill
\minipage{0.33\textwidth}%
  \includegraphics[width=\linewidth]{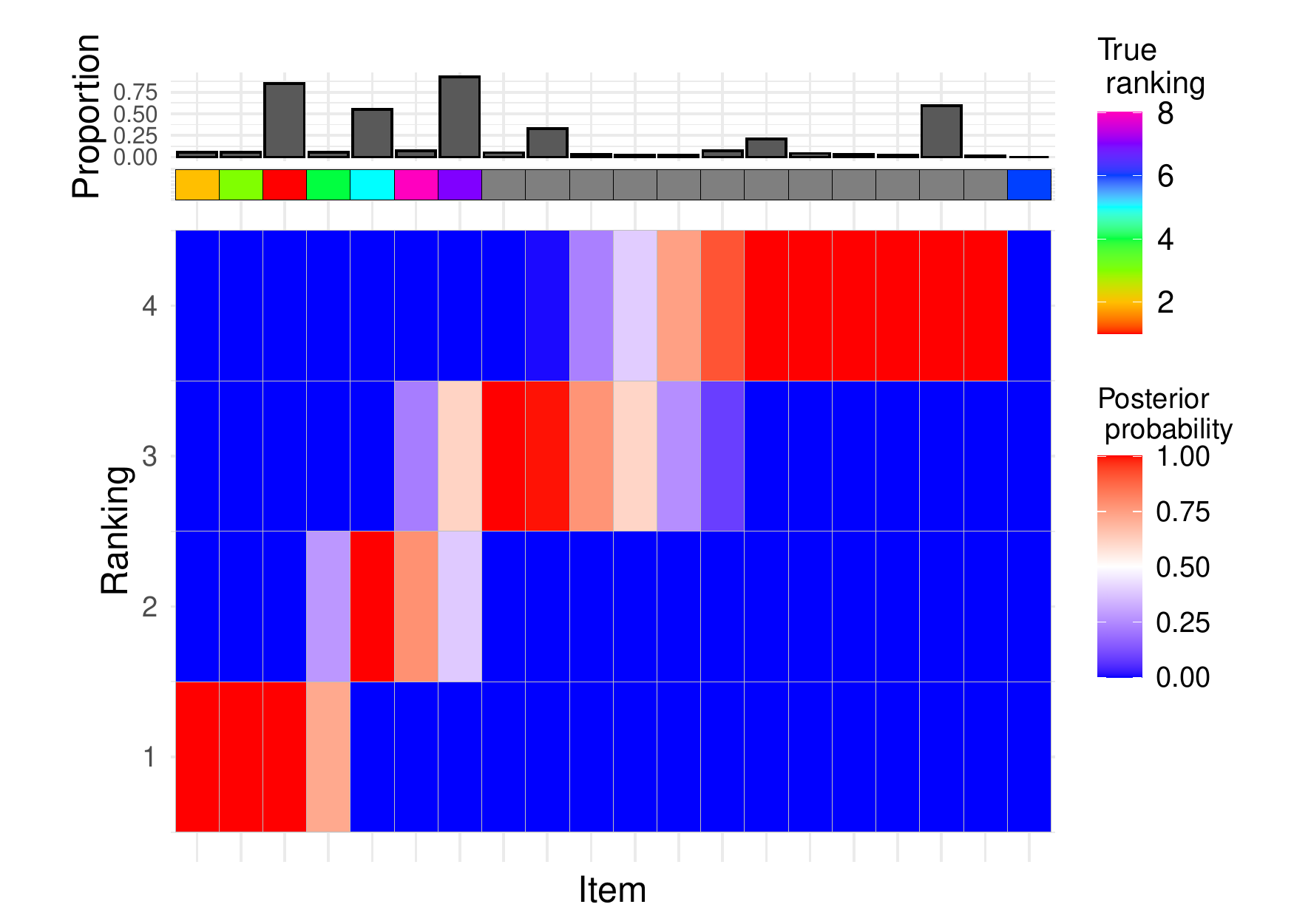}
\endminipage
\caption{Results from the top-rank simulation experiments described in Section 3.2 with $n^*_{\textnormal{true}}=8$, $n^*_{\textnormal{guess}}=4$: each panel displays the marginal posterior distribution of $\bm{\rho}$, where the items have been ordered according to $\bm{\Hat{\rho}}_{\mathcal{A}^*}$ on the x-axis. From left to right $l=1,2,3$, and from top to bottom $L=1,2,3$. The rainbow grid indicates the true $\bm{\rho}_{\mathcal{A}^*}$, and the bar plot indicates the proportion of times the items were selected in $\mathcal{A}^*$ over all MCMC iterations. $n=20$, $N=50$, $\alpha=3$.}
\end{figure}

\section{}\label{supp:alpha10_nstar_sub_four}
\begin{figure}[H]
\minipage{0.33\textwidth}
  \includegraphics[width=\linewidth]{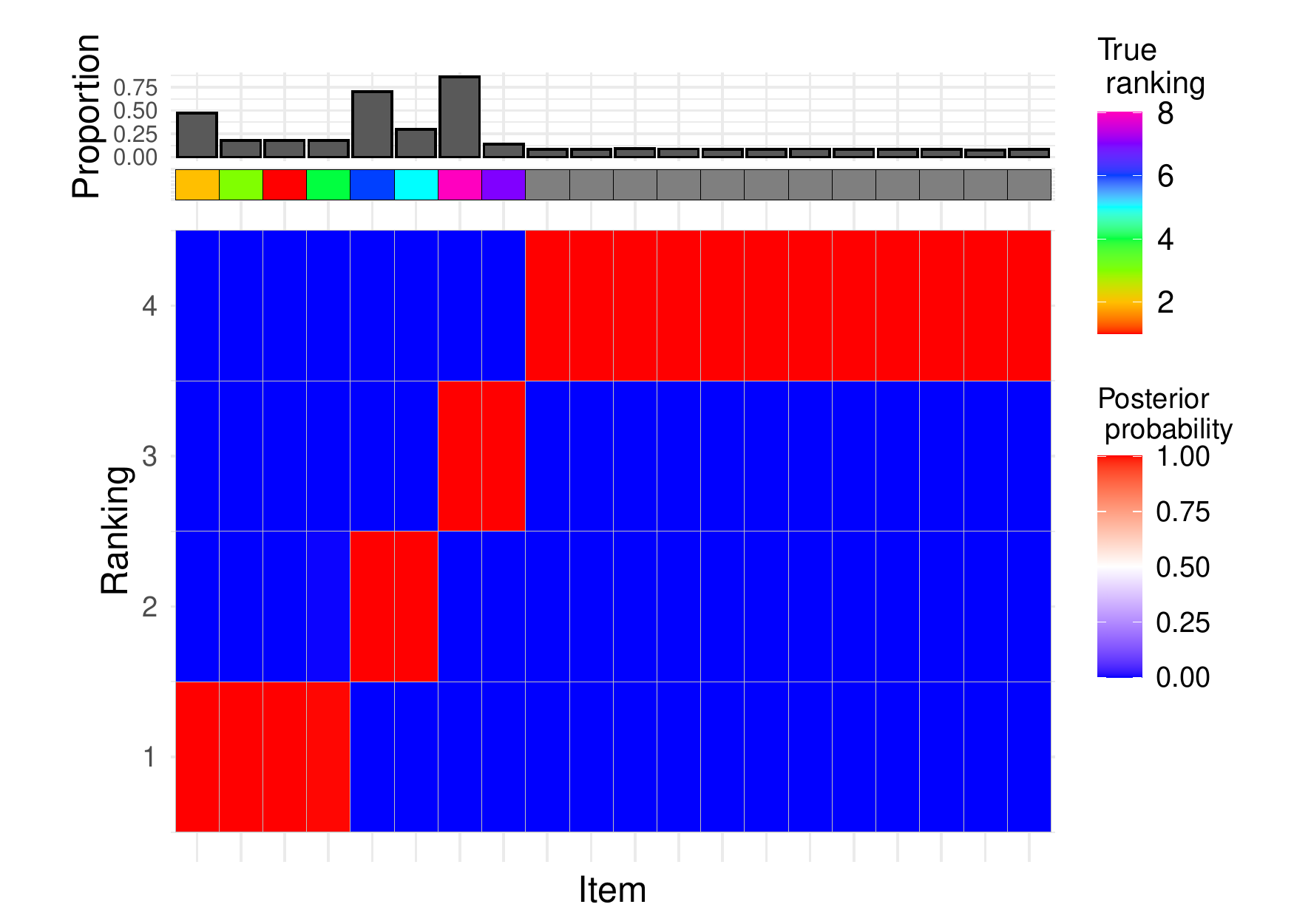}
\endminipage\hfill
\minipage{0.33\textwidth}
  \includegraphics[width=\linewidth]{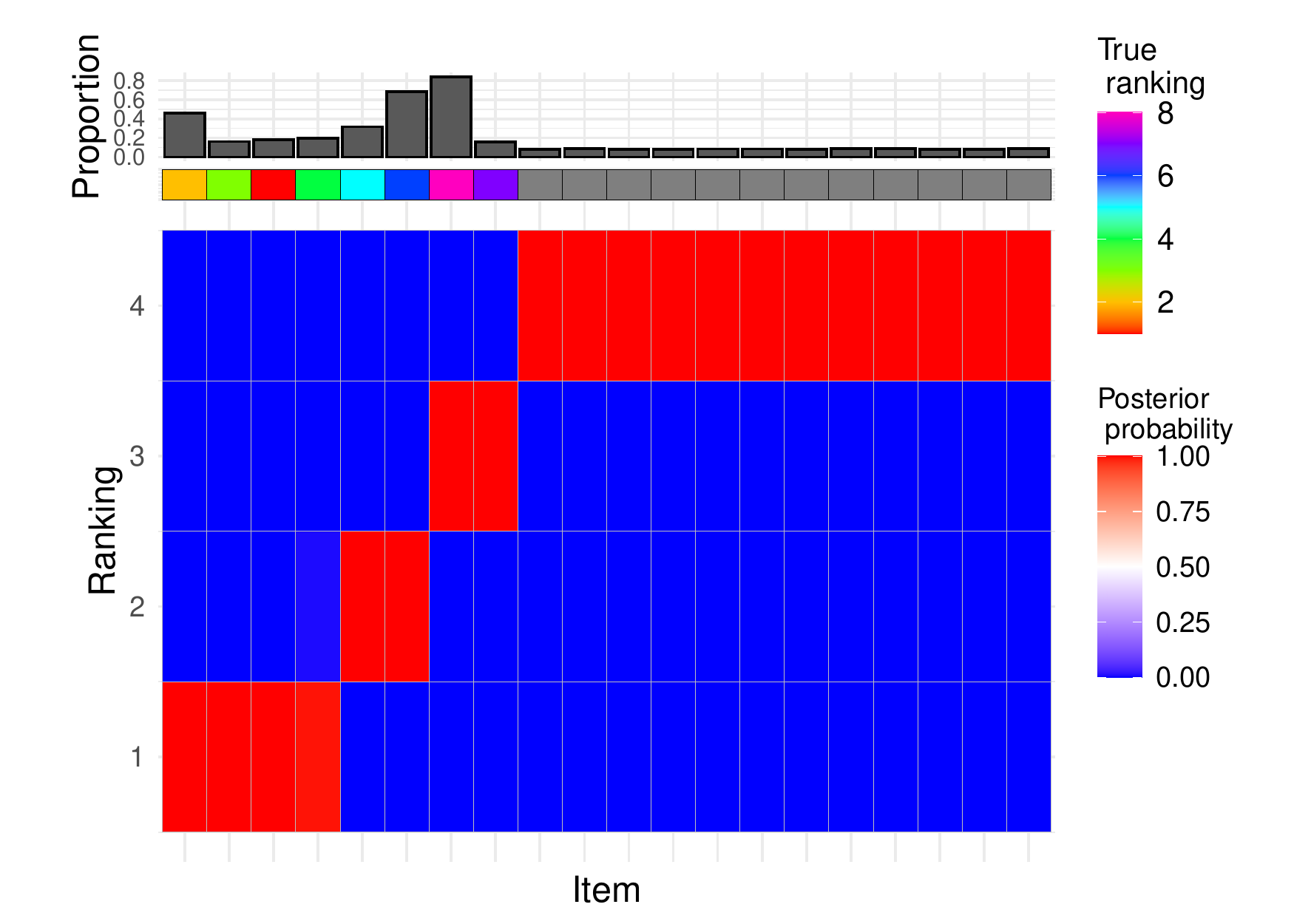}
\endminipage\hfill
\minipage{0.33\textwidth}%
  \includegraphics[width=\linewidth]{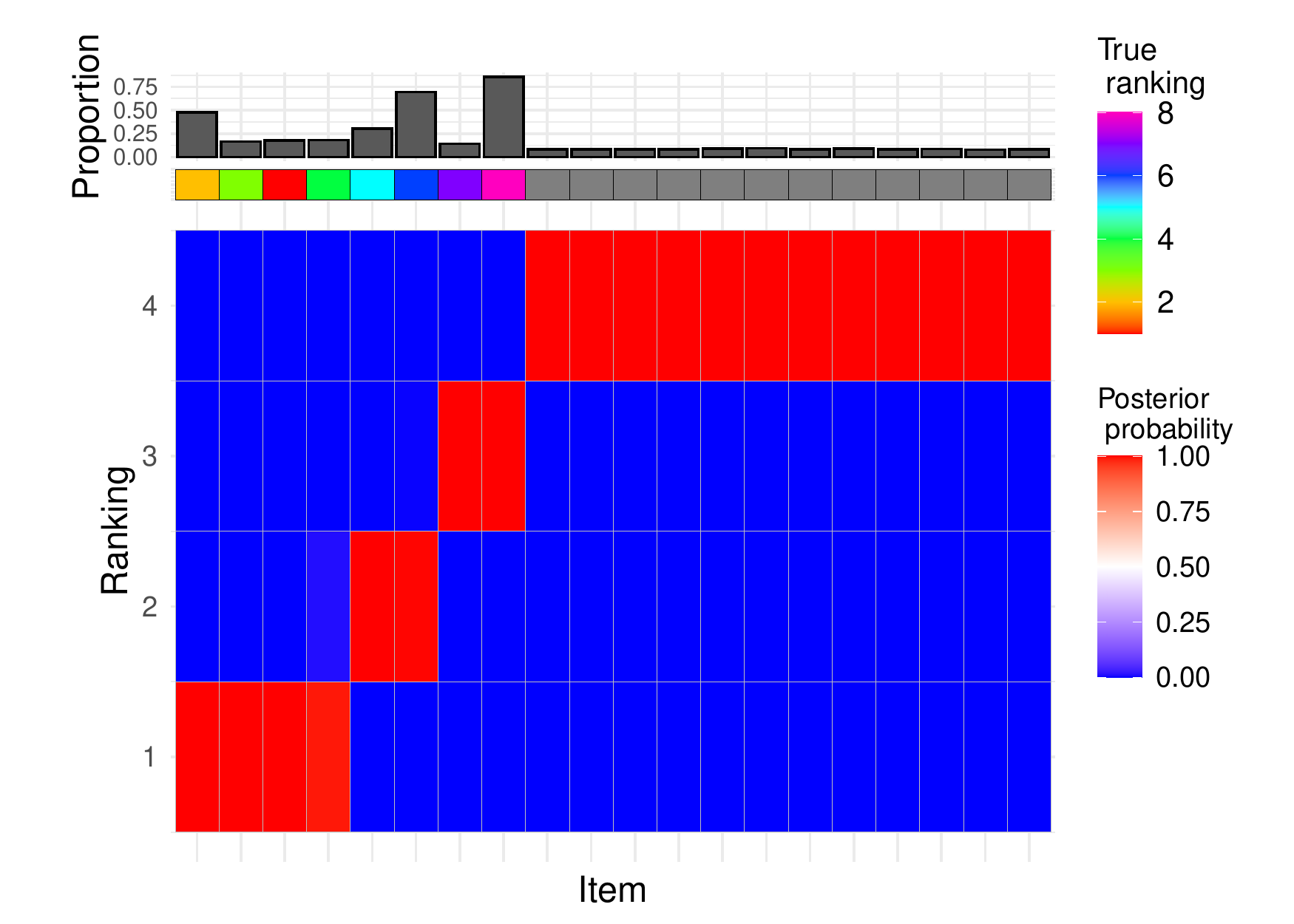}
\endminipage

\minipage{0.33\textwidth}
  \includegraphics[width=\linewidth]{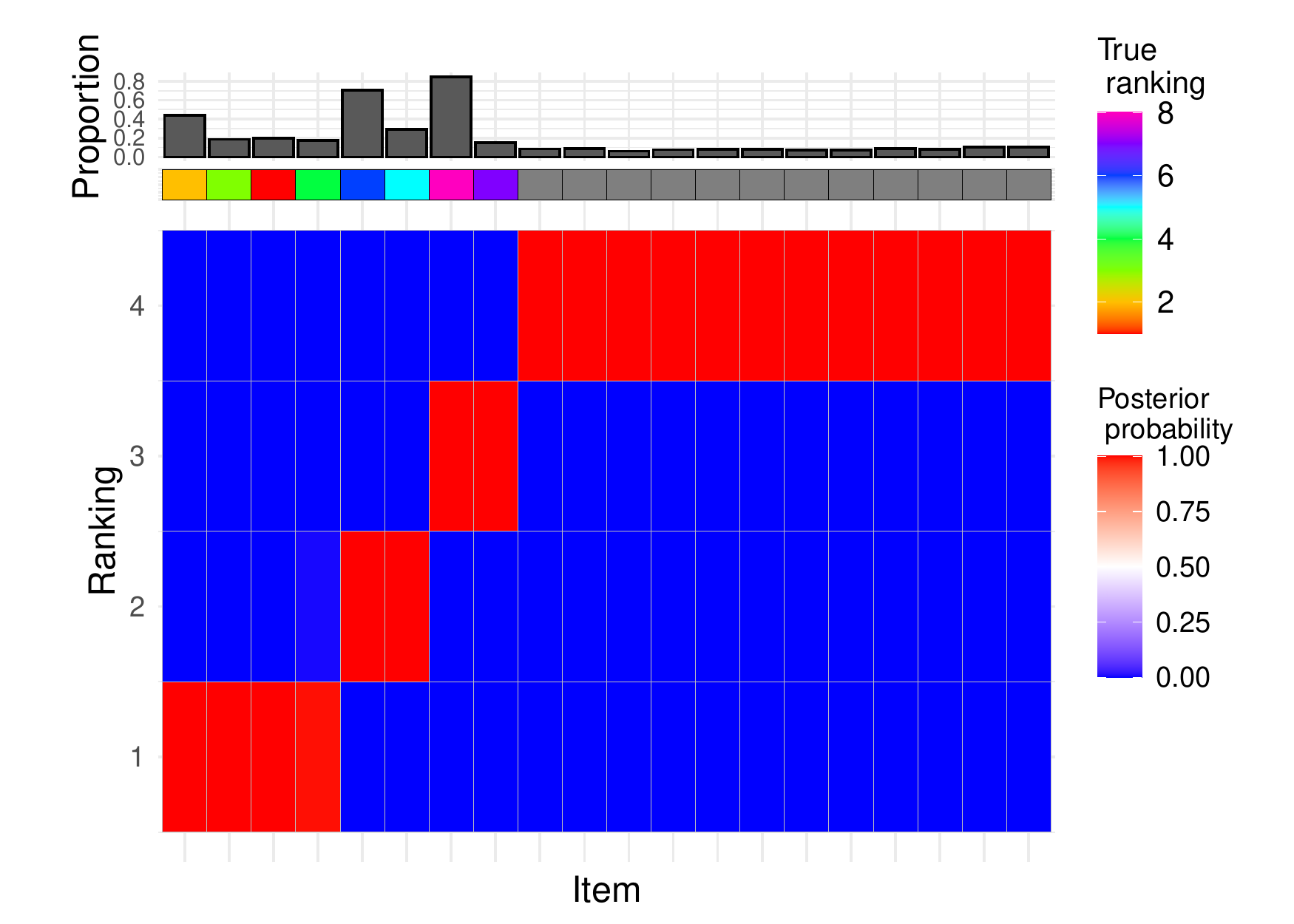}
\endminipage\hfill
\minipage{0.33\textwidth}
  \includegraphics[width=\linewidth]{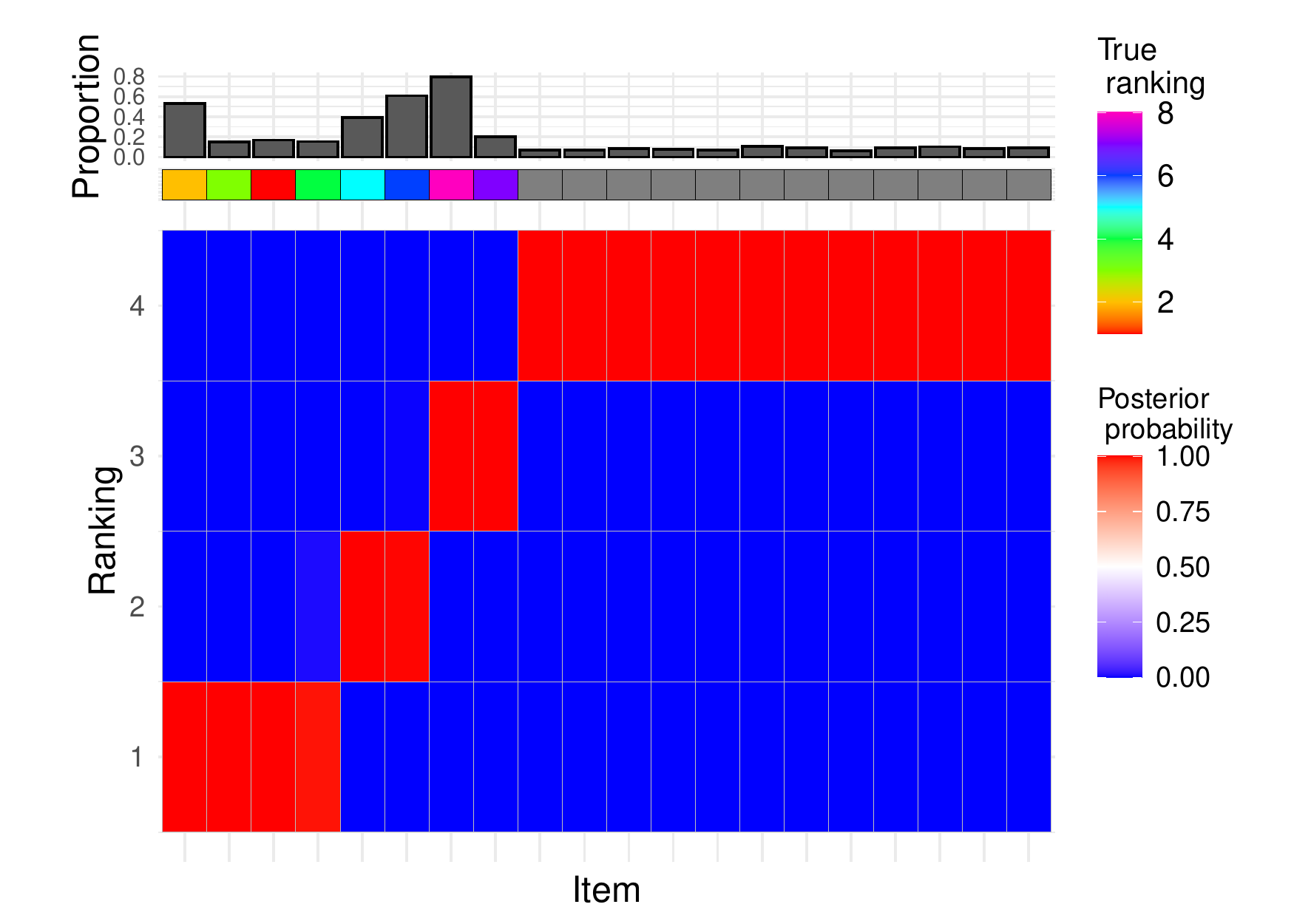}
\endminipage\hfill
\minipage{0.33\textwidth}%
  \includegraphics[width=\linewidth]{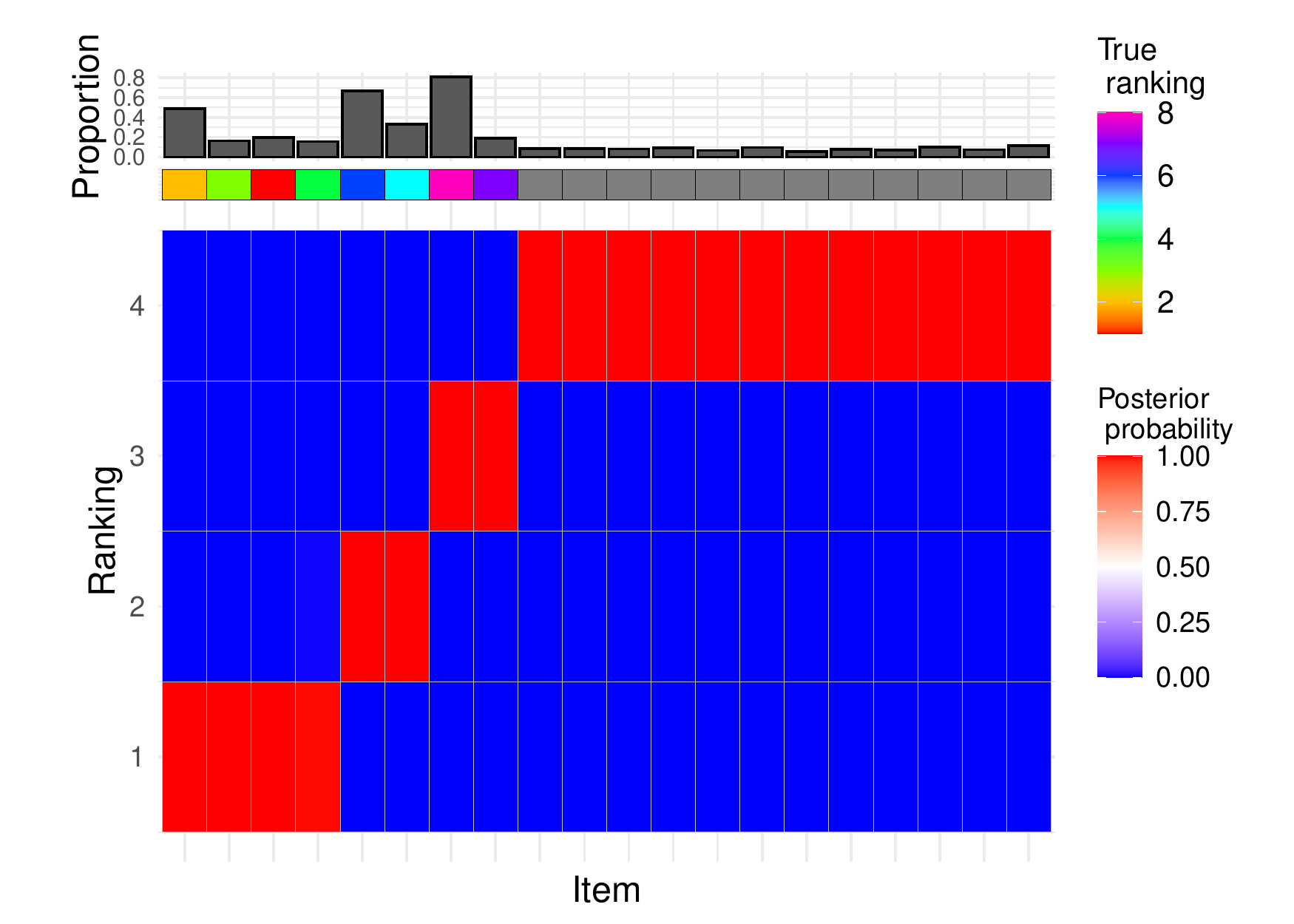}
\endminipage

\minipage{0.33\textwidth}
  \includegraphics[width=\linewidth]{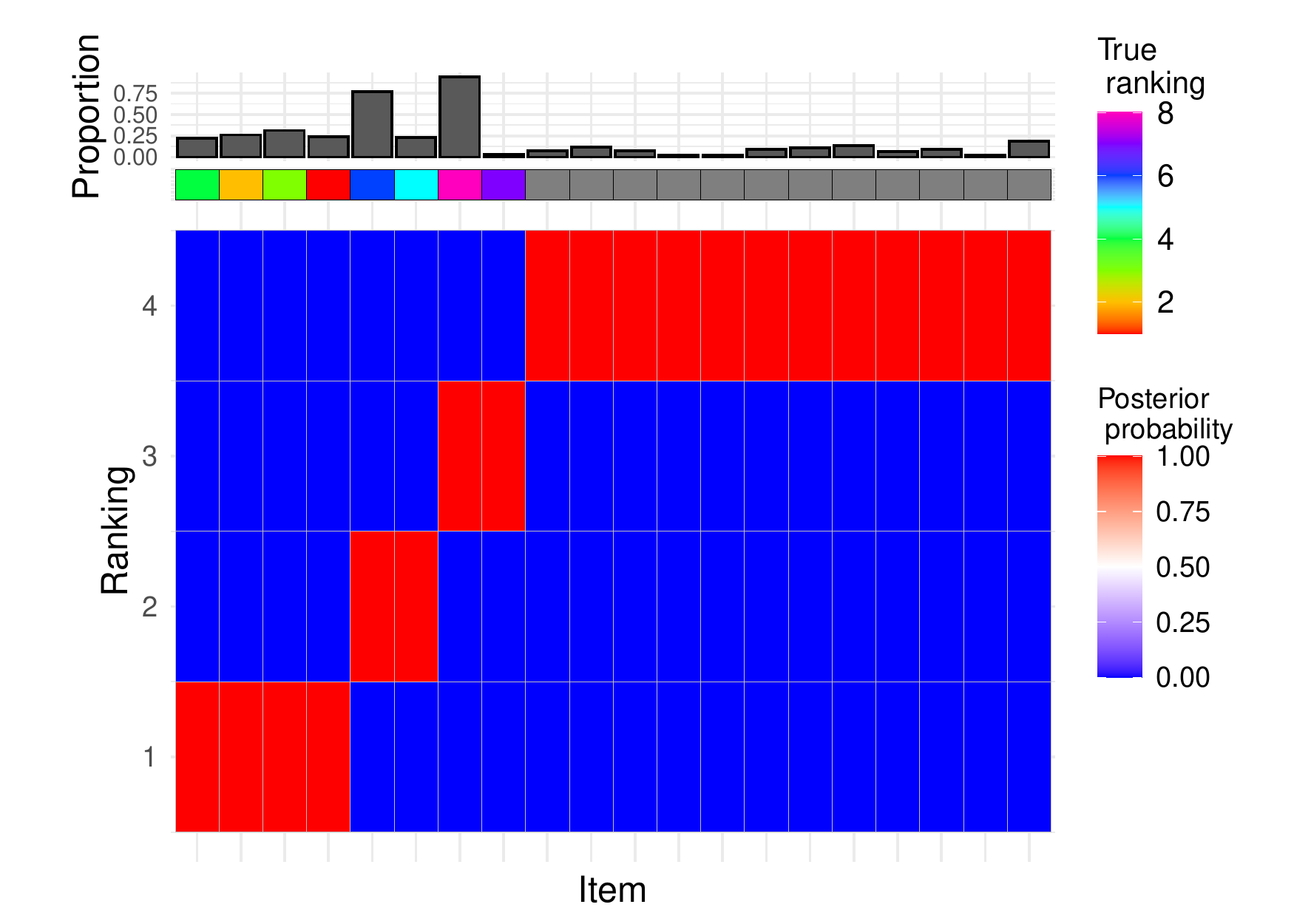}
\endminipage\hfill
\minipage{0.33\textwidth}
  \includegraphics[width=\linewidth]{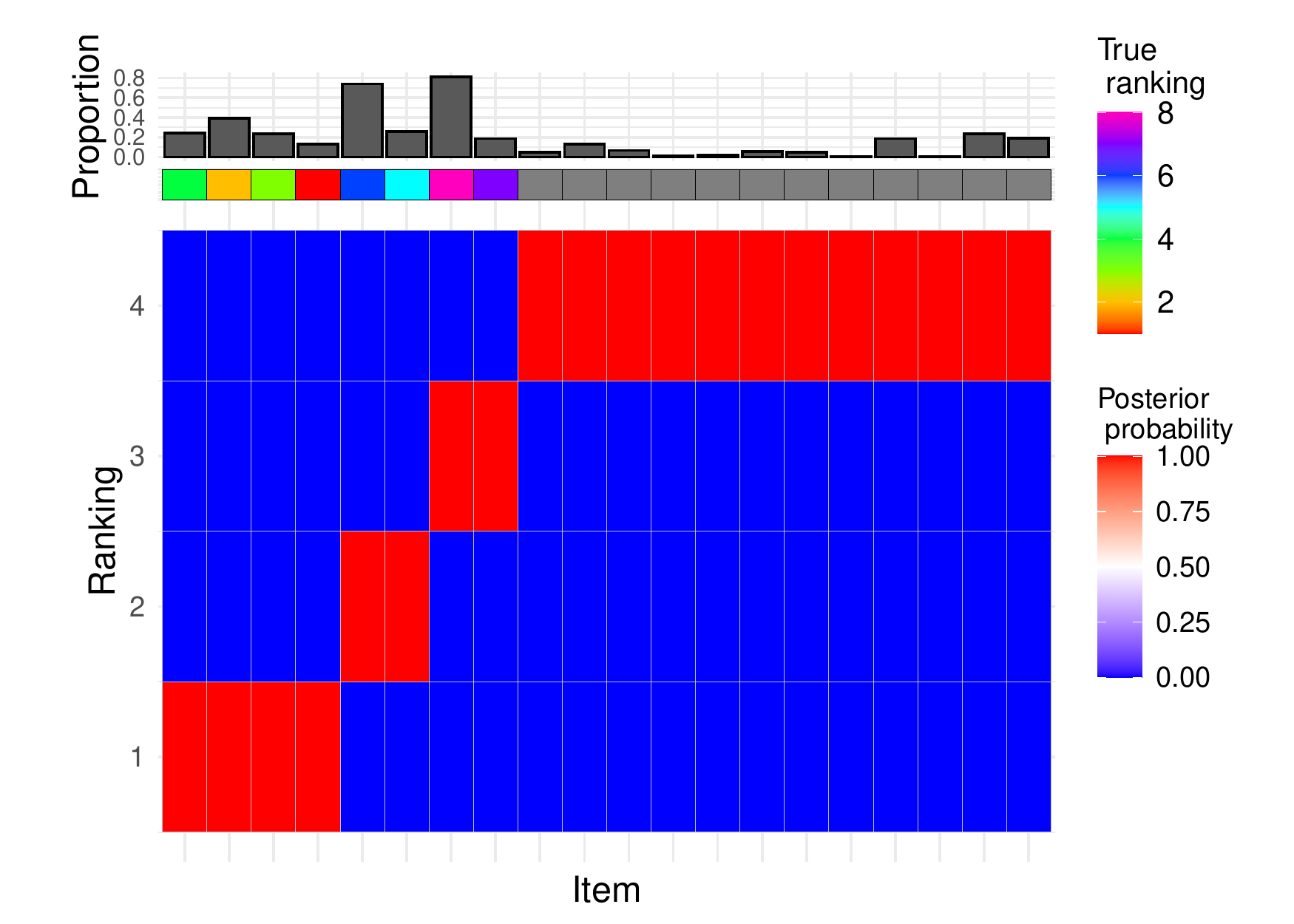}
\endminipage\hfill
\minipage{0.33\textwidth}%
  \includegraphics[width=\linewidth]{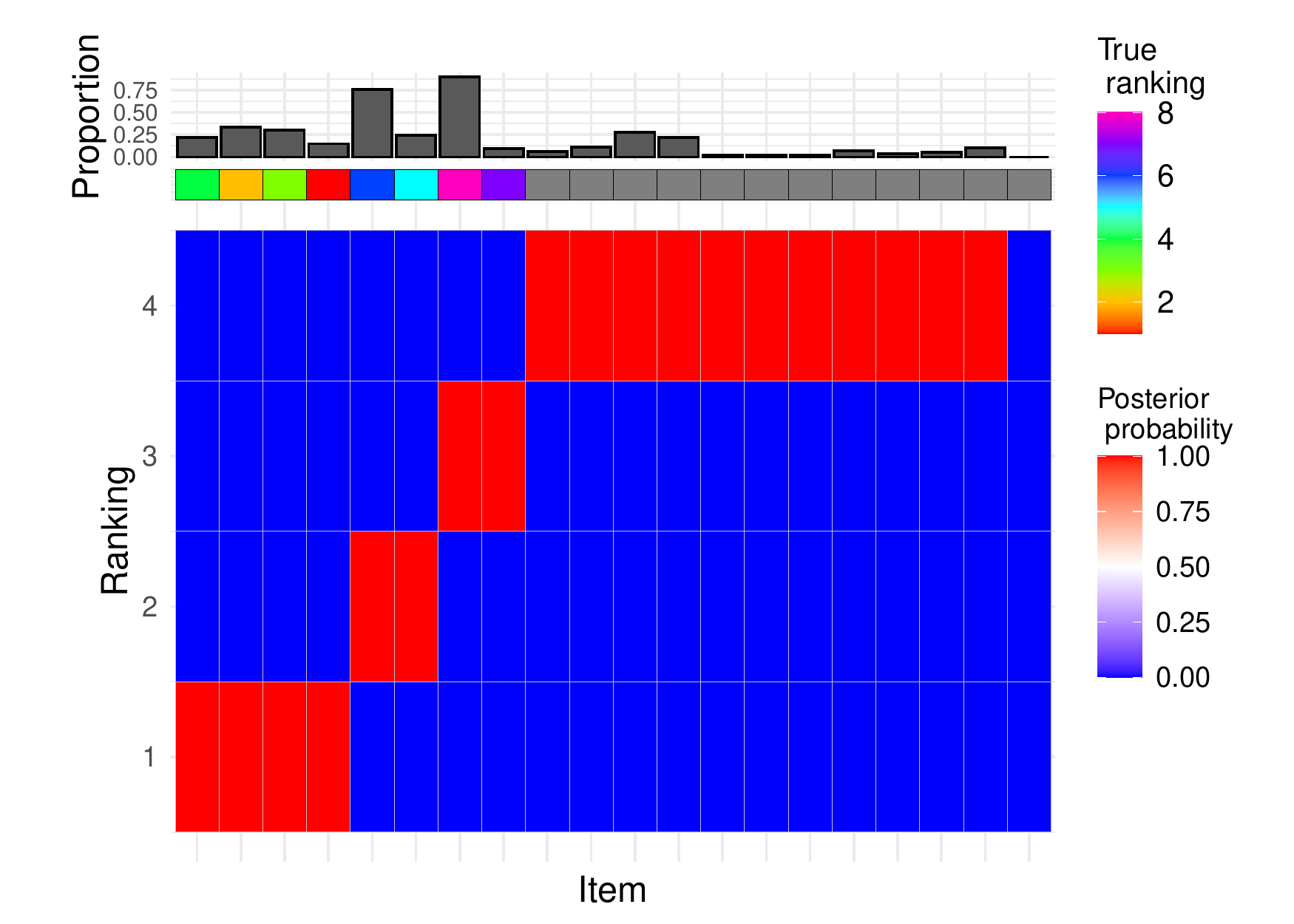}
\endminipage
\caption{Results from the top-rank simulation experiments described in Section 3.2 with $n^*_{\textnormal{true}}=8$, $n^*_{\textnormal{guess}}=4$: each panel displays the marginal posterior distribution of $\bm{\rho}$, where the items have been ordered according to $\bm{\Hat{\rho}}_{\mathcal{A}^*}$ on the x-axis. From left to right $l=1,2,3$, and from top to bottom $L=1,2,3$. The rainbow grid indicates the true $\bm{\rho}_{\mathcal{A}^*}$, and the bar plot indicates the proportion of times the items were selected in $\mathcal{A}^*$ over all MCMC iterations. $n=20$, $N=50$, $\alpha=10$.}
\end{figure}

\section{}\label{supp:alpha3_nstar_add_four}
\begin{figure}[H]
\minipage{0.33\textwidth}
  \includegraphics[width=\linewidth]{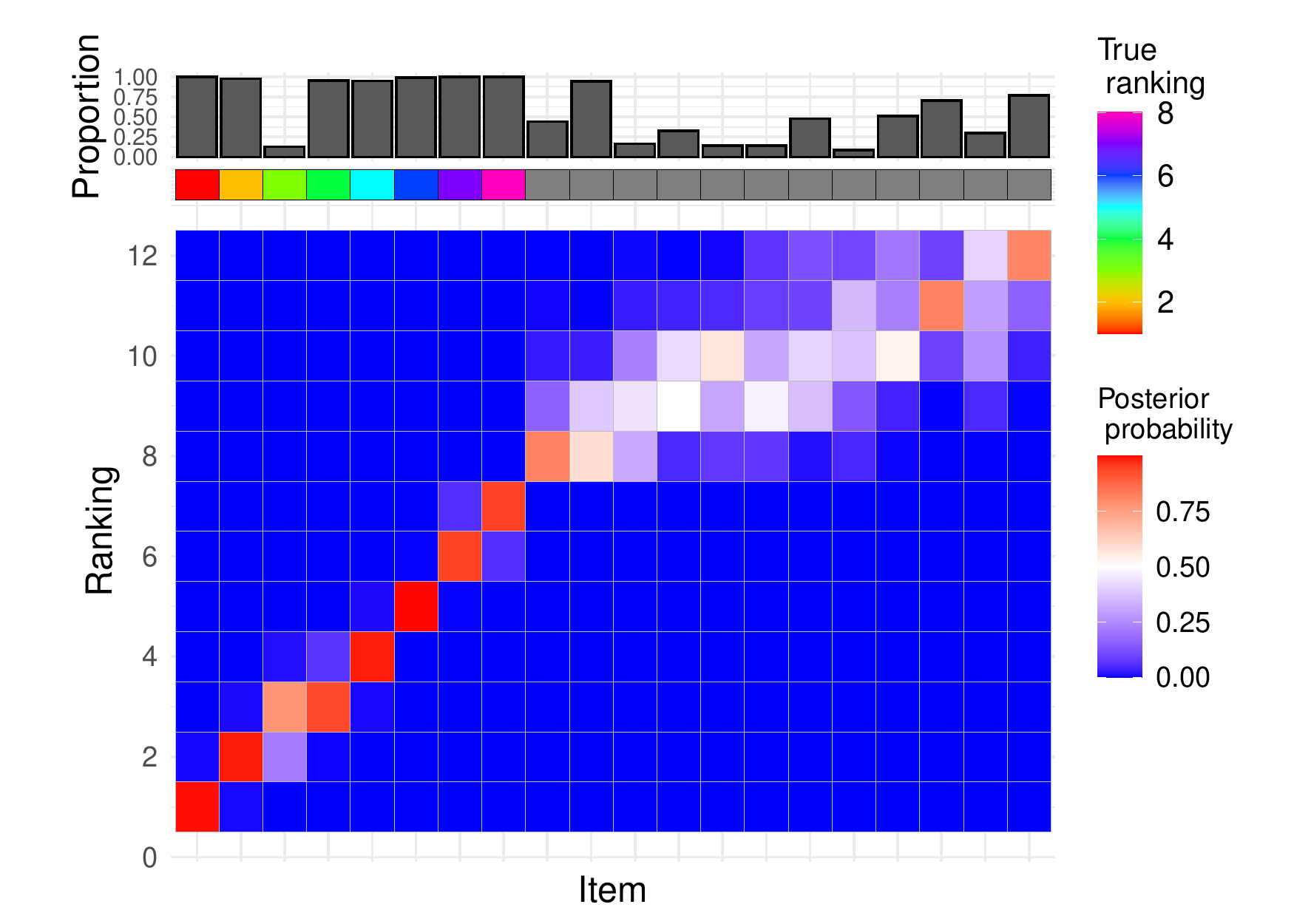}
\endminipage\hfill
\minipage{0.33\textwidth}
  \includegraphics[width=\linewidth]{fig/probitems_top20_item_selection_alphafixed3_simulation_top_items_nstartrue8_nstar12_L1_leap2_acceptance_v1.pdf}
\endminipage\hfill
\minipage{0.33\textwidth}%
  \includegraphics[width=\linewidth]{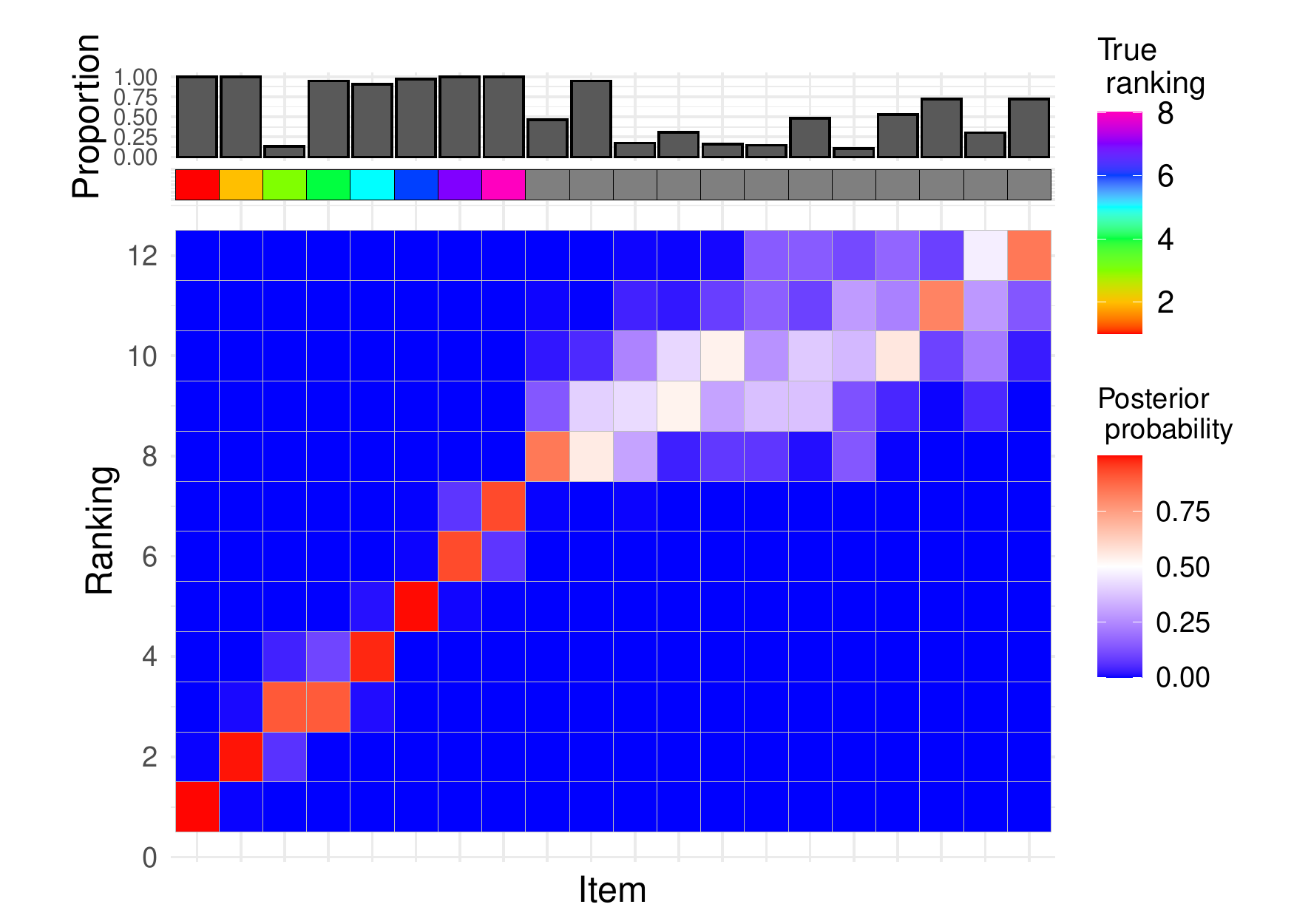}
\endminipage

\minipage{0.33\textwidth}
  \includegraphics[width=\linewidth]{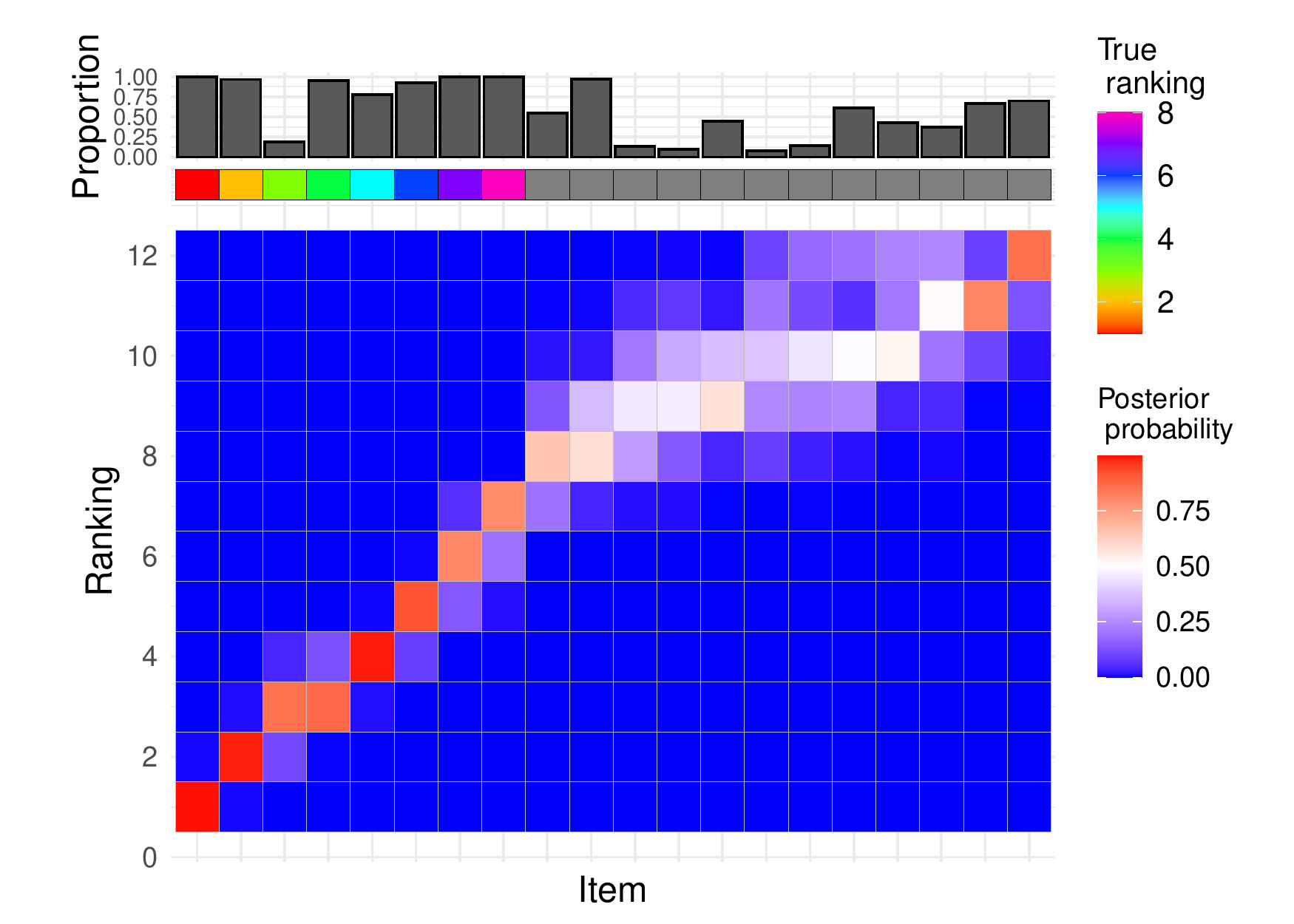}
\endminipage\hfill
\minipage{0.33\textwidth}
  \includegraphics[width=\linewidth]{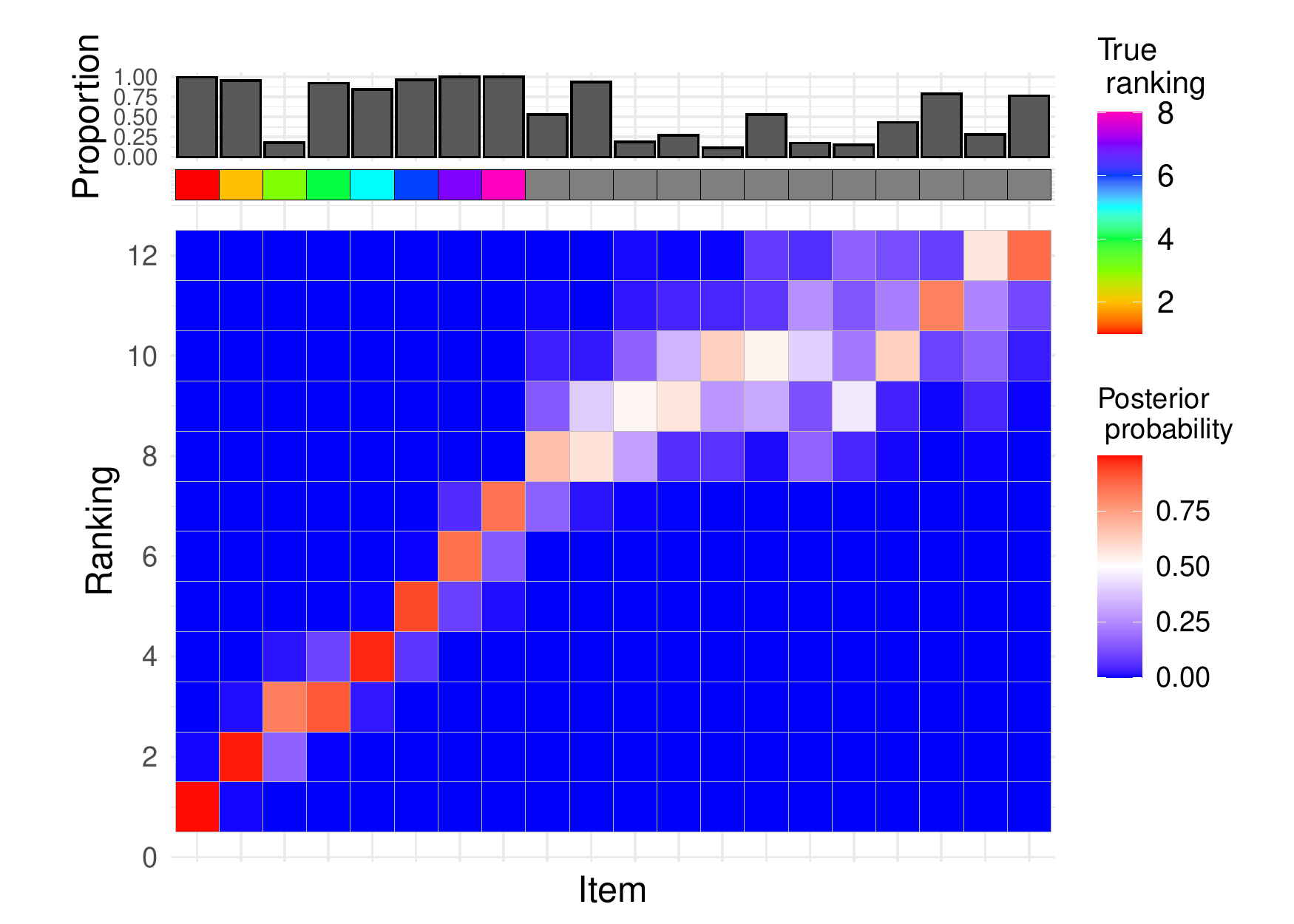}
\endminipage\hfill
\minipage{0.33\textwidth}%
  \includegraphics[width=\linewidth]{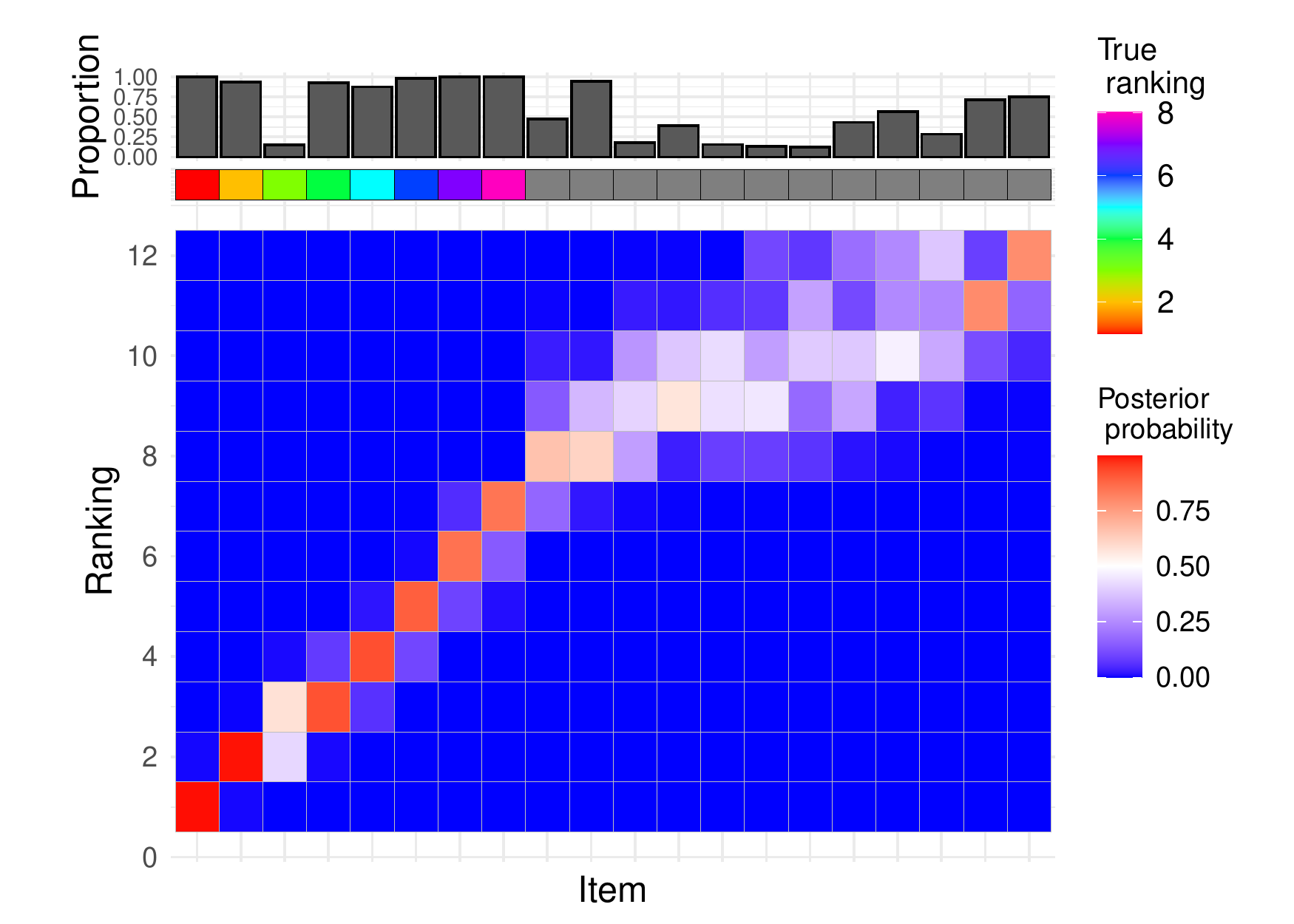}
\endminipage

\minipage{0.33\textwidth}
  \includegraphics[width=\linewidth]{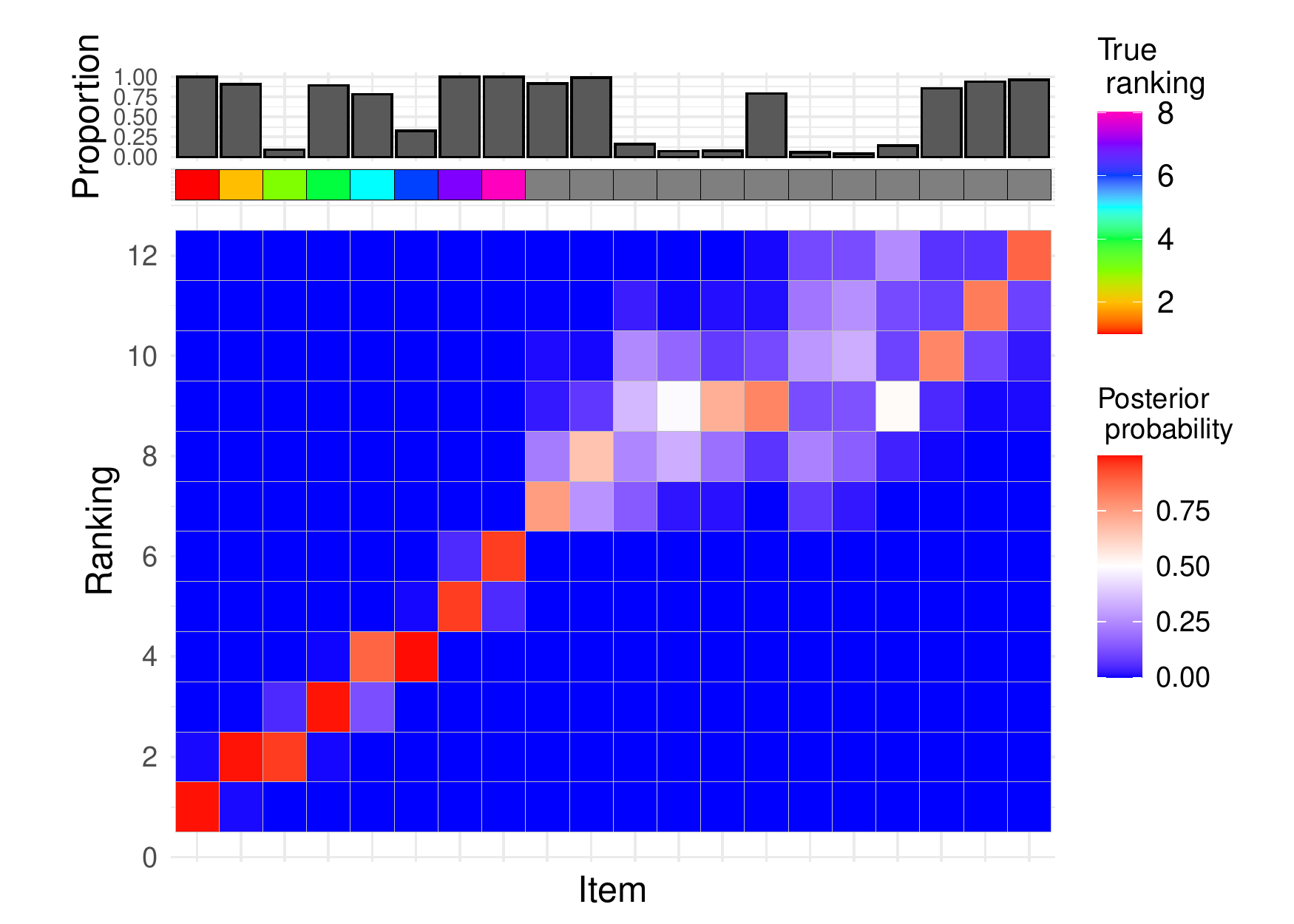}
\endminipage\hfill
\minipage{0.33\textwidth}
  \includegraphics[width=\linewidth]{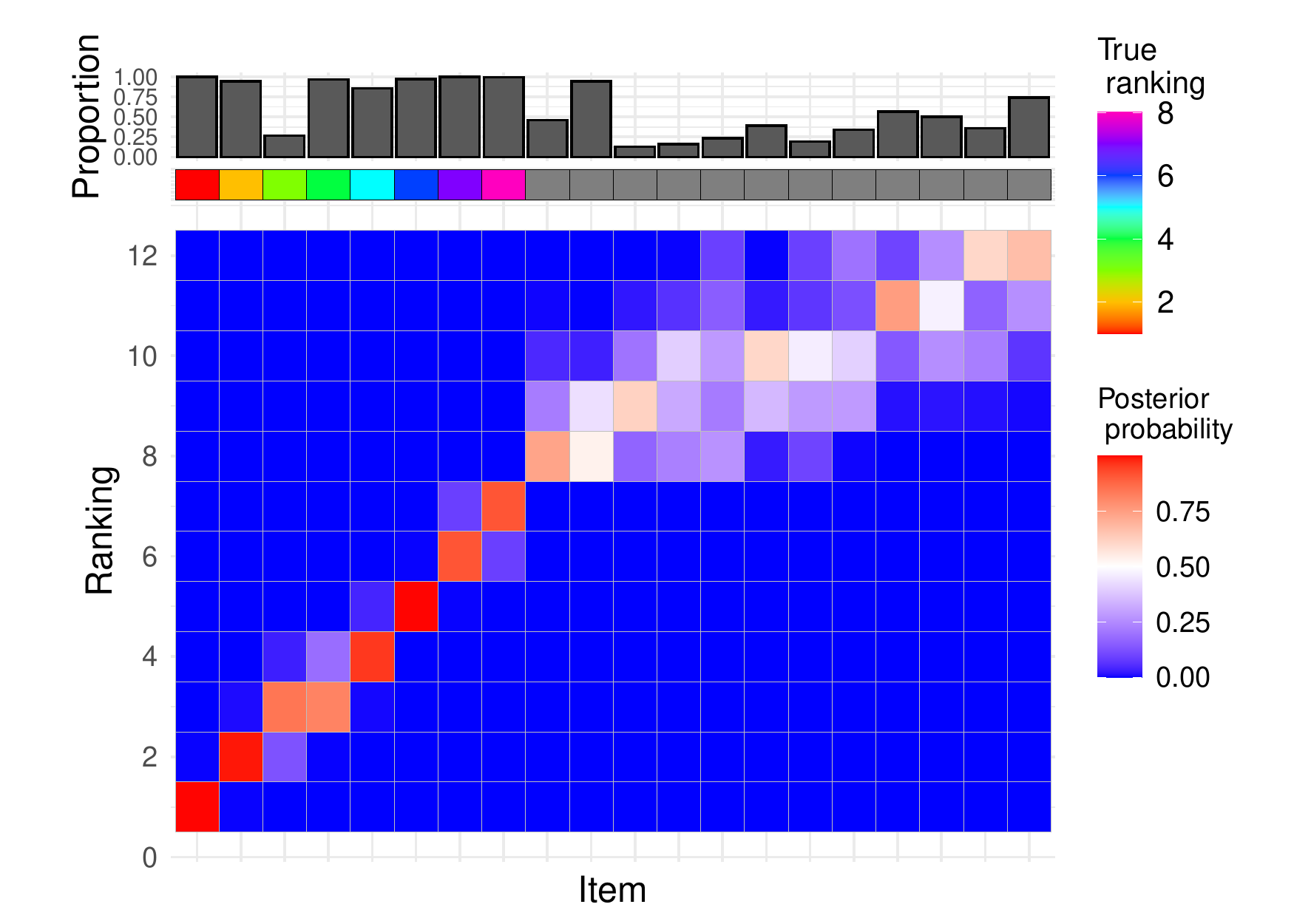}
\endminipage\hfill
\minipage{0.33\textwidth}%
  \includegraphics[width=\linewidth]{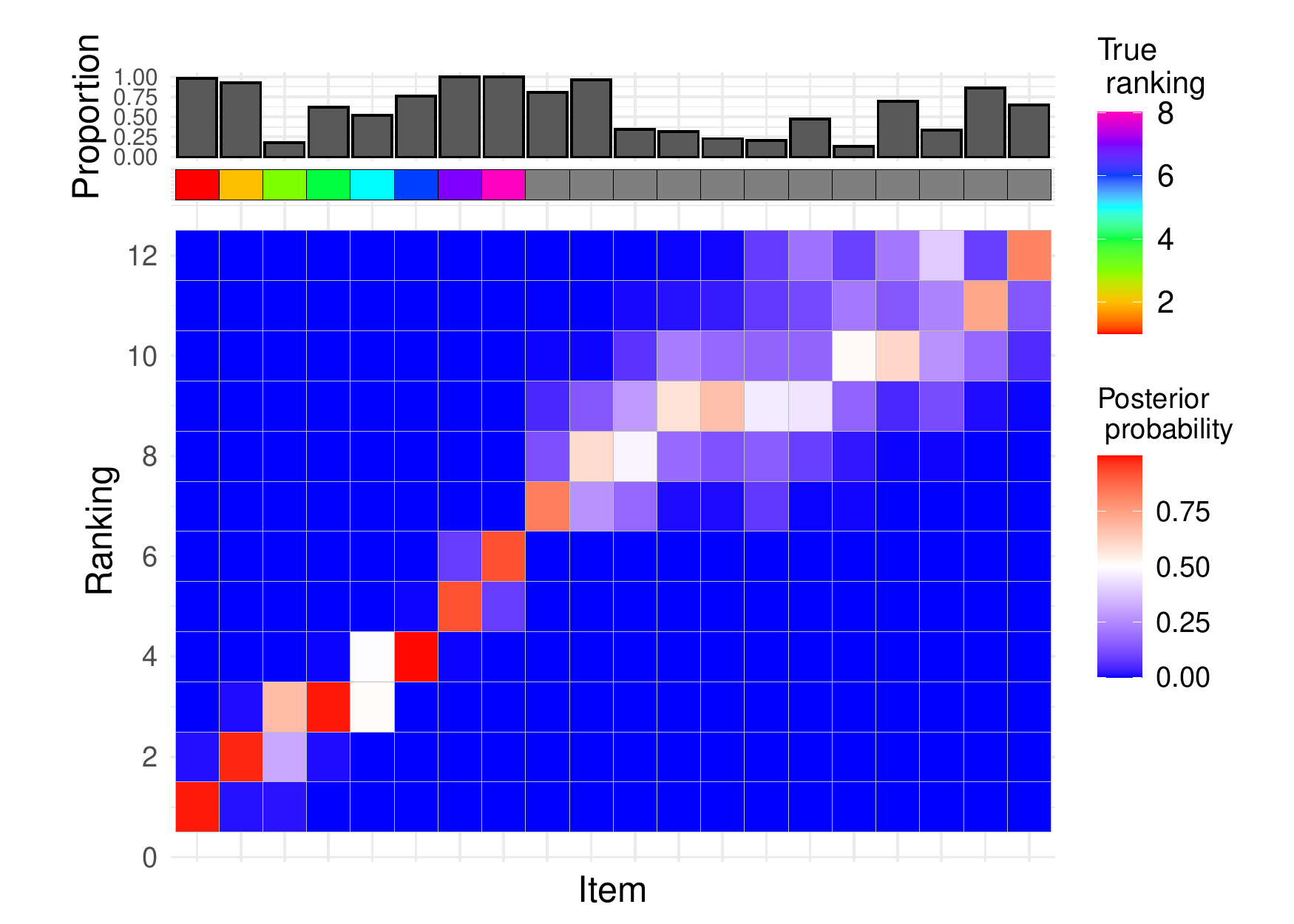}
\endminipage
\caption{Results from the top-rank simulation experiments described in Section 3.2 with $n^*_{\textnormal{true}}=8$, $n^*_{\textnormal{guess}}=12$: each panel displays the marginal posterior distribution of $\bm{\rho}$, where the items have been ordered according to $\bm{\Hat{\rho}}_{\mathcal{A}^*}$ on the x-axis. From left to right $l=1,2,3$, and from top to bottom $L=1,2,3$. The rainbow grid indicates the true $\bm{\rho}_{\mathcal{A}^*}$, and the bar plot indicates the proportion of times the items were selected in $\mathcal{A}^*$ over all MCMC iterations. $n=20$, $N=50$, $\alpha=3$.}
\end{figure}

\section{}\label{supp:alpha10_nstar_add_four}
\begin{figure}[H]
\minipage{0.33\textwidth}
  \includegraphics[width=\linewidth]{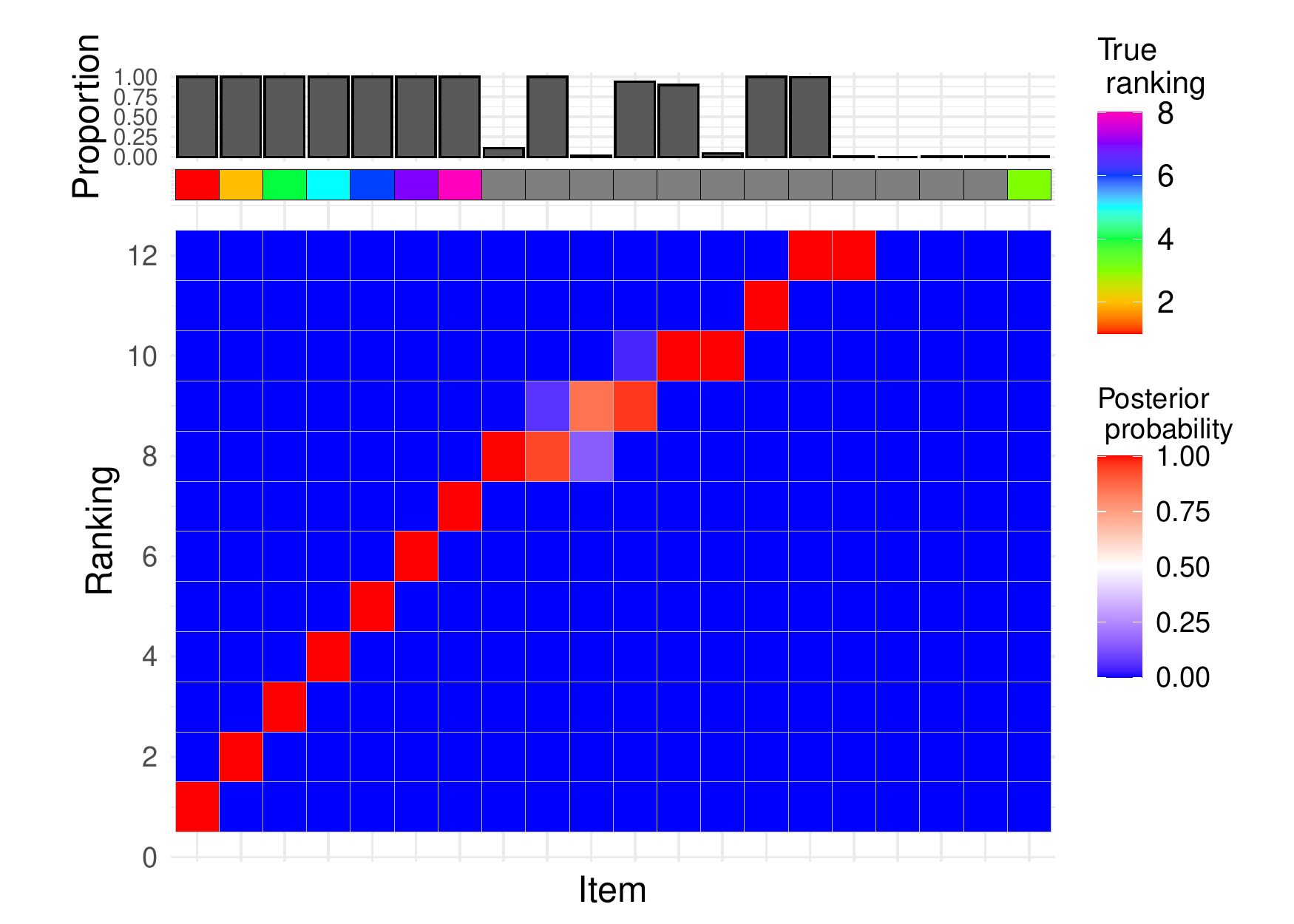}
\endminipage\hfill
\minipage{0.33\textwidth}
  \includegraphics[width=\linewidth]{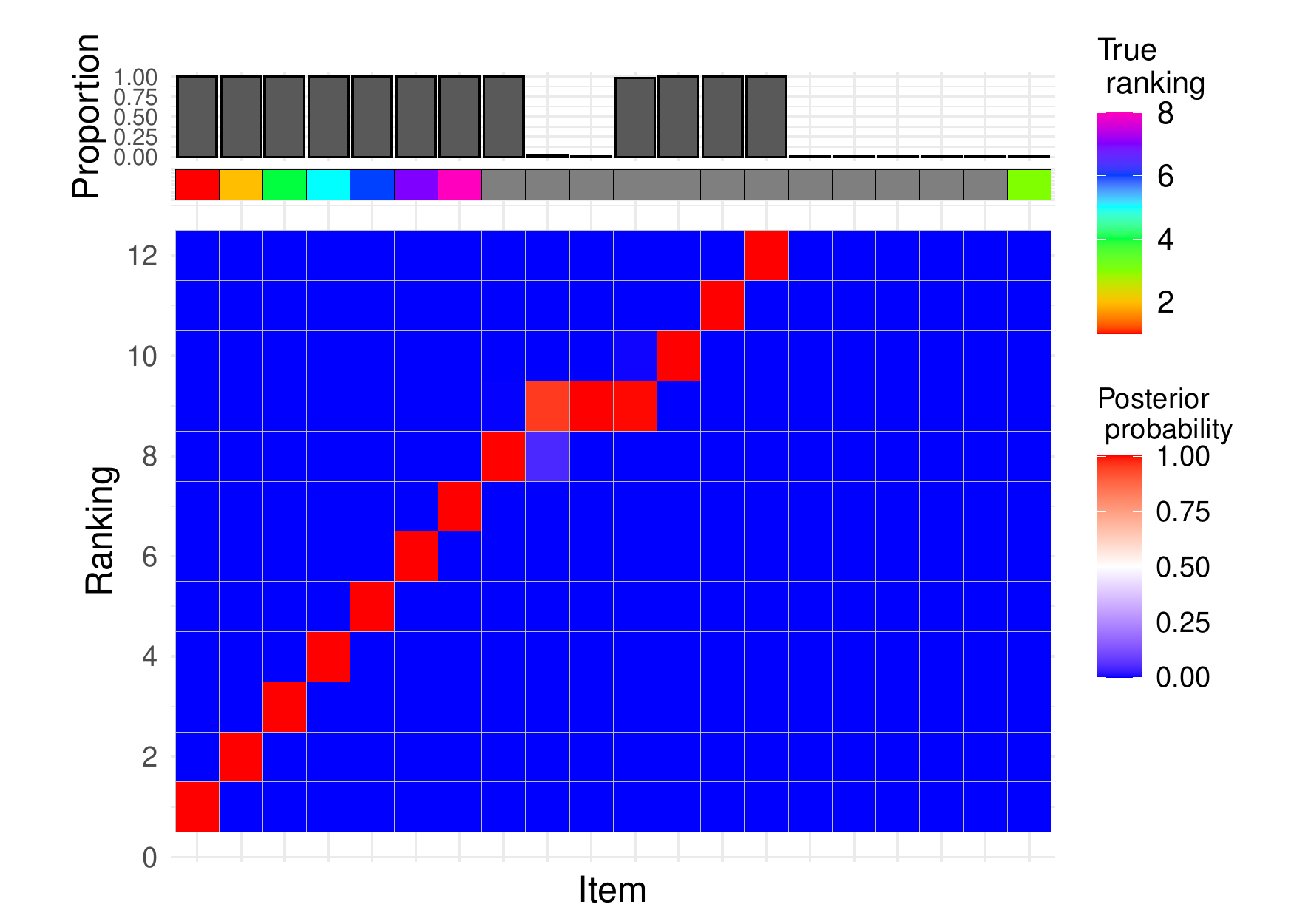}
\endminipage\hfill
\minipage{0.33\textwidth}%
  \includegraphics[width=\linewidth]{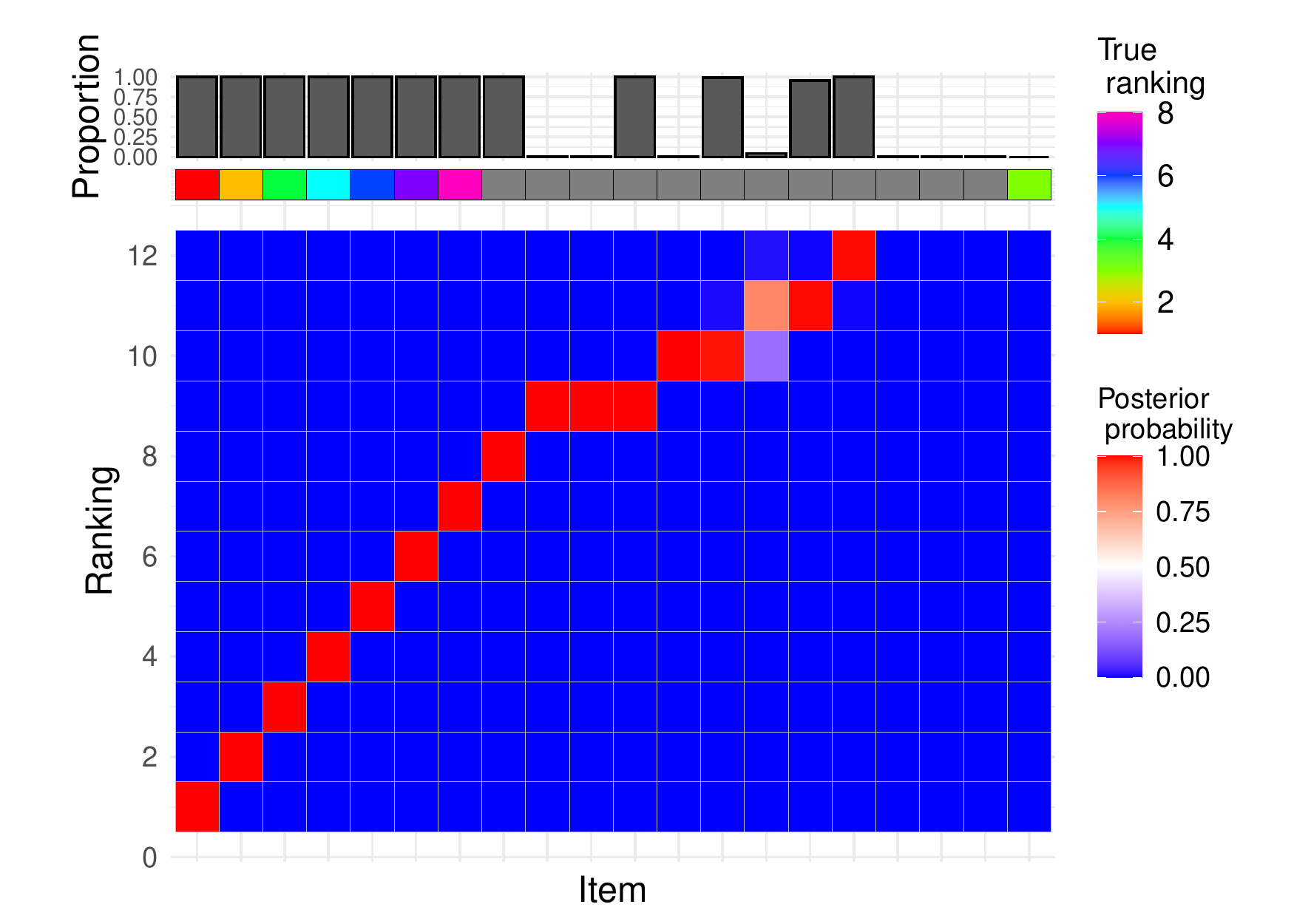}
\endminipage

\minipage{0.33\textwidth}
  \includegraphics[width=\linewidth]{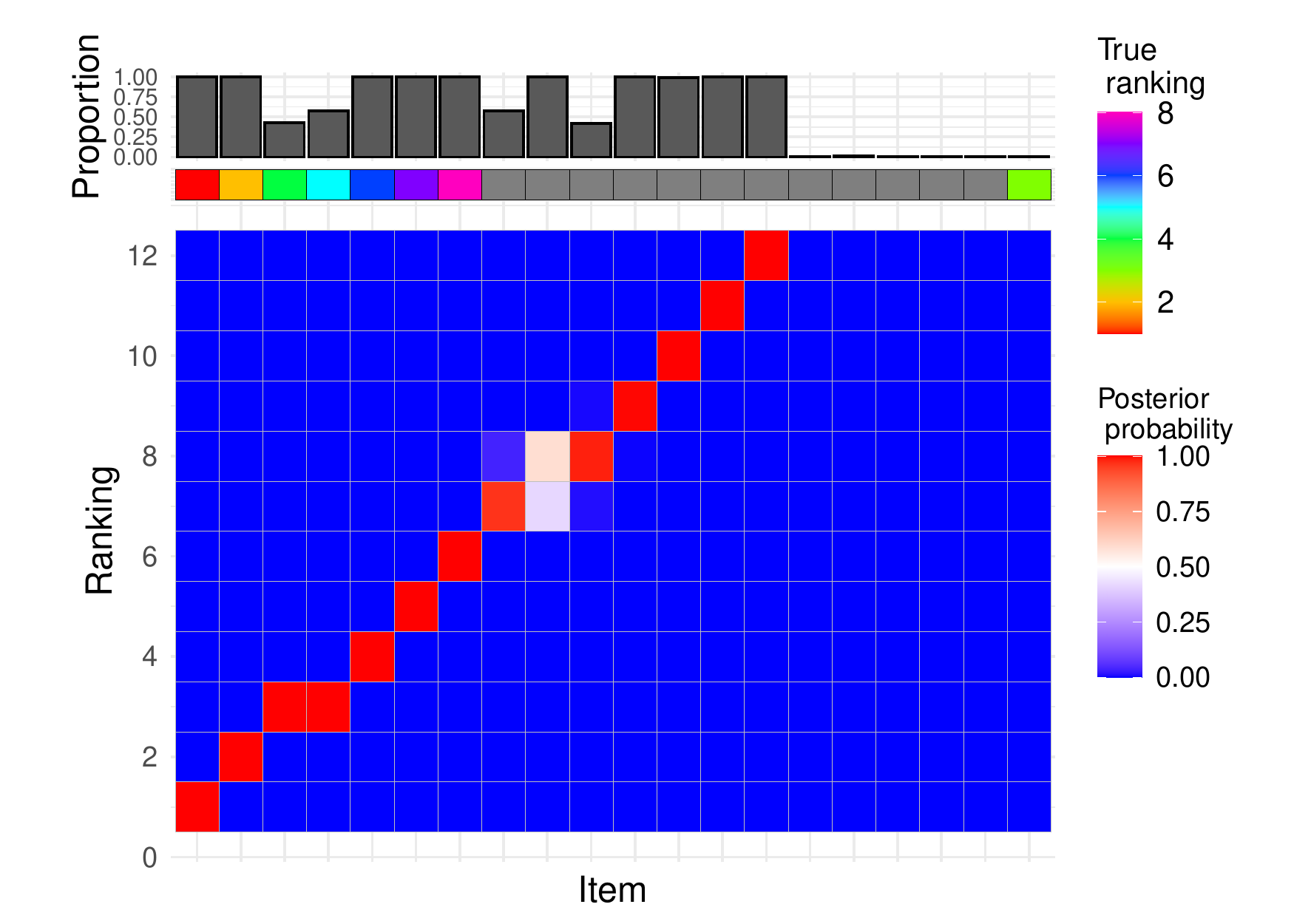}
\endminipage\hfill
\minipage{0.33\textwidth}
  \includegraphics[width=\linewidth]{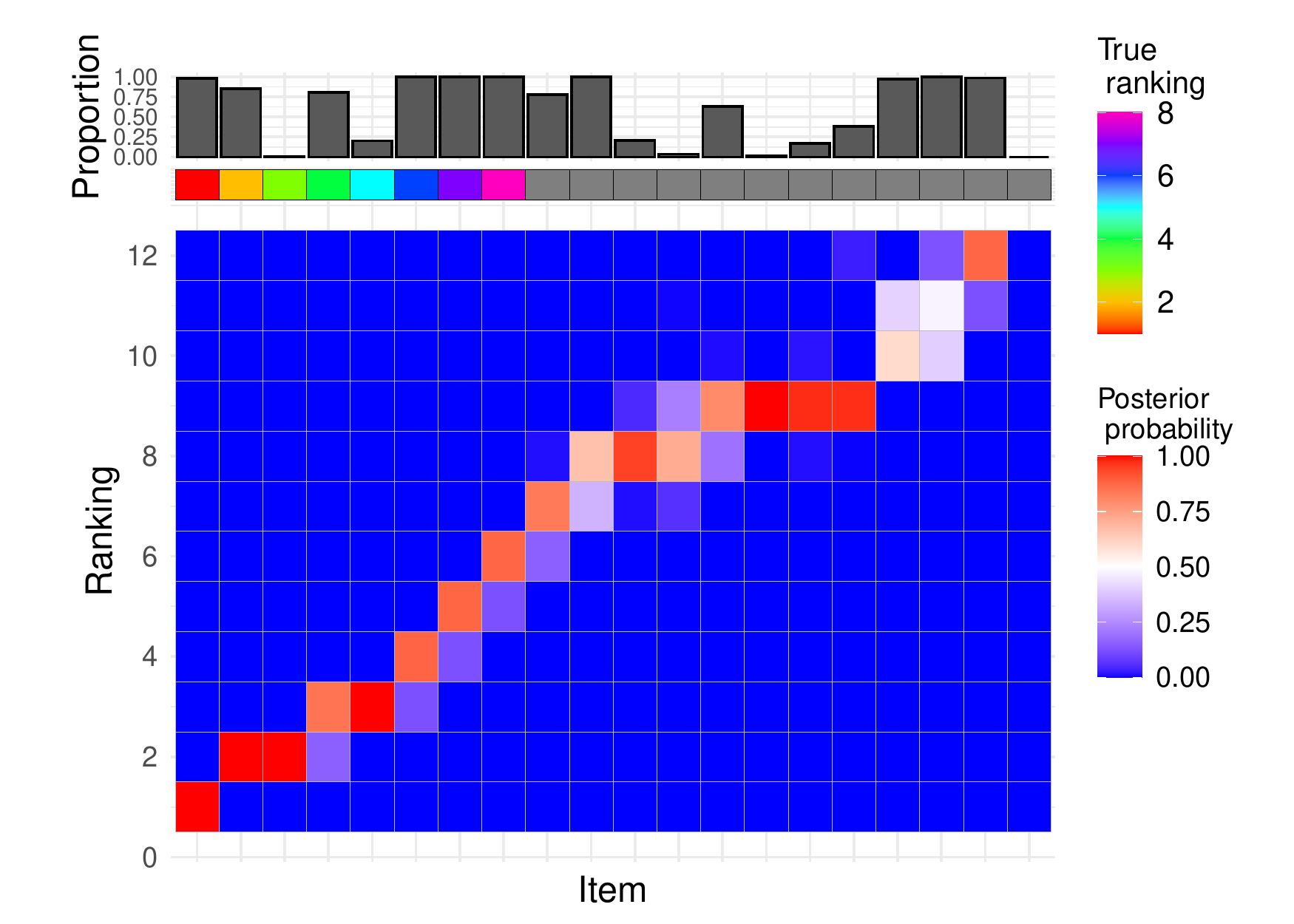}
\endminipage\hfill
\minipage{0.33\textwidth}%
  \includegraphics[width=\linewidth]{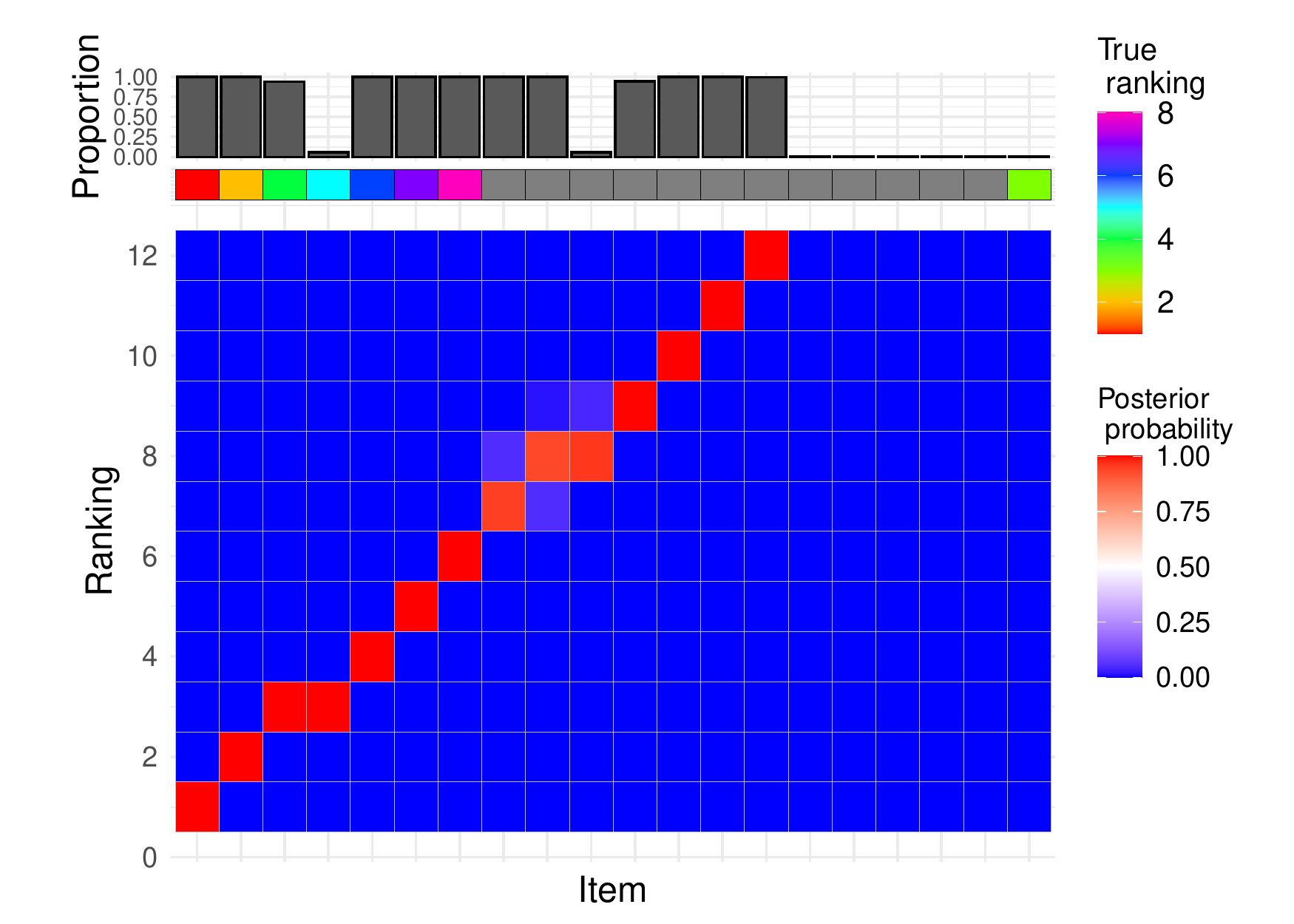}
\endminipage

\minipage{0.33\textwidth}
  \includegraphics[width=\linewidth]{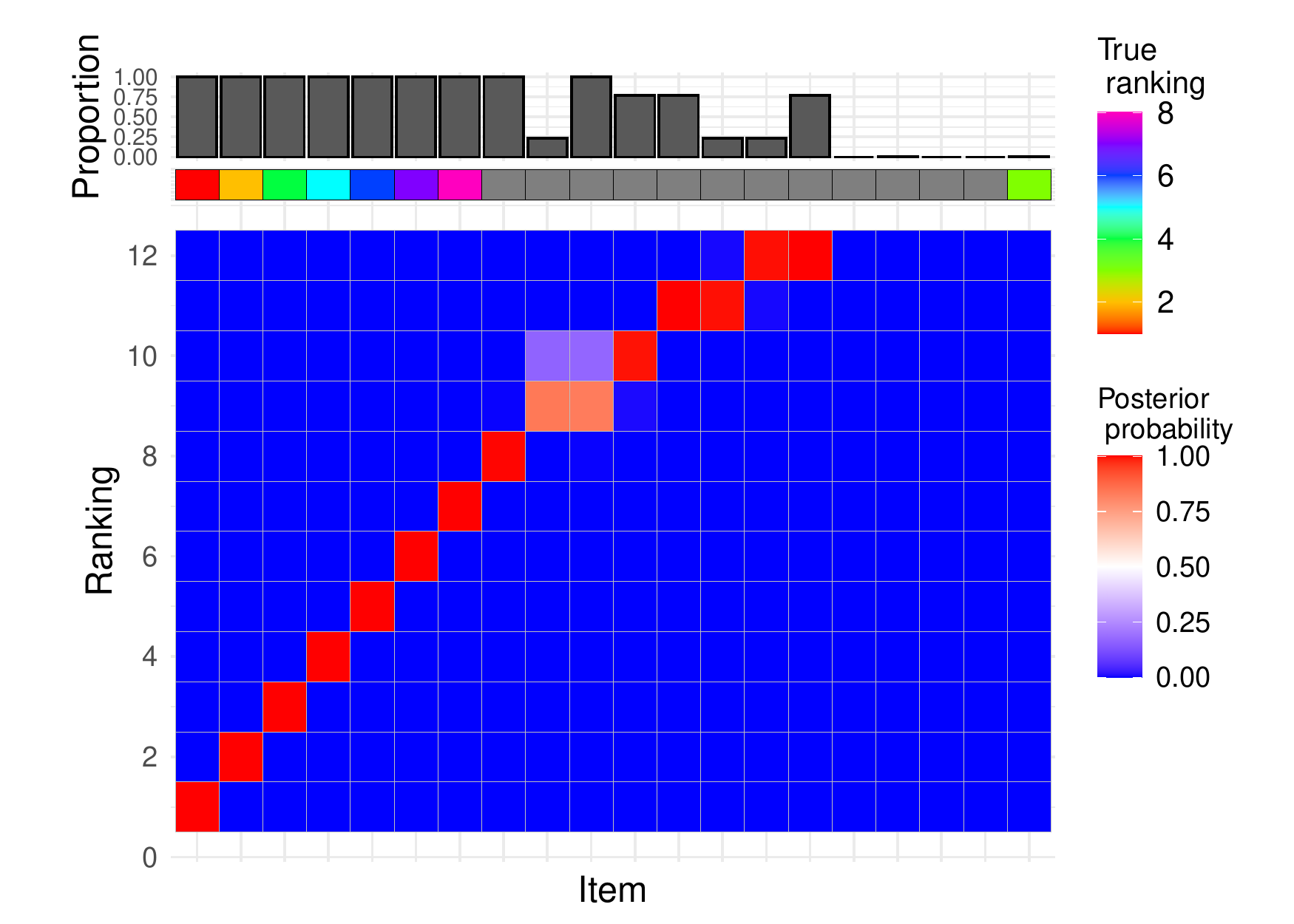}
\endminipage\hfill
\minipage{0.33\textwidth}
  \includegraphics[width=\linewidth]{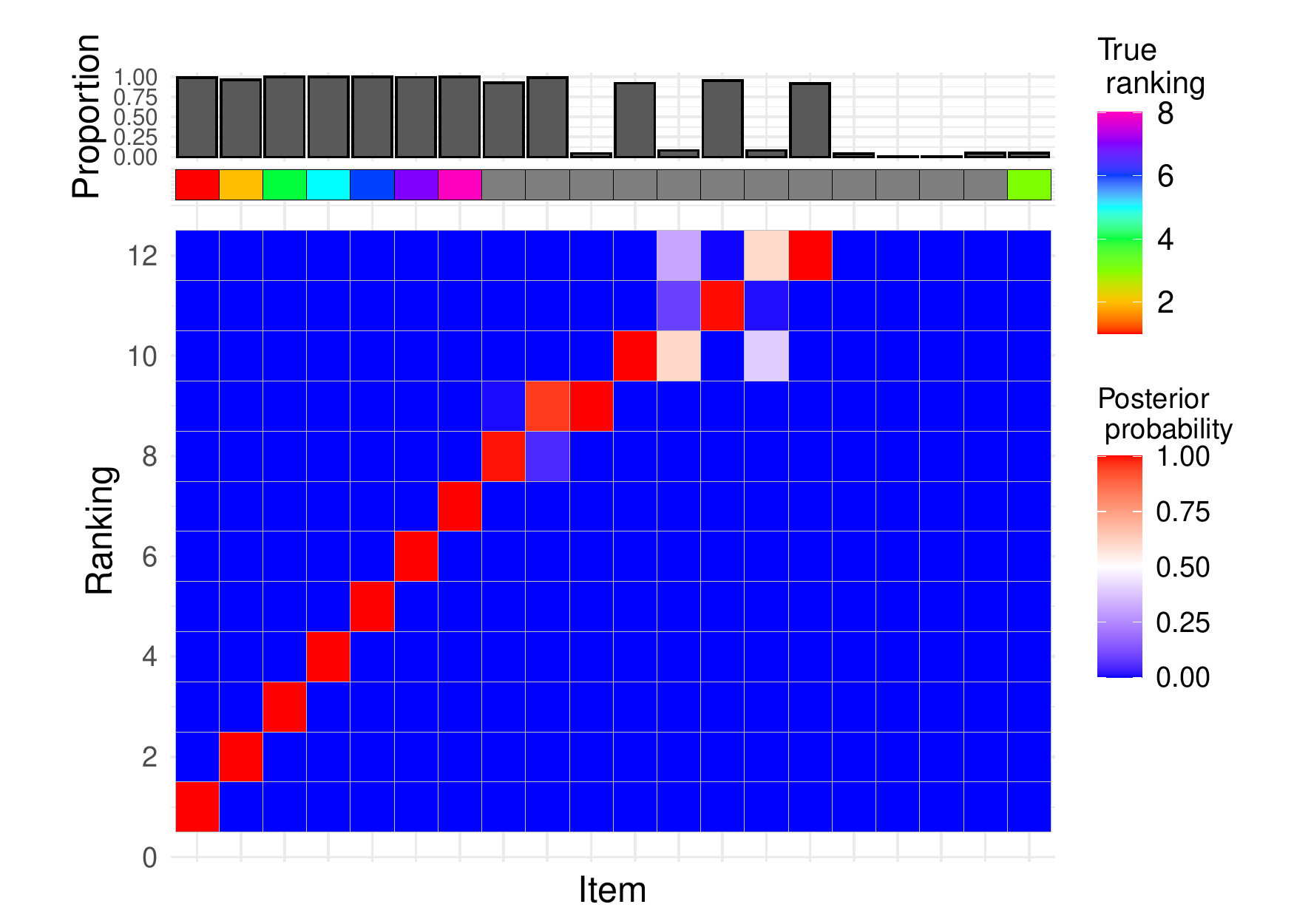}
\endminipage\hfill
\minipage{0.33\textwidth}%
  \includegraphics[width=\linewidth]{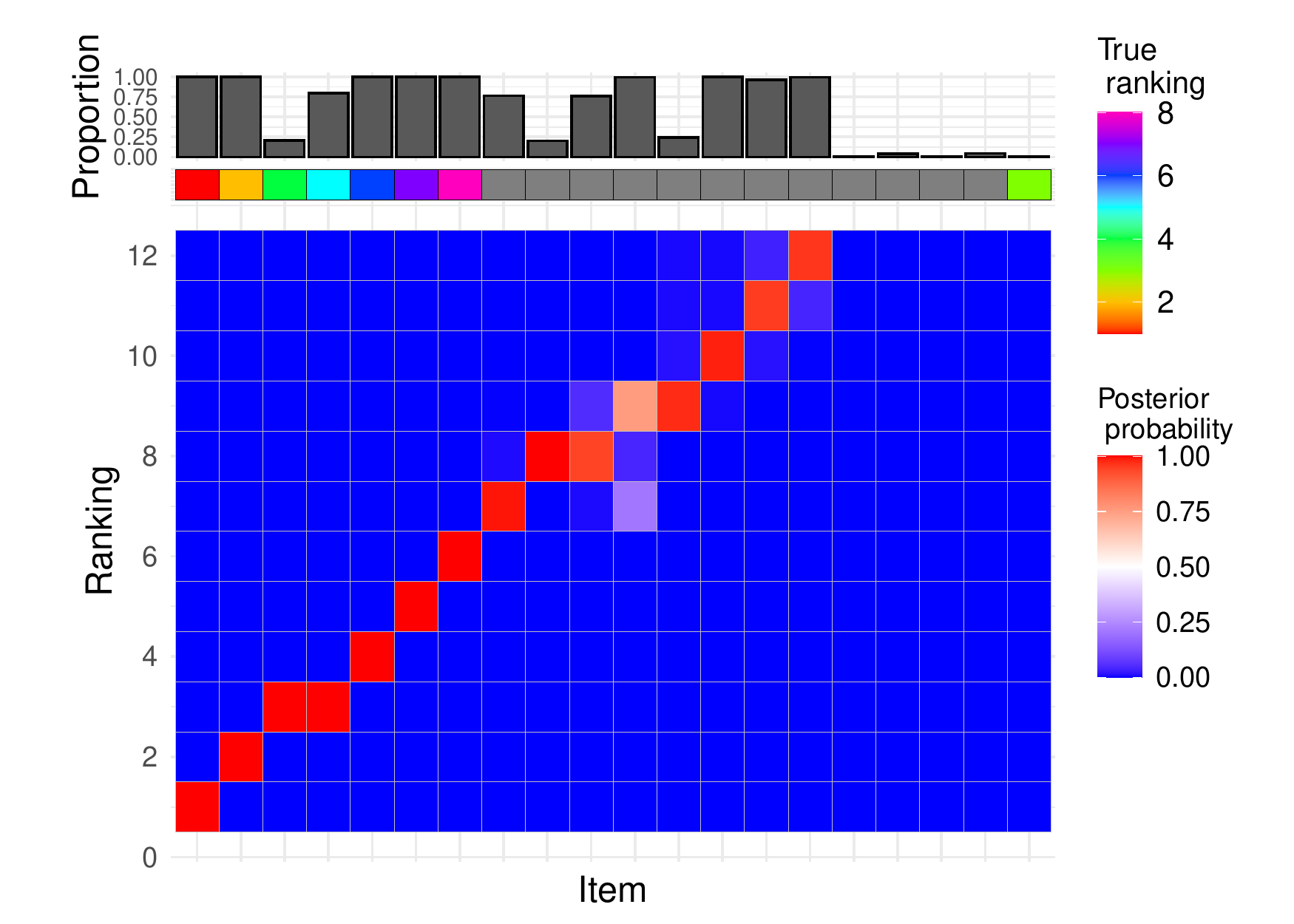}
\endminipage
\caption{Results from the top-rank simulation experiments described in Section 3.2 with $n^*_{\textnormal{true}}=8$, $n^*_{\textnormal{guess}}=12$: each panel displays heatplots of the marginal posterior distribution of $\bm{\rho}$, where the items have been ordered according to $\bm{\Hat{\rho}}_{\mathcal{A}^*}$ on the x-axis. From left to right $l=1,2,3$, and from top to bottom $L=1,2,3$. The rainbow grid indicates the true $\bm{\rho}_{\mathcal{A}^*}$, and the bar plot indicates the proportion of times the items were selected in $\mathcal{A}^*$ over all MCMC iterations. $n=20$, $N=50$, $\alpha=10$.}
\end{figure}

\section{}\label{supp:alpha10_sim_big_trace}
\begin{figure}[H]
\minipage{0.48\textwidth}
  \includegraphics[width=\linewidth]{fig/traceplot_top15_items_sim_study_big_alphafixed10_nstartrue50_nstar50_L1_leap10_acceptance_v1.pdf}
\endminipage\hfill
\minipage{0.48\textwidth}
  \includegraphics[width=\linewidth]{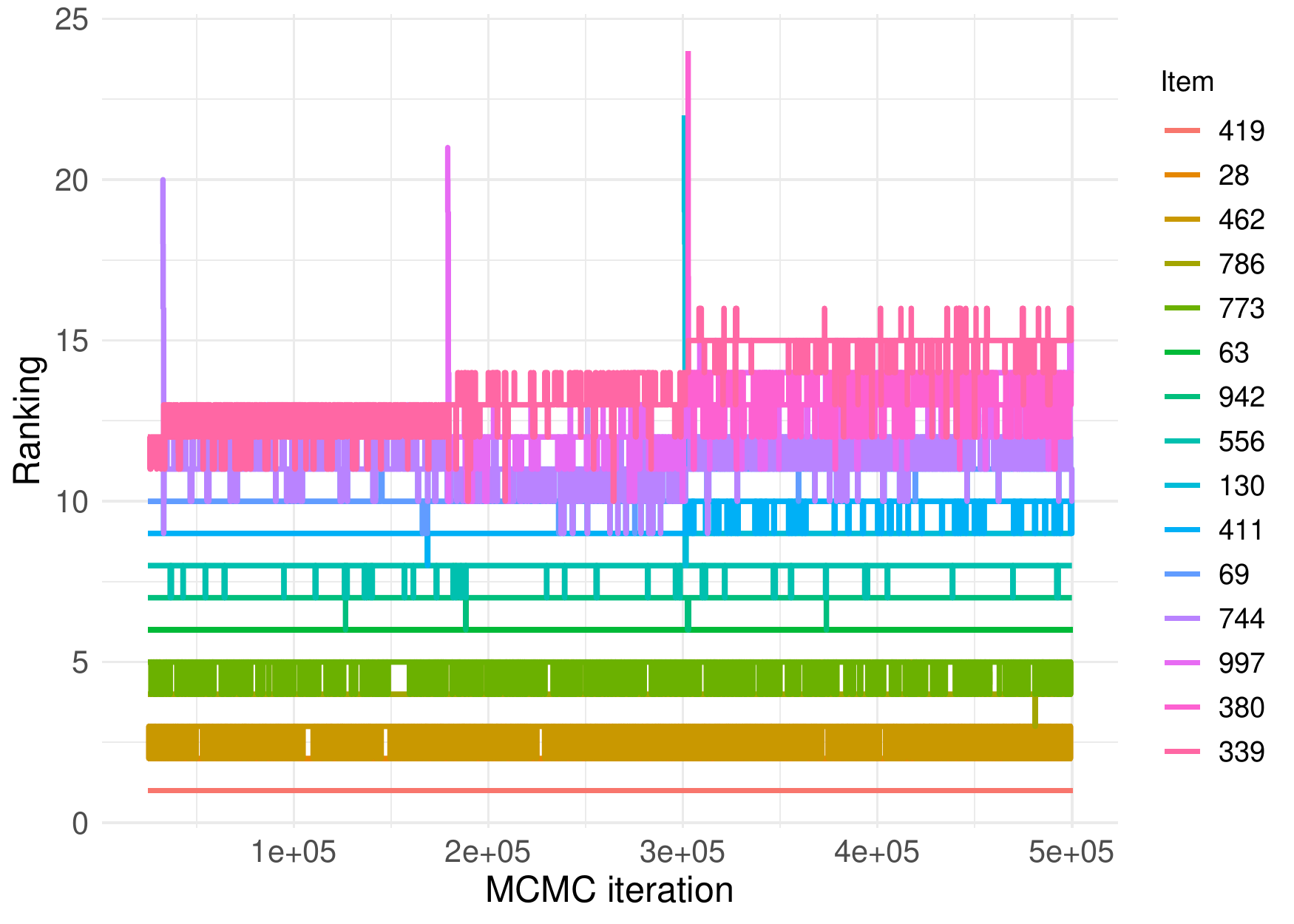}
\endminipage
\caption{Results from the top-rank simulation experiments described in Section 3.1 with $n^*=50$. Each panel displays the trace plot of the top-15 items in $\bm{\rho}_{\mathcal{A}^*}$. From left to right: $L=1$ and $L=5$. $n=1000$, $N=50$, $\alpha=10$ and $l=\textnormal{round}(n^*/5)$. }
\end{figure}

\section{}\label{supp:alpha3_fullmod_vs_subset}
\begin{figure}[H]
\minipage{0.48\textwidth}
  \includegraphics[width=\linewidth]{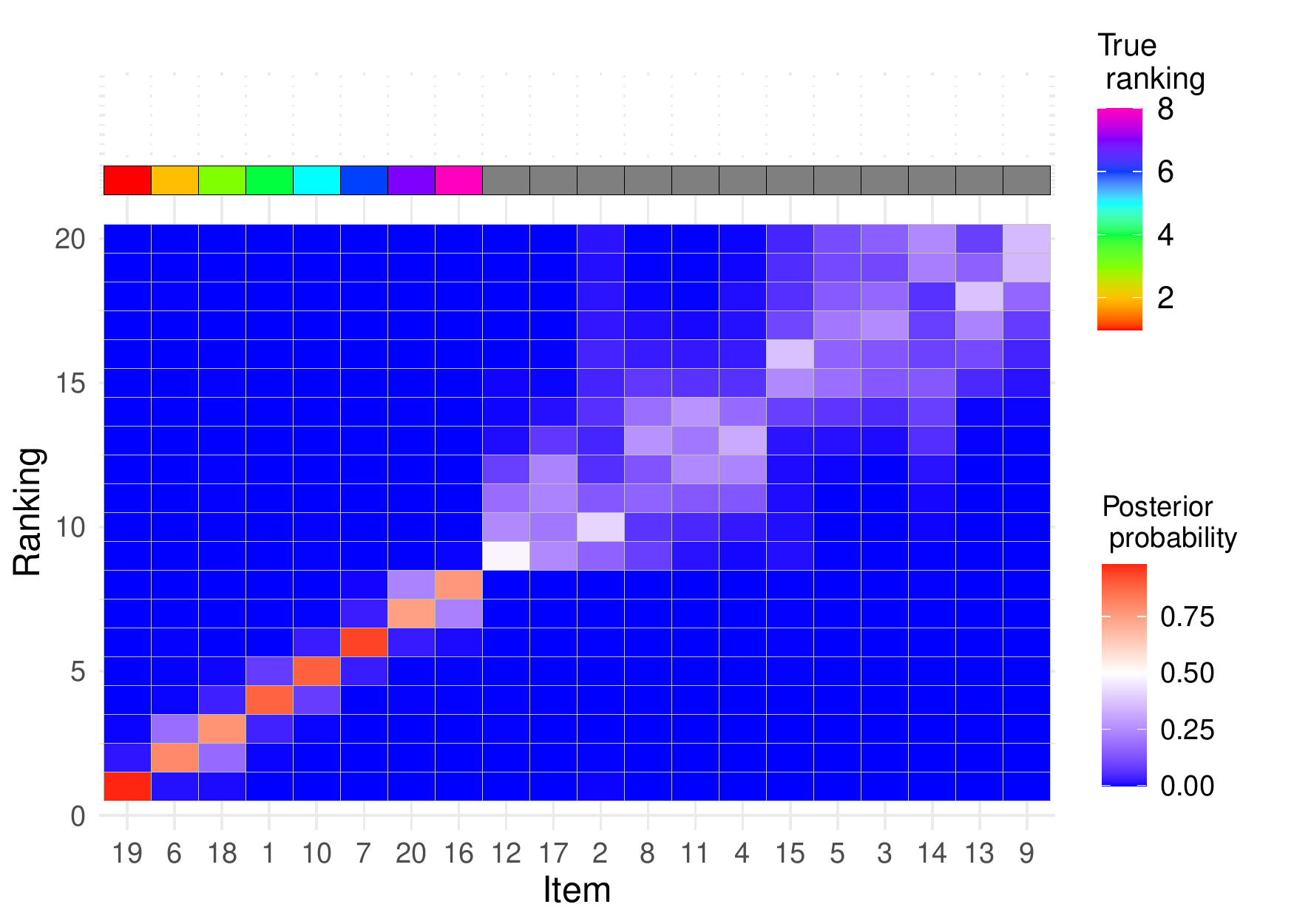}
\endminipage\hfill
\minipage{0.48\textwidth}
  \includegraphics[width=\linewidth]{fig/probitems_top20_item_selection_alphafixed3_simulation_top_items_nstartrue8_nstar8_L1_leap2_acceptance_v1_with_itemlabels.pdf}
\endminipage
\caption{Results from the simulation experiments described in Section 3.4: comparison of the \texttt{BayesMallows} model on the complete set of items to lowBMM with $n^*=8$, $n=20$, $N=50$. Heatplots of the marginal posterior distribution of $\bm{\rho}$, where the items have been ordered according to $\bm{\Hat{\rho}}_{\mathcal{A}^*}$ on the x-axis, for BMM on the left and lowBMM on the right. The rainbow grid indicates the true $\bm{\rho}_{\mathcal{A}^*}$, and the bar plot indicates the proportion of times the items were selected in $\mathcal{A}^*$ over all MCMC iterations. Parameters for lowBMM: $\alpha=3$, $L=1$, $l=2$. }
\end{figure}